\let\ORIlabel\label
\let\ORIrefstepcounter\refstepcounter
\AddToHook{package/hyperref/before}{%
  \let\label\ORIlabel
  \let\refstepcounter\ORIrefstepcounter
}
\documentclass[hidelinks,onefignum,onetabnum]{siamart220329}

\usepackage{amsmath,amsfonts,bm}

\def\eqref#1{(\ref{#1})}

\newcommand{\dif}{{\mathrm{d}}}

\def\mI{{\bm{I}}}

\DeclareMathAlphabet{\mathsfit}{\encodingdefault}{\sfdefault}{m}{sl}
\SetMathAlphabet{\mathsfit}{bold}{\encodingdefault}{\sfdefault}{bx}{n}

\makeatletter
\DeclareRobustCommand{\cev}[1]{%
  {\mathpalette\do@cev{#1}}%
}
\newcommand{\do@cev}[2]{%
  \vbox{\offinterlineskip
    \sbox\z@{$\m@th#1 x$}%
    \ialign{##\cr
      \hidewidth\reflectbox{$\m@th#1\vec{}\mkern4mu$}\hidewidth\cr
      \noalign{\kern-\ht\z@}
      $\m@th#1#2$\cr
    }%
  }%
}
\makeatother

\usepackage{lipsum}
\usepackage{amsfonts}
\usepackage{graphicx}
\usepackage{epstopdf}
\usepackage{algorithmic}
\ifpdf
  \DeclareGraphicsExtensions{.eps,.pdf,.png,.jpg}
\else
  \DeclareGraphicsExtensions{.eps}
\fi

\usepackage{enumitem}
\setlist[enumerate]{leftmargin=.5in}
\setlist[itemize]{leftmargin=.5in}

\newsiamremark{remark}{Remark}
\newsiamremark{hypothesis}{Hypothesis}
\crefname{hypothesis}{Hypothesis}{Hypotheses}
\newsiamthm{claim}{Claim}

\headers{Interacting AIS}{H. Chen, and L. Ying}
\title{Interacting Annealed Importance Sampling via Population-Based MCMC Kernels\thanks{Accepted by SIAM Journal on Scientific Computing (SISC)
\funding{The first author was partly supported by the T. S. Lo Graduate Fellowship Fund and the second author was partly supported by the National Science Foundation (NSF) under award number DMS-2011699 and DMS-2208163.}}}

\author{Haoxuan Chen\thanks{Institute for Computational and Mathematical Engineering (ICME), Stanford University, Stanford, CA, 94305, USA
(\email{haoxuanc@stanford.edu}).}
\and Lexing Ying\thanks{Department of Mathematics and Institute for Computational and Mathematical Engineering (ICME), Stanford University, Stanford, CA, 94305, USA 
(\email{lexing@stanford.edu}, \url{https://web.stanford.edu/\string~lexing/}).}}

\usepackage{amsopn}

\ifpdf
\hypersetup{
  pdftitle={Interacting Annealed Importance Sampling (AIS)},
  pdfauthor={H. Chen, and L. Ying}
}
\fi

\usepackage{hyperref}
\usepackage{caption}
\usepackage{subcaption}
\usepackage{float}

\begin{document}

\maketitle

\begin{abstract}
Sampling from a multimodal distribution is a fundamental and challenging problem in computational science and statistics. Among various approaches proposed for this task, one popular method is Annealed Importance Sampling (AIS). In this paper, we propose an interaction-based version of AIS by combining it with population-based Monte Carlo methods to improve its efficiency. By keeping track of an ensemble instead of a single sampler along some continuation path between the starting distribution and the target distribution, we take advantage of the interaction within the ensemble to encourage the exploration of undiscovered modes. Specifically, our main idea is to utilize either the snooker algorithm or the genetic algorithm used in Evolutionary Monte Carlo. We discuss how the proposed algorithm can be implemented and derive a partial differential equation governing the evolution of the ensemble under the continuous time and mean-field limit. We also test the efficiency of the proposed algorithm on various continuous and discrete distributions.
\end{abstract}




\section{Introduction}
Sampling from an unnormalized distribution numerically is a fundamental problem with applications in various fields, such as computational physics, Bayesian inference, uncertainty quantification, data assimilation, and machine learning. The most popular class of algorithms used for tackling this problem is the Markov Chain Monte Carlo (MCMC), whose thorough review can be found in ~\cite{liang2011advanced, liu2001monte}. Common examples of MCMC methods include Unadjusted Langevin Algorithm (ULA) or Metropolis-Adjusted Langevin Algorithm (MALA)~\cite{roberts1998optimal, roberts2002langevin, roberts1996exponential}, Random-Walk Metropolis Algorithm~\cite{gelman1997weak, roberts2001optimal}, Bouncy Particle Sampler~\cite{bouchard2018bouncy}, Hamiltonian or Hybrid Monte Carlo~\cite{andersen1980molecular, duane1987hybrid, neal2011mcmc, neal1996monte, rossky1978brownian}, Zig-Zag Sampler~\cite{bierkens2019zig}, etc. When the target distribution is (strongly) log-concave, it has been proved that most MCMC methods listed above can sample from the distribution efficiently~\cite{cheng2018underdamped, dalalyan2017theoretical, durmus2017nonasymptotic, dwivedi2018log, livingstone2019geometric, lu2022complexity, mangoubi2017rapid}. 

In contrast, when the target distribution is multimodal, the efficiency of standard MCMC methods is often limited by metastability, which refers to the demand of transiting between different high-probability modes separated by low-probability regions frequently. The main reason is that common MCMC methods are designed to sample a region based on its probability. Hence, they are able to explore a local high-probability mode efficiently, but they will only transit to another high-probability mode occasionally. To resolve metastability, numerous approaches have been proposed, such as mode-jumping (or mode-hopping) methods~\cite{andricioaei2001smart, ibrahim2009new, pompe2020framework, sminchisescu2007generalized, sminchisescu2003mode, tjelmeland2004use, tjelmeland2001mode}, cluster algorithms~\cite{swendsen1987nonuniversal, wolff1989collective}, umbrella sampling methods~\cite{dinner2020stratification, matthews2018umbrella, thiede2016eigenvector}, leap-point samplers~\cite{tan2023accelerate, tawn2021annealed}, transport map based methods~\cite{marzouk2016sampling, parno2018transport}, machine learning based methods~\cite{albergo2023stochastic, cao2022learning, gabrie2022adaptive, midgley2022flow, vargas2023denoising}, to name a few.

Among various approaches proposed for sampling multimodal distributions, our main focus is on two classes of methods. The first class is the continuation (or tempering) method, whose main idea is to start with an easy-to-sample distribution and construct a continuous path between the initial distribution and the target distribution. Metropolis-Hastings (MH) Algorithm is then applied for continuation methods to move along the path. Compared to classical MCMC methods, continuation methods gradually transform the starting distribution into the target distribution when moving along the continuous path, which allows the samplers to explore the whole state space and transit between high-probability modes in the target distribution. One other advantage of continuation methods over traditional MCMC methods is that they are run over a finite time horizon instead of an infinite time horizon. Influential examples of continuation methods include Tempered Transitions~\cite{neal1996sampling}, Simulated Tempering~\cite{geyer1995annealing, marinari1992simulated, neal1996sampling}, Iterated Batch Importance Sampling (IBIS)~\cite{chopin2002sequential}, Sequential Monte Carlo (SMC)~\cite{dai2022invitation,del2006sequential, doucet2009tutorial, maceachern1999sequential} and Annealed Importance Sampling (AIS)~\cite{neal2001annealed}. 

The second class is the population-based methods, which keep track of multiple samplers instead of a single sampler throughout the sampling process. The invariant distributions associated with distinct samplers can be either the same or different as long as they are all related to the target distribution. Consequently, these methods utilize the interaction between different samplers as a way to speed up its convergence towards the target distribution. Examples of interaction-based methods include (Sequential) Parallel Tempering~\cite{geyer1991markov, hukushima1996exchange, liang2003use}, Conjugate Gradient Monte Carlo~\cite{liu2000multiple} and equi-energy sampler~\cite{kou2006equi}. Some other interaction-based methods proposed in recent studies include affine invariant and gradient-free sampling methods~\cite{carrillo2022consensus, coullon2021ensemble, dunlop2022gradient, garbuno2020interacting, garbuno2020affine, goodman2010ensemble, karamanis2021ensemble, leimkuhler2018ensemble, pidstrigach2023affine}, Stein Variational Gradient Descent and its variants~\cite{liu2017stein, liu2016stein}, etc. As the size of the ensemble approaches infinity, many studies~\cite{carrillo2022consensus, chen2023sampling, ding2021ensemble, garbuno2020interacting, garbuno2020affine, lindsey2022ensemble, lu2019scaling, lu2019accelerating, lu2023birth, maurais2024sampling, pidstrigach2023affine, sprungk2023metropolis} have also investigated the mean-field limit of some ensemble-based sampling algorithms and obtained new insights. 

This paper aims to improve the efficiency of Annealed Importance Sampling (AIS), which is one of the continuation methods and has been applied to many fields including computational physics~\cite{lyman2007annealed, yasuda2021spatial, yasuda2022free, ying2022annealed} and machine learning~\cite{carbone2023efficient, ding2019learning, doucet2022score, zhang2021differentiable}. As the standard AIS generates independent samples, it is natural to incorporate interaction-based updates that share information across samplers to improve efficiency. Similar to other continuation methods, standard AIS uses MH algorithms to sample from all the intermediate distributions along the continuation path connecting the initial distribution and the target distribution. Motivated by empirical success of population-based methods, our main idea is to replace the MH algorithm used in AIS with a composition of standard MCMC algorithms and interaction-based methods, which can exploit discovered modes locally, explore undiscovered modes globally, and simultaneously balance the weights of the samplers in different modes. As a result, each intermediate distribution on the continuation path can be sampled in a more effective way, which leads to a more efficient AIS overall. 

\subsection{Related Work}
The interaction-based algorithm we proposed to replace the MH Algorithm used in AIS consists of three components: a local exploitation part, a global exploration part, and a reweighting part that utilizes the birth-death process. The local exploitation part can be taken to be any standard MCMC algorithm, while our choices of the other two parts are based on previous work. 

On the one hand, the reweighting part is motivated from the sequential Monte Carlo (SMC) method~\cite{del2006sequential} and its variants~\cite{carbone2023efficient,stordal2015iterative}, which are the most relevant to our work among all related studies. Along the continuation path, SMC propagates each particle via a fixed MH algorithm and then resamples the ensemble of particles according to the importance weights. Therefore, SMC can be viewed as AIS augmented with a resampling step, which naturally contains the local exploitation component and the reweighting component mentioned above. We note that an alternative approach to perform the reweighting step is to utilize the birth-death dynamics, which randomly replicates particles with large importance weights and removes particles with small importance weights within the ensemble. For a detailed description of how the resampling step in SMC relates to the birth-death processes, we refer the readers to~\cite{del2004feynman}. The practical implementation based on the birth-death dynamics adopted in our work, whose detailed description can be found in subsection~\ref{subsec: birth death and resampling} below, is inspired by a few recent works like~\cite{lu2019accelerating,lindsey2022ensemble}. 

The global exploration part, on the other hand, is used for eliminating the impact of metastability on sampling the intermediate distributions. Here, we select either the snooker algorithm~\cite{gilks1994adaptive} or the genetic algorithm~\cite{holland1992adaptation} to be the global exploration part, which has been used in previous work~\cite{braak2006markov, ter2008differential} to encourage exploration of missed modes and improve the efficiency of other sampling algorithms. In fact, both the snooker and the genetic algorithm are two important units of the Evolutionary Monte Carlo (EMC) method~\cite{liang2001evolutionary, liang2001real}, which belongs to the class of population-based Monte Carlo methods. 

\subsection{Contributions}
Our main contributions are summarized below:

\begin{itemize}
    \item We propose a novel sampling algorithm by combining Annealed Importance Sampling (AIS) with interaction-based sampling methods. For the task of sampling continuous distributions, we prove that the empirical density formed by the ensemble converges to some target density and derive the partial differential equation (PDE) governing the evolution of the target density under the continuous time and mean-field limit. 
    \item We provide numerical experiments to demonstrate that the new algorithm outperforms standard AIS on both continuous and discrete distributions with different dimensions.  
\end{itemize}

\subsection{Organization of the paper}
This paper is organized as follows. In
\cref{sec:overview of sampling}, we discuss a few existing sampling methods related to our work. In \cref{sec: sampling via annealing}, we discuss the two interacting AIS algorithms we developed. In \cref{sec: numerics}, we present numerical results on both continuous and discrete distributions to demonstrate the practical performance of the proposed algorithms. Conclusions and discussions for future work can be found in \cref{sec:conclusions}.

\subsection{Preliminaries and Notations}
Below, we introduce a few preliminaries and list some basic notations used throughout this paper and its supplement. Let $p(\cdot) \propto e^{-U(\cdot)}$ be the target distribution we aim to sample from, where $U: \mathcal{X} \rightarrow \mathbb{R}$ is the associated energy function and $\mathcal{X}$ is the state space. Let $d \in \mathbb{N}$ be some fixed integer. For continuous target distributions, we consider the case when $\mathcal{X} = \mathbb{R}^d$. For discrete target distributions, we consider the case when $\mathcal{X} = \{-1,1\}^d$, \emph{i.e.}, $\mathcal{X}$ is binary-coded. For any domain $\Omega \subset \mathbb{R}^d$, the boundary of $\Omega$ is denoted by $\partial \Omega$. 

We use $\langle \cdot, \cdot \rangle$ and $\|\cdot\| = \sqrt{\langle \cdot, \cdot \rangle}$ to denote the standard dot product and the Euclidean norm in $\mathbb{R}^d$, respectively. The standard normal distribution in $\mathbb{R}^d$ is represented by $\mathcal{N}(\boldsymbol{0}, \boldsymbol{I}_d)$, where $\boldsymbol{I}_d$ is the identity matrix of size $d \times d$. For any $a \in \mathbb{R}^d$, we use $\delta_{a}(\cdot)$ to denote the Dirac delta distribution at point $a$. For any function $f: \mathbb{R}^d \rightarrow \mathbb{R}$, the support of $f$ is defined as $\text{supp}(f)$. For any time-dependent path $x=x(t): [0,1] \rightarrow \mathbb{R}^d$ in $\mathbb{R}^d$, we use $\dot{x} = \frac{\dif x}{\dif t}$ to denote the time derivative of the path. For any time-dependent function $g=g(x,t):\mathbb{R}^d \times \mathbb{R} \rightarrow \mathbb{R}$, we use $\nabla_{x}g, \nabla_{x} \cdot g$ and $\Delta_{x} g$ to denote the spatial derivative, divergence and laplacian of $g$, \emph{i.e.}, the derivatives are taken with respect to the spatial variable $x$. The space of all the probability measures on $\mathbb{R}^d$ is defined as $\mathcal{P}(\mathbb{R}^d)$. Furthermore, we use $\mathcal{B}(\mathcal{P}(\mathbb{R}^d), \mathbb{R})$ to represent the space of all the functionals $F: \mathcal{P}(\mathbb{R}^d) \rightarrow \mathbb{R}$. For any $N \in \mathbb{N}$, the space of all the empirical probability measures that can be written as a weighted sum of $N$ Dirac delta distributions is denoted by
\begin{equation*}
\mathcal{P}_N(\mathbb{R}^d) := \left\{\sum_{i=1}^{N}\gamma_i \delta_{a_i}: \sum_{i=1}^{N}\gamma_i = 1, \ \gamma_i > 0, a_i \in \mathbb{R}^d, \ \forall \ 1 \leq i \leq N\right\}.    
\end{equation*}

\section{Review of Related Sampling Methods}
\label{sec:overview of sampling}
This section is devoted to investigating several examples of continuation and ensemble-based sampling methods that are most relevant to our work. In the first subsection, we describe Annealed Importance Sampling (AIS)~\cite{neal2001annealed}. The second subsection is devoted to discussing the resampling step in Sequential Monte Carlo, with a particular focus on its relation to the birth death dynamics. In the third subsection, we provide a review of the snooker algorithm~\cite{gilks1994adaptive}, and the genetic algorithm~\cite{holland1992adaptation}. 

\subsection{Annealed Importance Sampling (AIS)}
\label{subsec: finite time}

Let $U_0$ be some appropriately chosen energy function such that the distribution proportional to $e^{-U_0}$ can be easily sampled. The idea behind Annealed Importance Sampling (AIS)~\cite{neal1996sampling} is to continue between $U_0$ and the target energy function $U$. Specifically, let $c(t): [0,1] \rightarrow [0,1]$ be some strictly monotone function with $c(0) = 0$ and $c(1) = 1$. Also, let $(L-1)$ denote the number of intermediate energy functions between $U_0$ and $U$. For $1 \leq l \leq L-1$, the $l$-th intermediate energy $U_l$ is defined as:
\begin{equation}
\label{eqn: defn of intermediate potential}
\begin{aligned}
U_l(s) := \Big(1-c\Big(\frac{l}{L}\Big)\Big)U_0(s)  + c\Big(\frac{l}{L}\Big)U(s).  
\end{aligned}    
\end{equation}

One typical example of $c$ used in practice is $c(t) = t$. Furthermore, for $1 \leq l \leq L-1$, we assume that each distribution $p_{l}(\cdot) \propto e^{-U_l(\cdot)}$ is associated with a transition kernel $T_{l}(\cdot, \cdot)$ satisfying the detailed balance condition, \emph{i.e.},
\begin{equation}
\label{eqn: detailed balance}
\begin{aligned}
p_l(x)T_{l}(x,y) = p_l(y)T_l(y,x).    
\end{aligned}
\end{equation}

Using the building blocks given above, AIS follows the procedure described in Algorithm \ref{alg: AIS} to obtain independent weighted samples from each $p_l$ step by step. Note that we use the symbol $\propto$ in (\ref{eqn: weight of AIS}) to hide the normalization constant, which remains the same for all independent samples generated by AIS.  

\begin{algorithm}
\caption{Annealed Importance Sampling (AIS)}
\label{alg: AIS}
\begin{algorithmic}
\STATE{\textbf{Initialization:} initial configuration $s_{\frac{1}{2}}$ sampled from $p_0(s) \propto e^{-U_0(s)}$}.
\FOR{$l = 1:(L-1)$}
\STATE{Take one step (or a few steps) of the transition kernel $T_l(\cdot,\cdot)$ associated with the distribution $p_l$ from $s_{l-\frac{1}{2}}$. Let $s_{l+\frac{1}{2}}$ be the resulting configuration.} 
\ENDFOR
\STATE{Set $s:= s_{L-\frac{1}{2}}$.}
\STATE{Compute the associated weight
\begin{equation}
\label{eqn: weight of AIS}
w_s = \frac{p_1(s_{\frac{1}{2}})}{p_0(s_{\frac{1}{2}})} \cdots \frac{p_L(s_{L-\frac{1}{2}})}{p_{L-1}(s_{L-\frac{1}{2}})} \propto \frac{e^{-U_1(s_{\frac{1}{2}})}}{e^{-U_0(s_{\frac{1}{2}})}} \cdots \frac{e^{-U_L(s_{L-\frac{1}{2}})}}{e^{-U_{L-1}(s_{L-\frac{1}{2}})}}.  
\end{equation}} 
\RETURN$(s,w_s)$
\end{algorithmic}
\end{algorithm}

Compared to classical MCMC methods, which need to be run over an infinite time horizon to converge to the desired invariant distribution, one advantage of AIS is that it is implemented over a finite time horizon. However, one tradeoff is that the samples need to be reweighted. To examine why reweighting is crucial here, we will show that the distribution of the output configuration $s$ reweighted by $w_s$ is exactly the target distribution $p(s) \propto e^{-U(s)}$. 

Consider the path $(s_{\frac{1}{2}}, \cdots, s_{L-\frac{1}{2}})$ in Algorithm \ref{alg: AIS}, which is generated with probability $p_{0}(s_{\frac{1}{2}})T_1(s_{\frac{1}{2}},s_{\frac{3}{2}}) \cdots T_{L-1}(s_{L-\frac{3}{2}}, s_{L-\frac{1}{2}})$. Multiplying this probability by $w_s$ given in (\ref{eqn: weight of AIS}) and plugging in the detailed balance condition (\ref{eqn: detailed balance}) imply that
\begin{equation*}
\begin{aligned}
p_{0}&(s_{\frac{1}{2}})T_1(s_{\frac{1}{2}},s_{\frac{3}{2}}) \cdots T_{L-1}(s_{L-\frac{3}{2}}, s_{L-\frac{1}{2}}) \cdot \frac{p_1(s_{\frac{1}{2}})}{p_0(s_{\frac{1}{2}})} \cdots \frac{p_L(s_{L-\frac{1}{2}})}{p_{L-1}(s_{L-\frac{1}{2}})} \\
= p_L&(s_{L-\frac{1}{2}})T_{L-1}(s_{L-\frac{1}{2}}, s_{L-\frac{3}{2}}) \cdots T_{1}(s_{\frac{3}{2}}, s_{\frac{1}{2}}),
\end{aligned}    
\end{equation*}
which is equal to the probability of going backward from $s_{L-\frac{1}{2}}$ to $s_{\frac{1}{2}}$. By taking the marginal distribution of $s_{L-\frac{1}{2}}$ on both sides above, we obtain that the distribution of the output $s=s_{L-\frac{1}{2}}$ with weight $w_s$ is exactly $p_{L}(\cdot) = p(\cdot) \propto e^{-U(\cdot)}$.  

An alternative way to understand the reweighting in AIS is to use path integral under the continuum limit $L \rightarrow \infty$. Consider the task of sampling from some continuous distribution. Let $U_t(x):= (1-c(t))U_0(x) + c(t)U(x)$ be the corresponding time-dependent energy annealing between the two energy functions $U_0, U: \mathbb{R}^d \rightarrow \mathbb{R}$, where $t \in [0,1]$ is a continuous variable. For any $t \in [0,1]$, we pick the transition kernel $T_t(\cdot,\cdot)$ to be the overdamped Langevin dynamics associated with $U_t(\cdot)$. Given the conditions specified above, we can deduce that under the continuum limit, the dynamics of any sample generated by AIS are governed by the following stochastic differential equation (SDE):
\begin{equation}
\label{eqn: langevin continuous AIS}
\begin{aligned}
\dot{x} = -\nabla_{x}U_{t}(x) + \sqrt{2}\dot{W},    
\end{aligned}    
\end{equation}
where $W$ above denotes the standard Brownian motion in $\mathbb{R}^d$. Under the dynamics (\ref{eqn: langevin continuous AIS}), the logarithm of the probability of a specific continuous path $\{x(t)\}_{t \in [0,1]}$ is proportional to
\begin{equation}
\label{eqn: AIS prob of forward path}
-U_0(x(0)) -\frac{1}{2}\int_{0}^{1}\dot{W}^2\dif t
= -U_0(x(0))-\frac{1}{4}\int_{0}^{1}(\dot{x} + \nabla_xU_t(x))^2\dif t.   
\end{equation}

Note that above the normalizing constants are omitted. Moreover, since the reversed dynamics of (\ref{eqn: langevin continuous AIS}) is given by $\dot{x} = \nabla_{x}U_t(x) + \sqrt{2}\dot{W}$, we can deduce that the logarithm of the probability that going backward from $x(1)$ to $x(0)$ along the same path is proportional to
\begin{equation} 
\label{eqn: AIS prob of backward path}
-U(x(1))-\frac{1}{2}\int_{0}^{1}\dot{W}^2\dif t
= -U(x(1))-\frac{1}{4}\int_{0}^{1}(\dot{x} - \nabla_xU_t(x))^2\dif t. 
\end{equation}

Comparing (\ref{eqn: AIS prob of forward path}) with (\ref{eqn: AIS prob of backward path}), we note that one of the two probabilities has to be reweighted in order for them to have the same marginal distribution with respect to $x(1)$. As the normalizing constants omitted in (\ref{eqn: AIS prob of forward path}) and (\ref{eqn: AIS prob of backward path}) are all independent of the path $\{x(t)\}_{t \in [0,1]}$, we may add (\ref{eqn: AIS prob of forward path}) by some weight $w$ and equate the sum to (\ref{eqn: AIS prob of backward path}), which is equivalent to reweighting the corresponding probability by $e^w$. Then, we may express the weight $w$ as follows: 
\begin{equation}
\label{eqn: initial expression of AIS continuous weight}
\begin{aligned}
w &= U_0(x(0)) - U(x(1)) + \frac{1}{4}\int_{0}^{1}\Big((\dot{x} + \nabla_xU_t(x))^2-(\dot{x} - \nabla_xU_t(x))^2\Big)\dif t.    
\end{aligned}    
\end{equation}

Simplifying (\ref{eqn: initial expression of AIS continuous weight}) above yields $w=-\int_{0}^{1}\frac{\partial}{\partial t}U_t(x)\dif t$, whose detailed derivation is deferred to the lemma proved in the first section of the supplementary material. Under the limit $L \rightarrow \infty$, the logarithm of the weight $w_s$ given in (\ref{eqn: weight of AIS}) converges to $w$.

\begin{remark}
Generally, the weight $w:=w(\tau)$ can be treated as a time-dependent variable satisfying the relation $w(\tau) =-\int_{0}^{\tau}\frac{\partial}{\partial t}U_t(x)\dif t$ for any $\tau \in (0,1)$. In fact, at any fixed time $\tau \in (0,1)$, the distribution of a sample generated by AIS always deviates from the target distribution $p_\tau \propto e^{-U_\tau(\cdot)}$, which results in the necessity of reweighting the distribution of $x(\tau)$ by $e^{w(\tau)}$  to correct the bias induced by the deviation. Such deviation between the sampling and target distribution originates from the difference between their normalizing constants, which are often referred to as the free energies in the physics literature~\cite{lifshitz2013statistical}. More generally, for any
two distributions connected by some fixed continuous-time dynamics, the Jarzynski Equality (JE)~\cite{jarzynski1997nonequilibrium} was established to pinpoint the relation between their normalizing constants in non-equilibrium thermodynamics. Hence, AIS can also be thought of as a discrete version of JE, and it is sometimes referred to as sampling via JE in other work~\cite{carbone2023efficient}. 
\end{remark}

\begin{remark}
\label{rmk: 2.2 in AIS}
Assuming that the Markov chain associated with $T_l$ mixes perfectly for every $l$, \emph{i.e.}, each $T_l$ returns an independent sample $s_{l+\frac{1}{2}}$ from the intermediate distribution $p_{l} \propto e^{-U_l(\cdot)}$, ~\cite{neal2001annealed} has shown that the smaller the variance of the weight $w_s$ is, the more efficient AIS will be. One approach to control the variance of the weight is the resampling step adopted in the Sequential Monte Carlo method~\cite{dai2022invitation,del2006sequential} and its variants~\cite{carbone2023efficient, stordal2015iterative}, which replicates particles with larger weights and removes those with smaller weights at each intermediate distribution $p_l$. In this sense, SMC can be viewed as an extension of AIS equipped with an explicit resampling step. An alternative way to perform such global moves of the particles based on their weights is to utilize the birth-death process, which has been applied in recent work like~\cite{lu2019accelerating,lindsey2022ensemble}. In the following subsection, we will discuss the practical implementations of those two reweighting mechanisms. 
\end{remark}

\begin{algorithm}[H]
\caption{Resampling Step}
\label{alg: resampling}
\begin{algorithmic}
\STATE{\textbf{Initialization:} threshold $\kappa \in (0,1)$, ensemble of weighted particles $\{(x_j, w_j)\}_{j=1}^{N}$.}
\STATE{(I) Compute the effective sample size $ESS := \frac{\left(N^{-1}\sum_{i=1}^{N}w_i\right)^2}{N^{-1}\sum_{i=1}^{N}\left(w_i\right)^2}$.}
\STATE{(II) If $ESS < \kappa$, then sample $\left\{\widehat{x}_j\right\}_{j=1}^{N}$ with replacement from $\left\{x_j\right\}_{j=1}^{N}$ with probability $\left\{\frac{w_j}{\sum_{i=1}^{N}w_i}\right\}_{j=1}^{N}$ and set $\widehat{w}_j =1$ for $1 \leq j \leq N$.}
\STATE{(III) If $ESS \geq \kappa$, keep $\{(\widehat{x}_j, \widehat{w}_j)\}_{j=1}^{N} = \{(x_j, w_j)\}_{j=1}^{N}$ unchanged.}
\RETURN $\left\{\left(\widehat{x}_j,\widehat{w}_j\right)\right\}_{j=1}^{N}$
\end{algorithmic}
\end{algorithm}

\subsection{Resampling Step in Sequential Monte Carlo (SMC) and Birth Death Dynamics}
\label{subsec: birth death and resampling}
Since AIS is typically used with an ensemble of weighted samples, here we use $\{(x_j, w_j)\}_{j=1}^{N}$ to denote the ensemble tracked by AIS. Note that the importance weights $w_j$'s all evolve according to the dynamics $w(\tau) =-\int_{0}^{\tau}\frac{\partial}{\partial t}U_t(x)\dif t$ along the continuation path, which may become extremely small or large. Consequently, most particles can have nearly zero weight at the final time $t=1$, while only a small fraction of the particles carry almost all of the mass and lead to large weight variance. To address such issue caused by weight degeneracy, it is necessary to rebalance the weights of the particles. A standard approach is the resampling step in SMC, which eliminates low-weight particles and replicates high-weight ones. For a detailed  description of the resampling step, the readers may refer to Algorithm~\ref{alg: resampling}. 

An alternative way to implement the rebalancing step is to use the birth-death dynamics~\cite{del2004feynman,lu2019accelerating,lindsey2022ensemble} under the gradient-flow formulation of sampling, which is used in our work. Specifically, particles undergo stochastic replication or deletion events with rates determined by their importance weights, so that high-weight particles are replicated and low-weight particles are removed. Just as the resampling step in SMC, the birth-death dynamics also mitigates weight degeneracy by reducing variance of the importance weights. Using the same set of notations as the subsection above, we let $U_t \ (t \in [0,1])$ denote the annealing path between the initial distribution $U_0$ and target distribution $U_1 =U$. Take $\Delta t$ and $\{x_j\}_{j=1}^{N}$ to be some fixed time stepsize and ensemble of particles, where the empirical distribution $\frac{1}{N}\sum_{j=1}^{N}\delta_{x_j}$ associated with the ensemble approximates $p_{t} \propto e^{-U_t}$. We provide a detailed description of how to update some fixed particle $x_l$ from the ensemble $\{x_j\}_{j=1}^{N}$ by simulating the birth-death dynamics in Algorithm~\ref{alg: birth death step}. As the fixed particle $x_l$ ranges over $\{x_j\}_{j=1}^{N}$, the whole ensemble gets updated.  

\begin{algorithm}[H]
\caption{Reweighting Step based on the Birth Death Dynamics}
\label{alg: birth death step}
\begin{algorithmic}
\STATE{\textbf{Initialization:} time stepsize $\Delta t$, ensemble of particles $\{x_j\}_{j=1}^{N}$, chosen particle $x_l$, target distribution $p_{t} \propto e^{-U_{t}}$.}
\STATE{Compute $\beta_j := \frac{\partial}{\partial t}U_t(x_j)$ for $1 \leq j \leq N$ and the mean value $\overline{\beta} := \frac{1}{N}\sum_{j=1}^{N}\beta_j$.}
\IF{$\beta_l > \overline{\beta}$}
\STATE{Kill $x_l$ with probability $1-e^{-\left(\beta_l - \overline{\beta}\right)\Delta t}$.}
\STATE{Duplicate a particle uniformly chosen from $\{x_j\}_{j=1}^{N} \setminus \{x_l\}$.}
\ELSE
\STATE{Duplicate $x_l$ with probability $1-e^{\left(\beta_l - \overline{\beta}\right)\Delta t}$.}
\STATE{Kill a particle uniformly chosen from $\{x_j\}_{j=1}^{N} \setminus \{x_l\}$.}
\ENDIF
\RETURN the updated ensemble $\left\{\widehat{x}_j\right\}_{j=1}^{N}$
\end{algorithmic}
\end{algorithm} 

Beyond the reweighting component already used in SMC, a key contribution of our work is to augment AIS with interacting MCMC mutation kernels, which enable information sharing across the ensemble of particles and improve exploration of multimodal targets. In the next subsection, we will describe these interaction-based MCMC methods in detail.

\subsection{Interaction-Based Monte Carlo Methods}
\label{subsec: ensemble}
Instead of keeping track of only one sample, population-based Monte Carlo methods keep track of multiple samples $\{x_i\}_{i=1}^{N}$ with invariant distribution $f(x_1,x_2,\cdots,x_N) =\prod_{i=1}^{N}f_i(x_i)$, where the target distribution $p(\cdot) \propto e^{-U(\cdot)}$ coincides with $f_i(\cdot)$ for at least one $i \in \{1,2,\cdots,N\}$. In general, compared to MCMC methods using a single chain, population-based methods take advantage of the interaction between different particles, which encourages the particles to explore different modes of the target distribution in practice.

Among all the population-based methods, one well-known example is the Evolutionary Monte Carlo (EMC)~\cite{liang2001evolutionary, liang2001real}. In the following two subsections, we provide a review of the snooker algorithm~\cite{gilks1994adaptive}, and the genetic algorithm~\cite{holland1992adaptation}, which are two important building blocks of EMC. For both algorithms, we set the number of iterations to be $N$ such that one sample is updated in each iteration. For any $0 \leq t \leq N$, we use $\boldsymbol{X}^{(t)} = \{x^{(t)}_i\}_{i=1}^{N}$ to denote the ensemble kept by both algorithms at the $t$-th iteration. Moreover, for any $0 \leq t \leq N$ and subset $\mathcal{S} \subset \{1,2,\cdots,N\}$, we let $\boldsymbol{X}^{(t)}_{-\mathcal{S}} := \{x_{j}^{(t)}\}_{j=1}^{N} \setminus \{x_{i}^{(t)}\}_{i \in \mathcal{S}}$ be the subset formed by removing the samples $\{x_{i}^{(t)}\}_{i \in \mathcal{S}}$ from the ensemble $\boldsymbol{X}^{(t)}$.

\subsubsection{Snooker Algorithm}
\label{subsubsec: snooker algo}
Consider the task of sampling from a continuous distribution $p(\cdot)$ proportional to $e^{-U(\cdot)}$, where $U:\mathbb{R}^d \rightarrow \mathbb{R}$. The snooker algorithm follows the procedure described in Algorithm \ref{alg:snooker} to update one sample in each iteration. We note that for the empirical distribution formed by the ensemble to converge to the target distribution, Algorithm \ref{alg:snooker} often needs to be repeated multiple times. Regarding the correctness of the snooker algorithm, it suffices to show that the algorithm generates independently and identically distributed samples from the target distribution $p(\cdot)$ when it reaches equilibrium. Assuming that $\boldsymbol{X}^{(t-1)}$ are independently and identically distributed samples from $p(\cdot)$ for any fixed $t$, we only need to prove that the sample $y$ generated in step (III) of the iteration above is distributed as $p(\cdot) \propto e^{-U(\cdot)}$ and independent of $x^{(t-1)}_k$ for any $k \neq i$. Such a claim follows directly from Lemma \ref{lem: snooker alg correctness}, which was proved in~\cite{liu2000generalised}. 

\begin{algorithm}[H]
\caption{Snooker Algorithm}
\label{alg:snooker}
\begin{algorithmic}
\STATE{\textbf{Initialization:} chosen particle $x_l$; initialized ensemble $\boldsymbol{X}^{(I)} = \{x_i\}_{i=1}^{N}$}; target distribution $p(\cdot) \propto e^{-U(\cdot)}$.
\STATE{(I) Sample one other particle $x_{j}$ uniformly at random from the remaining particles $\boldsymbol{X}^{(I)}_{-\{l\}}$ and form the update direction $e = x_{l} - x_{j}$.}
\STATE{(II) Sample $r \in \mathbb{R}$ from the following density $\rho(r)$:
\begin{equation}
\label{eqn: density in snooker}
\begin{aligned}
\rho(r) \propto |r|^{d-1}p(x_{j}+re) = |r|^{d-1}p\Big((1-r)x_{j} + rx_{l}\Big),    
\end{aligned}    
\end{equation}
where $p(\cdot)$ above is the target distribution. Compute the new sample $y = x_{j}+re$.}
\STATE{(III) Form the new ensemble $\boldsymbol{X}^{(N)}$ by replacing $x_l$ with $y$, \emph{i.e.}, $\boldsymbol{X}^{(N)} := \boldsymbol{X}^{(I)}_{-\{l\}} \cup \{y\}$.}
\RETURN $\boldsymbol{X}^{(N)}$
\end{algorithmic}
\end{algorithm}

\begin{lemma}
\label{lem: snooker alg correctness}
Fix some probability distribution function $\pi:\mathbb{R}^d \rightarrow \mathbb{R}$ and some point $b \in \mathbb{R}^d$. Assume that $a \in \mathbb{R}^d$ is distributed as $\pi(\cdot)$ and let $e_a := a-b$. Furthermore, assume that $r \in \mathbb{R}$ is sampled from the distribution $\rho(r) \propto |r|^{d-1}\pi(b+re_a)$, then $c:=  b+re_a$ is also distributed as $\pi(\cdot)$. Moreover, if $b$ is independent of $a$, then $c$ is also independent of $a$.   
\end{lemma}

We note that it might be hard to sample from the density given in (\ref{eqn: density in snooker}) directly. Hence, one may also replace step (II) in the for loop with one or a few steps of the Metropolis-Hastings (MH) algorithm for practical use. One other approach is to utilize the stretch move proposed in~\cite{goodman2010ensemble}, which emphasizes affine invariance. Specifically, one may sample the scalar $r$ from some prescribed distribution $g(\cdot)$ satisfying the symmetry condition $g(\frac{1}{z}) = zg(z)$, compute the corresponding new sample $y = (1-r)x^{(t-1)}_{j} + rx^{(t-1)}_{l}$ and accept the move $x^{(t-1)}_l \rightarrow y$ with the following probability: 
\begin{equation}
\label{eqn: snooker accept ratio}
\begin{aligned}
\min\left\{1,\frac{\left\|y - x^{(t-1)}_j\right\|^{d-1} \cdot p(y)}{\left\|x^{(t-1)}_l - x^{(t-1)}_j\right\|^{d-1} \cdot p(x^{(t-1)}_l)}\right\}.    
\end{aligned}    
\end{equation}

\subsubsection{Genetic Algorithm}
Consider the task of sampling from a discrete distribution $p(\cdot)$ proportional to $e^{-U(\cdot)}$, where $U:\{-1,1\}^d \rightarrow \mathbb{R}$. The genetic algorithm we discuss here is motivated by the crossover operation between chromosome pairs in biology. Roughly speaking, in each iteration, a pair of samples' coordinates get mixed according to some crossover operator, which generates a pair of ``offsprings''. Specifically, the genetic algorithm follows the steps listed in Algorithm \ref{alg:genetic} to obtain samples from $p(\cdot)$. Similar to the snooker algorithm, the genetic algorithm also needs to be run multiple times to converge to equilibrium in practice. The correctness of the genetic algorithm can be verified via the fact that the transition probability given in (\ref{eqn: genetic acceptance ratio}) satisfies the detailed balance condition with invariant distribution $\rho(x,y):= p(x)p(y)$.

\begin{algorithm}[H]
\caption{Genetic Algorithm}
\label{alg:genetic}
\begin{algorithmic}
\STATE{\textbf{Initialization:} particle $x_l = (x_{l,1}, x_{l,2}, \cdots, x_{l,d})$; initialized ensemble $\boldsymbol{X}^{(I)} = \{x_i\}_{i=1}^{N}$; target distribution $p(\cdot) \propto e^{-U(\cdot)}$.}
\STATE{(I) Sample one other particle $x_{j} = (x_{j,1}, x_{j,2}, \cdots, x_{j,d})$ from $\boldsymbol{X}^{(I)}_{-\{l\}}$ uniformly at random.}
\STATE{(II) Mix the $2d$ coordinates of $x_{l}$ and $x_{j}$ to generate two new samples $y_{l} = (y_{l,1}, y_{l,2}, \cdots, y_{l,d})$ and $y_{j} = (y_{j,1}, y_{j,2}, \cdots, y_{j,d})$, such that for any $1 \leq s \leq d$, $y_{l,s}$ is uniformly sampled from $\{x_{l,s}, x_{j,s}\}$ and $y_{j,s}$ is assigned to the remaining coordinate, \emph{i.e.},
\begin{equation*}
\begin{aligned}
\mathbb{P}\left(x_{l,s} = y_{l,s}, x_{j,s} = y_{j,s}\right) = \mathbb{P}\left(x_{l,s} = y_{j,s}, x_{j,s} = y_{l,s}\right) = \frac{1}{2}.  
\end{aligned}    
\end{equation*}
}
\STATE{(III) Form the new ensemble $\boldsymbol{X}^{(N)}$ by replacing $\{x_l, x_j\}$ with $\{y_l, y_j\}$, \emph{i.e.}, set $\boldsymbol{X}^{(N)} := \boldsymbol{X}^{(I)}_{-\{l,j\}} \cup \{y_l, y_j\}$, with the following acceptance ratio:
\begin{equation}
\label{eqn: genetic acceptance ratio}
\begin{aligned}
\min\left\{1,\frac{p(y_l)p(y_j)}{p(x_l)p(x_j)}\right\} = \min\left\{1, \frac{e^{-U(y_l)}e^{-U(y_j)}}{e^{-U(x_l)}e^{-U(x_j)}}\right\}.    
\end{aligned}    
\end{equation}
Set $\boldsymbol{X}^{(N)} = \boldsymbol{X}^{(I)}$ if the move $\{x_l, x_j\} \rightarrow \{y_l, y_j\}$ above is rejected.}
\RETURN $\boldsymbol{X}^{(N)}$
\end{algorithmic}
\end{algorithm}

\section{Interacting Annealed Importance Sampling (IAIS)}
\label{sec: sampling via annealing}
This section discusses two hybrid algorithms that combine AIS with population-based Monte Carlo methods for sampling both continuous and discrete target distributions. To improve the efficiency of AIS, there are two main questions to be addressed. One question is how to control the variance of the importance weights, while the other one is how to accelerate the mixing of the Markov chain $T_l$ associated with each intermediate distribution $p_l$. For a broader discussion of issues in sampling multimodal distributions, the readers may refer to~\cite{tan2023accelerate}. 

The first question can be handled by the two reweighting mechanisms included in Algorithms~\ref{alg: resampling} and~\ref{alg: birth death step} above, which are already adopted in SMC~\cite{del2006sequential} and other related work like~\cite{del2004feynman,lu2019accelerating,lindsey2022ensemble}. The second bottleneck is closely related to exploration of multimodal landscapes and can benefit from interacting MCMC kernels that enable information sharing across an ensemble of samplers. Below we will describe different kinds of interaction-based MCMC methods and how they are incorporated with both AIS and the reweighting step for different distributions. Specifically, the first subsection focuses on the case of continuous distributions, while the second subsection considers the case of discrete distributions.

\subsection{Interacting AIS for Continuous Distributions}
\label{subsec: interacting AIS for cont distribution}
For the task of sampling continuous distributions, each $T_l$ is picked to be a composition of the Unadjusted Langevin Algorithm (ULA) and the snooker algorithm described in Algorithm \ref{alg:snooker} above, where the ULA exploits locally, and the snooker algorithm explores globally. Moreover, we use Algorithm~\ref{alg: birth death step} to adjust the weights of the particles returned by each $T_l$. With all the building blocks specified above, we provide a detailed description of the Interacting AIS (IAIS) method with exploration component for sampling continuous distributions in Algorithm \ref{alg: AIS + snooker}.

\begin{algorithm}[H]
\caption{Interacting AIS with Exploration for Continuous Distributions}
\label{alg: AIS + snooker}
\begin{algorithmic}
\STATE{\textbf{Initialization:} number of intermediate distributions $L$ and time stepsize $\Delta t = \frac{1}{L}$}; intermediate distributions $p_l(\cdot) \propto e^{-U_l(\cdot)}$, where $U_l(\cdot) = \Big(1-c(l\Delta t)\Big)U_0(\cdot)  + c(l\Delta t)U(\cdot) \ (1 \leq l \leq L-1)$; an initial ensemble of particles $\boldsymbol{X}^{(0)} = \{x^{(0)}_i\}_{i=1}^{N}$ sampled identically and independently from $p_0(\cdot) \propto e^{-U_0(\cdot)}$.
\FOR{$l = 1:L$}
\FOR{$i = 1:N$}
\STATE{\textbf{Langevin Dynamics:}}
\STATE{
Update $x^{(l-1+\frac{3i-2}{3N})}_i = x^{(l-1+\frac{i-1}{N})}_{i} - \Delta t \nabla U_l(x^{(l-1+\frac{i-1}{N})}_{i}) + \sqrt{2\Delta t}\xi_i$, where $\xi_i \ (1 \leq i \leq N)$ are identically and independently sampled from $\mathcal{N}(\boldsymbol{0}, \boldsymbol{I_d})$. Keep all the other particles unchanged by setting $x^{(l-1+\frac{3i-2}{3N})}_j = x^{(l-1+\frac{i-1}{N})}_j \ (j \neq i)$. 
}
\STATE{\textbf{Snooker Algorithm:}}
\STATE{Apply Algorithm \ref{alg:snooker} with the chosen particle being  $x^{(l-1+\frac{3i-2}{3N})}_i$, the initialized ensemble being $\boldsymbol{X}^{(l-1+\frac{3i-2}{3N})} = \{x^{(l-1+\frac{3i-2}{3N})}_j\}_{j=1}^{N}$ and the target density being $p_l(\cdot) \propto e^{-U_l(\cdot)}$. Let $\boldsymbol{X}^{(l-1+\frac{3i-1}{3N})} = \{x^{(l-1+\frac{3i-1}{3N})}_j\}_{j=1}^{N}$ be the returned ensemble.}
\STATE{\textbf{Birth-Death Dynamics:}}
\STATE{Apply Algorithm~\ref{alg: birth death step} with the time stepsize being $\Delta t$, initialized particle being $x^{(l-1+\frac{3i-1}{3N})}_i$, input ensemble of particles being $\boldsymbol{X}^{(l-1+\frac{3i-1}{3N})} = \{x^{(l-1+\frac{3i-1}{3N})}_j\}_{j=1}^{N}$ and target distribution being $p_l \propto e^{-U_l}$. Let $\boldsymbol{X}^{(l-1+\frac{i}{N})} = \{x^{(l-1+\frac{i}{N})}_j\}_{j=1}^{N}$ be the returned ensemble.}
\ENDFOR
\ENDFOR
\RETURN{$\boldsymbol{X}^{(L)} = \{x^{(L)}_i\}_{i=1}^{N}$}
\end{algorithmic}
\end{algorithm}

\subsection{Interacting AIS for Discrete Distributions}
For the task of sampling discrete distributions, each $T_l$ is selected to be a composition of the Glauber dynamics~\cite{glauber1963time} and the genetic algorithm, which are described in Algorithms~\ref{alg:glauber} and~\ref{alg:genetic} respectively. In particular, the Glauber dynamics exploit locally, while the genetic algorithm explores globally. Again, we use Algorithm~\ref{alg: birth death step} to balance the particles' weights. Combining all the building blocks listed above gives us the Interacting AIS (IAIS) method with exploration part for sampling discrete distributions, whose detailed description is given in Algorithm \ref{alg: AIS + genetic}.  

\begin{algorithm}[H]
\caption{Glauber Dynamics}
\label{alg:glauber}
\begin{algorithmic}
\STATE{\textbf{Initialization:} particle $x_l = (x_{l,1}, x_{l,2}, \cdots, x_{l,d})$; target distribution $p_l(\cdot) \propto e^{-U_l(\cdot)}$.}
\STATE{(I) Choose a component $x_{l,j}$ of $x_{l}$ uniformly at random and consider the new sample $y_l$ formed by flipping the sign of $x_{l,j}$, \emph{i.e.},
$y_l= (x_{l,1},\cdots, -x_{l,j}, \cdots, x_{l,d})$.}
\STATE{(II) Accept the move $x_l \rightarrow y_l$ (\emph{i.e.}, $\widehat{x}_l = y_l$) with probability $r:= \frac{e^{-U_l(y_l)}}{e^{-U_l(x_l)} + e^{-U_l(y_l)}}$. Otherwise keep $x_l$ unchanged (\emph{i.e.}, $\widehat{x}_l = x_l$).}
\RETURN{$\widehat{x}_l$}
\end{algorithmic}
\end{algorithm}

\begin{algorithm}[H]
\caption{Interacting AIS with Exploration for Discrete Distributions}
\label{alg: AIS + genetic}
\begin{algorithmic}
\STATE{\textbf{Initialization:} number of intermediate distributions $L$ and time stepsize $\Delta t = \frac{1}{L}$}; intermediate distributions $p_l(\cdot) \propto e^{-U_l(\cdot)}$, where $U_l(\cdot) = \Big(1-c(l\Delta t)\Big)U_0(\cdot)  + c(l\Delta t)U(\cdot) \ (1 \leq l \leq L-1)$; an initial ensemble of particles $\boldsymbol{X}^{(0)} = \{x^{(0)}_i\}_{i=1}^{N}$ sampled identically and independently from $p_0(\cdot) \propto e^{-U_0(\cdot)}$.
\FOR{$l = 1:L$}
\FOR{$i = 1:N$}
\STATE{\textbf{Glauber Dynamics:}}
\STATE{Apply Algorithm~\ref{alg:glauber} with the chosen particle being $x^{(l-1+\frac{i-1}{N})}_i$ and the target distribution being $p_l \propto e^{-U_l}$. Let $\widehat{x}^{(l-1+\frac{i-1}{N})}_i$ be the returned particle and set $x^{(l-1+\frac{3i-2}{3N})}_i = \widehat{x}^{(l-1+\frac{i-1}{N})}_i$. Keep all the other particles unchanged by setting $x^{(l-1+\frac{3i-2}{3N})}_j = x^{(l-1+\frac{i-1}{N})}_j$ for $j \neq i$.}
\STATE{\textbf{Genetic Algorithm:}\\
Apply Algorithm \ref{alg:genetic} with the chosen particle being  $x^{(l-1+\frac{3i-2}{3N})}_i$, the initialized ensemble being $\boldsymbol{X}^{(l-1+\frac{3i-2}{3N})} = \{x^{(l-1+\frac{3i-2}{3N})}_j\}_{j=1}^{N}$ and the target density being $p_l(\cdot) \propto e^{-U_l(\cdot)}$. Let $\boldsymbol{X}^{(l-1+\frac{3i-1}{3N})} = \{x^{(l-1+\frac{3i-1}{3N})}_j\}_{j=1}^{N}$ be the returned ensemble.}
\STATE{\textbf{Birth-Death Dynamics:}}
\STATE{Apply Algorithm~\ref{alg: birth death step} with the time stepsize being $\Delta t$, initialized particle being $x^{(l-1+\frac{3i-1}{3N})}_i$, input ensemble of particles being $\boldsymbol{X}^{(l-1+\frac{3i-1}{3N})} = \{x^{(l-1+\frac{3i-1}{3N})}_j\}_{j=1}^{N}$ and target distribution being $p_l \propto e^{-U_l}$. Let $\boldsymbol{X}^{(l-1+\frac{i}{N})} = \{x^{(l-1+\frac{i}{N})}_j\}_{j=1}^{N}$ be the returned ensemble.}
\ENDFOR
\ENDFOR
\RETURN{$\boldsymbol{X}^{(L)} = \{x^{(L)}_i\}_{i=1}^{N}$}
\end{algorithmic}
\end{algorithm}

\begin{remark}
In practice, the ULA step in the first component of Algorithm~\ref{alg: AIS + snooker} can be replaced by the Metropolis-adjusted Langevin algorithm (MALA). In general, each mutation kernel $T_l$ can be implemented as a composition of several MALA updates and snooker moves. Similarly for the case of discrete distributions, $T_l$ can be taken as a combination of multiple steps of Glauber dynamics and the genetic algorithm. 

To investigate the theoretical properties of Algorithm~\ref{alg: AIS + snooker} in the continuous setting, we also develop a mean-field analysis in the continuous-time limit. For brevity, we defer the technical details to the second section of the supplementary material.
\end{remark}

\begin{remark}
Given that each Markov transition kernel $T_l$ can be picked to be an arbitrary Markov chain with invariant distribution $p_l$, we can obtain a truncated version of Algorithm \ref{alg: AIS + snooker} by removing the Snooker Algorithm and keeping the other two parts. Similarly, a truncated form of Algorithm \ref{alg: AIS + genetic} can be obtained by removing the Genetic Algorithm in each iteration. 

Therefore, for the following section on experiments, we denote each tested algorithm by an abbreviation consisting of its components. For instance, the two truncated algorithms mentioned above (also referred to as interacting AIS without exploration) will be denoted by ``MALA+BD” and ``Glauber +BD" respectively, where the abbreviation ``BD" corresponds to Algorithm~\ref{alg: birth death step}. When Algorithm~\ref{alg: birth death step} is replaced by explicit time-evolving importance weights used in standard AIS (Algorithm~\ref{alg: AIS}), we replace “BD” with “Reweight” in the abbreviation.
\end{remark}

\section{Numerical Experiments}
\label{sec: numerics}
In this section, we exhibit the effectiveness of the interacting AIS with exploration via a series of numerical examples. From the numerical tests presented below, we can see that Algorithm \ref{alg: AIS + snooker} and \ref{alg: AIS + genetic} outperform both interacting AIS without exploration and standard AIS that involves reweighting, neither of which contains any part encouraging global moves towards the undiscovered modes. The testing cases below are designed based on related examples used in previous work~\cite{lindsey2022ensemble, peng2023generative, ren2023high, tan2023accelerate, liang2024smc}. For all the AIS-related algorithms tested in this section, we pick $c(t) = t:[0,1] \rightarrow [0,1]$ to be the function used for interpolating between the starting and the target distribution. 

\subsection{Continuous Distributions}
The first subsection focuses on testing Algorithm \ref{alg: AIS + snooker} and its variants on a few continuous distributions with varying dimensions. To illustrate the efficacy of adding the exploration part, we make a comparison between Algorithm \ref{alg: AIS + snooker}, interacting AIS without exploration, and standard AIS. Regarding the implementation of Algorithm \ref{alg: AIS + snooker}, we use the stretch move proposed in~\cite{goodman2010ensemble} to sample from the density (\ref{eqn: density in snooker}) in the snooker algorithm. Moreover, we use MALA to replace ULA in our implementation of all the algorithms we test. Furthermore, two versions of standard AIS are implemented. One of them uses MALA as the transition kernel, while the other one's transition kernel is selected to be the MH Algorithm associated with a Gaussian proposal density.

Let $p(x) = \frac{1}{Z}e^{-U(x)}$ denote the target continuous distribution, where $Z$ is the normalizing constant. Also, we use $\{s_i\}_{i=1}^{N}$ and $\{w_i\}_{i=1}^{N}$ to denote the samples generated by an algorithm, where the $\{s_i\}_{i=1}^{N}$ are the locations and $\{w_i\}_{i=1}^{N}$ are the weights. Note that the weights are all equal for interacting AIS algorithms utilizing the birth death dynamics, \emph{i.e.}, $w_i = \frac{1}{N} \ (1 \leq i \leq N)$. We take two approaches to evaluate the quality of the empirical distribution $\hat{p}$ formed by the generated samples $\{s_i\}_{i=1}^{N}$ and $\{w_i\}_{i=1}^{N}$. On the one hand, a direct computation implies that the Kullback–Leibler (KL) divergence between $\hat{p}$ and $p$ can be explicitly written as $D_{\text{KL}}(\hat{p} \ \| \ p) = \int_{\text{supp}(\hat{p})}\hat{p}(x)\log\Big(\frac{\hat{p}(x)}{p(x)}\Big)\dif x = \sum_{i=1}^{N}w_iU(s_i) + \sum_{i=1}^{N}w_i\log(w_i) - \log(Z)$. As the normalizing constant $Z$ is independent of the generated samples and weights, we may drop off the $-\log(Z)$ term above to obtain the expression of the loss function $\mathcal{L}_{\text{KL}}(\{w_i\}_{i=1}^{N}, \{s_i\}_{i=1}^{N}) = \sum_{i=1}^{N}w_iU(s_i) + \sum_{i=1}^{N}w_i\log(w_i)$. We remark that the loss function is not guaranteed to be always positive since the $-\log(Z)$ term gets dropped in the expression above. Throughout this subsection, the loss function $\mathcal{L}_{\text{KL}}(\cdot)$ is referred to as the empirical KL loss. On the other hand, to confirm which algorithms can discover the modes of the underlying distribution more effectively, we also plot the marginal distribution of certain coordinates formed by the samples generated by all four algorithms listed above and make a comparison between the plots.

\subsubsection{Gaussian Mixture Model}
\label{example: 2D Gaussian Mixture}
We first test the performance of the four algorithms on a $2$-dimensional Gaussian mixture model, which is of similar form as the testing cases used in~\cite{lu2019accelerating, lu2023birth, tan2023accelerate}. Specifically, the target density can be written as $\pi(x,y) := \sum_{i=1}^{4}w_i\mathcal{N}(x,y;\mu_i,\Sigma_i)$, where $\mu_i \in \mathbb{R}^{2}, \Sigma_i \in \mathbb{R}^{2 \times 2}$ and $\mathcal{N}(x,y;\mu_i,\Sigma_i)$ denotes the Gaussian distribution in $\mathbb{R}^2$ with mean vector $\mu_i$ and covariance matrix $\Sigma_i$ for any $1 \leq i \leq 4$. The numerical values of the parameters used in our experiment are given by
\begin{equation}
\label{eqn: 2D Gaussian Mixture parameters}
\begin{aligned}
w_i &= \frac{1}{4} \ (1 \leq i \leq 4), \ \mu_1 = [0, -3]^T, \ \mu_2 = [0, 8]^T, \ \mu_3 = [-4, 4]^T, \mu_4 = [4, 4]^T\\
\Sigma_1 &= \begin{bmatrix}
1.2 & 0 \\
0 & 0.01     
\end{bmatrix}, \  
\Sigma_2 = \begin{bmatrix}
0.01 & 0 \\
0 & 2     
\end{bmatrix}, \ 
\Sigma_3 = \Sigma_4 = \begin{bmatrix}
0.2 & 0 \\
0 & 0.2     
\end{bmatrix}.
\end{aligned}    
\end{equation}

\begin{figure}[H]
\centering
\begin{subfigure}[b]{0.475\textwidth}
\centering
\includegraphics[width=\textwidth]{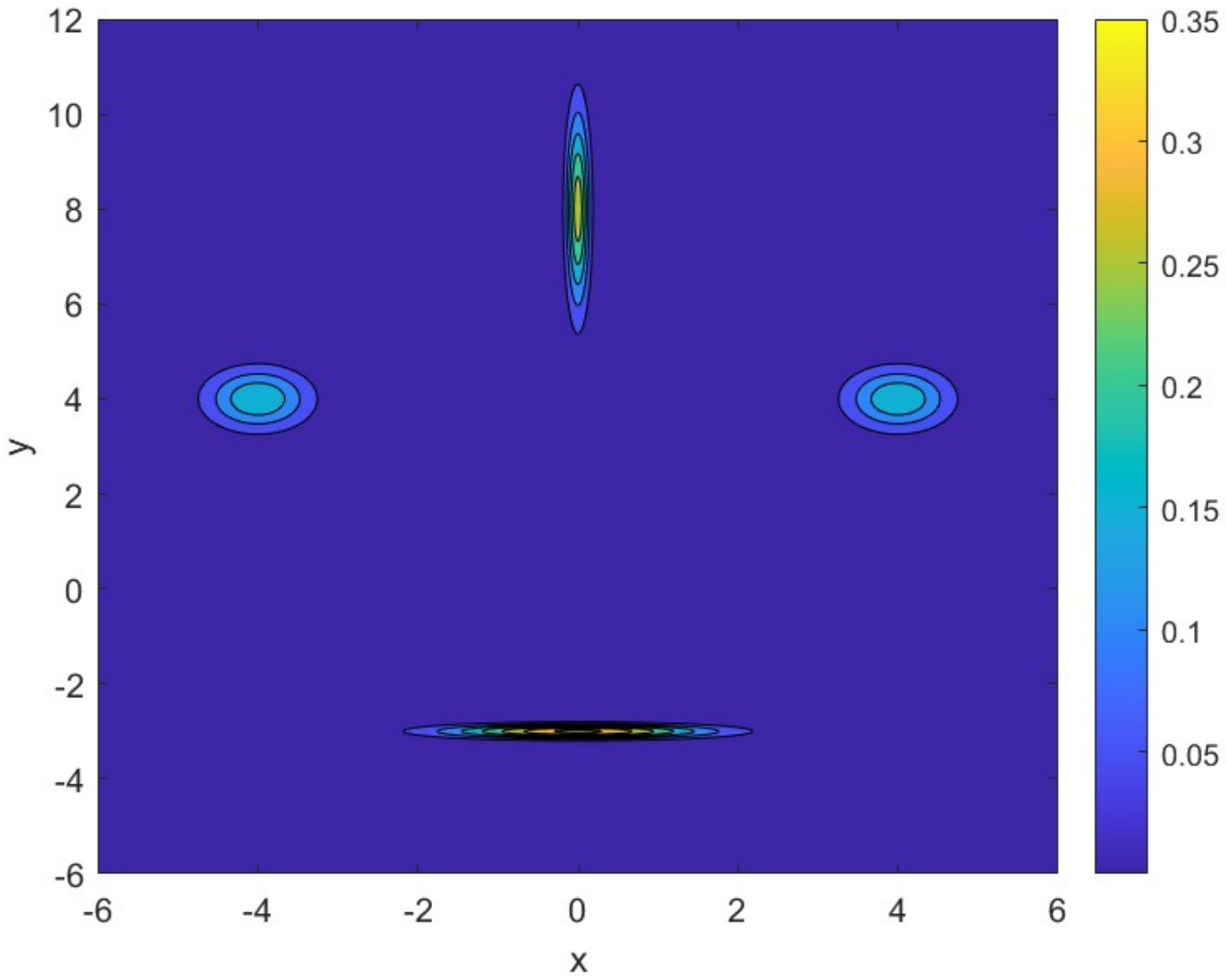}
\caption[]
{{\small Target Density}}    
\label{fig: 2d gaussian mixture target density contour plot}
\end{subfigure}
\hfill
\begin{subfigure}[b]{0.475\textwidth}  
\centering 
\includegraphics[width=\textwidth]{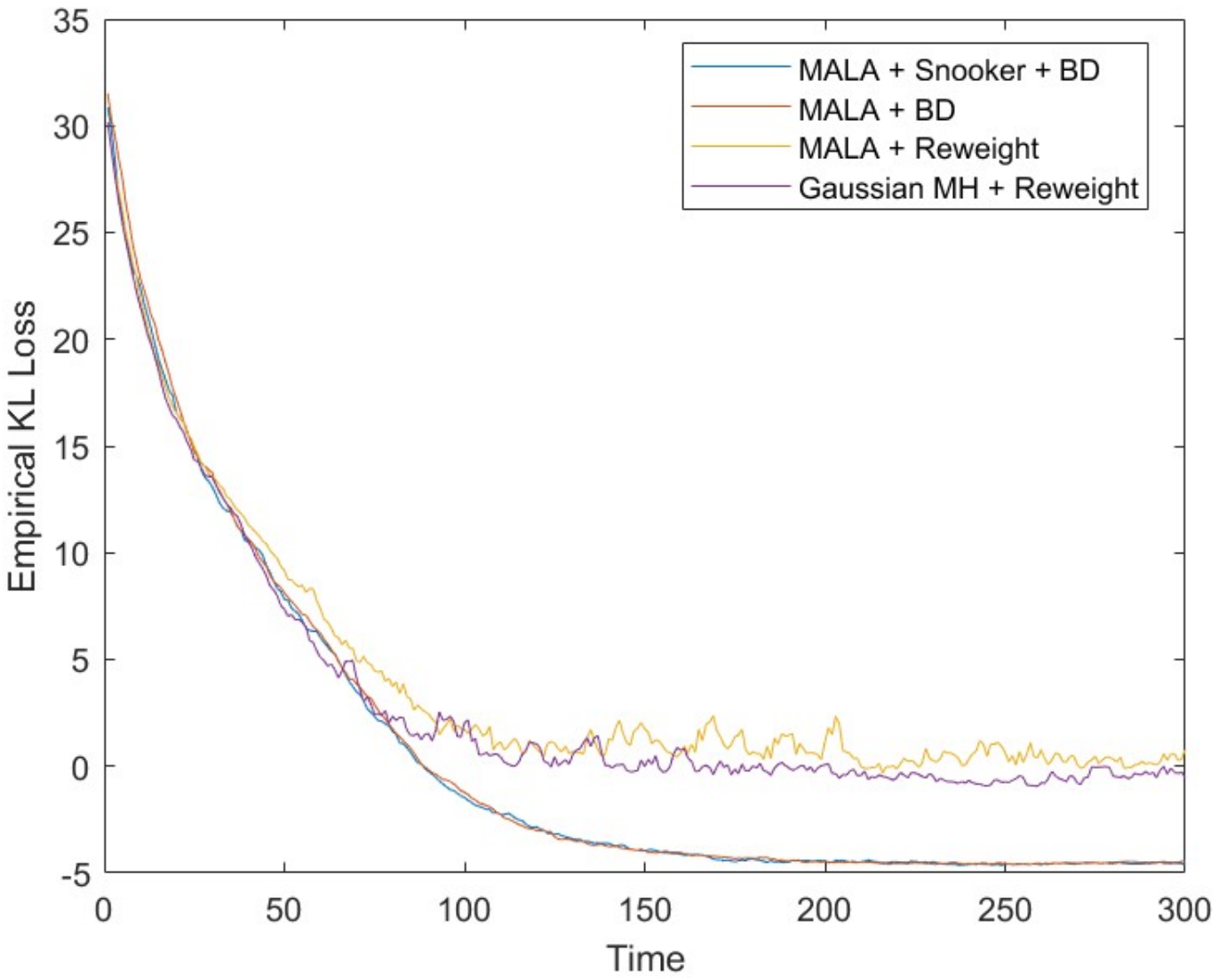}
\caption[]
{{\small Empirical KL Loss}}    
\label{fig: 2d gaussian mixture empirical kl loss plot}
\end{subfigure}
\caption[]
{(a) Contour plot of the target density in \ref{example: 2D Gaussian Mixture}; (b) Plots of $\mathcal{L}_{\text{KL}}$ evaluated at evolving weighted samples generated by different testing algorithms in \ref{example: 2D Gaussian Mixture}.} 
\label{fig: 2d gaussian mixture target density contour plot and empirical kl loss plot}
\end{figure}

A contour plot of the target density is provided in Figure (\ref{fig: 2d gaussian mixture target density contour plot}). In our implementation, we set the number of samplers and time steps to be $N=1000$ and $L = 300$, respectively. Moreover, for the standard AIS using an MH Algorithm as the transition kernel, we pick its Gaussian proposal density to be $\mathcal{N}(\boldsymbol{0}, 0.01\boldsymbol{I}_2)$. The starting distribution is picked to be the normal distribution $\mathcal{N}(\boldsymbol{0},\boldsymbol{I}_2)$ in $\mathbb{R}^2$. We track values of the empirical KL loss evaluated at weighted samples generated by the four testing algorithms with respect to time and plot them in Figure (\ref{fig: 2d gaussian mixture empirical kl loss plot}). These plots reveal that the two Interacting AIS algorithms (``MALA+BD” and ``MALA+Snooker+BD”, \emph{i.e.}, variants of Algorithm~\ref{alg: AIS + snooker}) converge faster and generate better samples than the two standard AIS algorithms (``MALA+Reweight” and ``Gaussian MH + Reweight”, \emph{i.e.}, variants of Algorithm~\ref{alg: AIS}) in terms of the loss function $\mathcal{L}_{\text{KL}}$. 

\begin{figure}[H]
\centering
\begin{subfigure}[b]{0.475\textwidth}   
\centering 
\includegraphics[width=\textwidth]{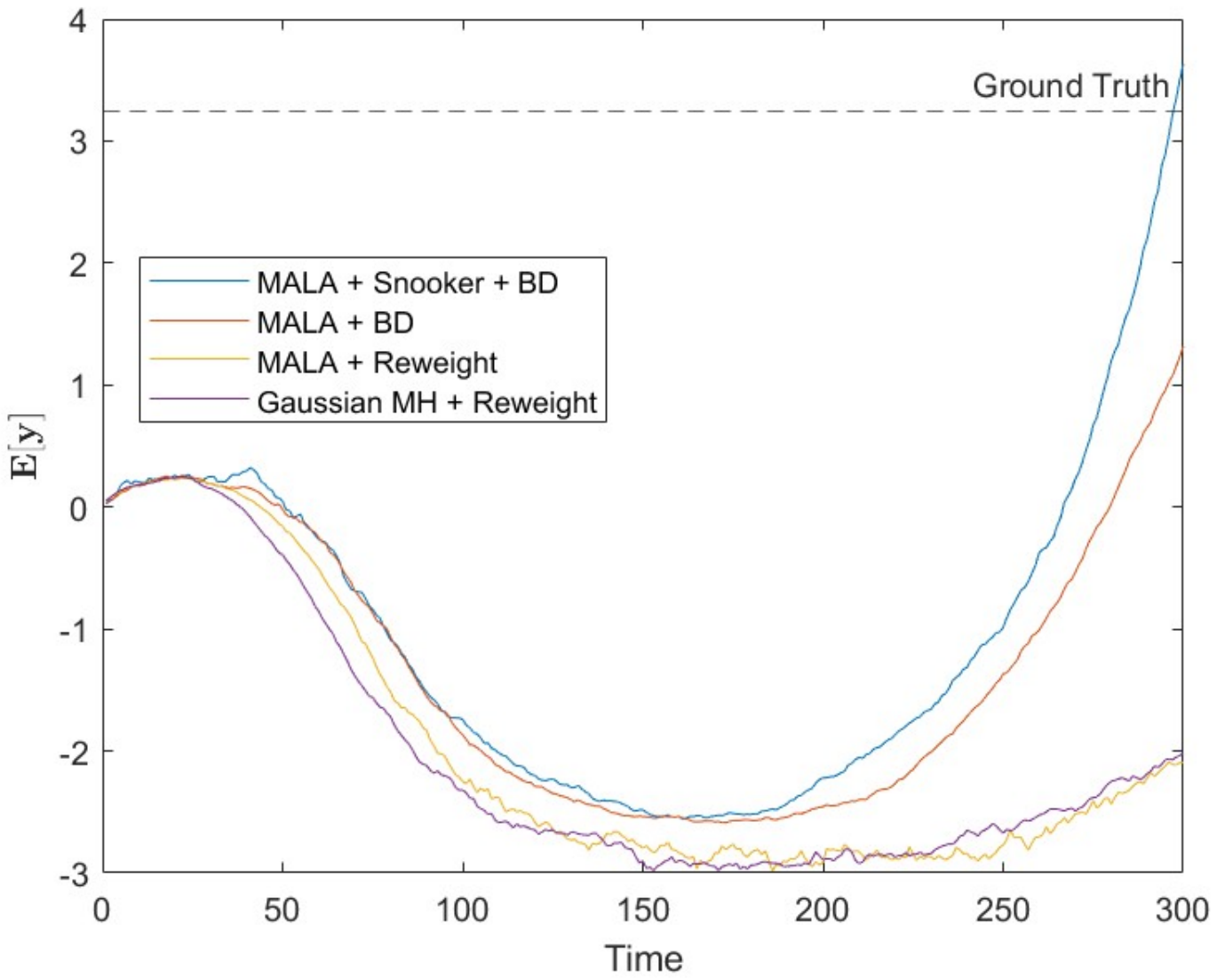}
\caption[]%
{{\small Estimation of $\mathbb{E}[y]$}}    
\label{fig: 2D Gaussian Mixture Expectation y small}
\end{subfigure}
\hfill
\begin{subfigure}[b]{0.475\textwidth}   
\centering 
\includegraphics[width=\textwidth]{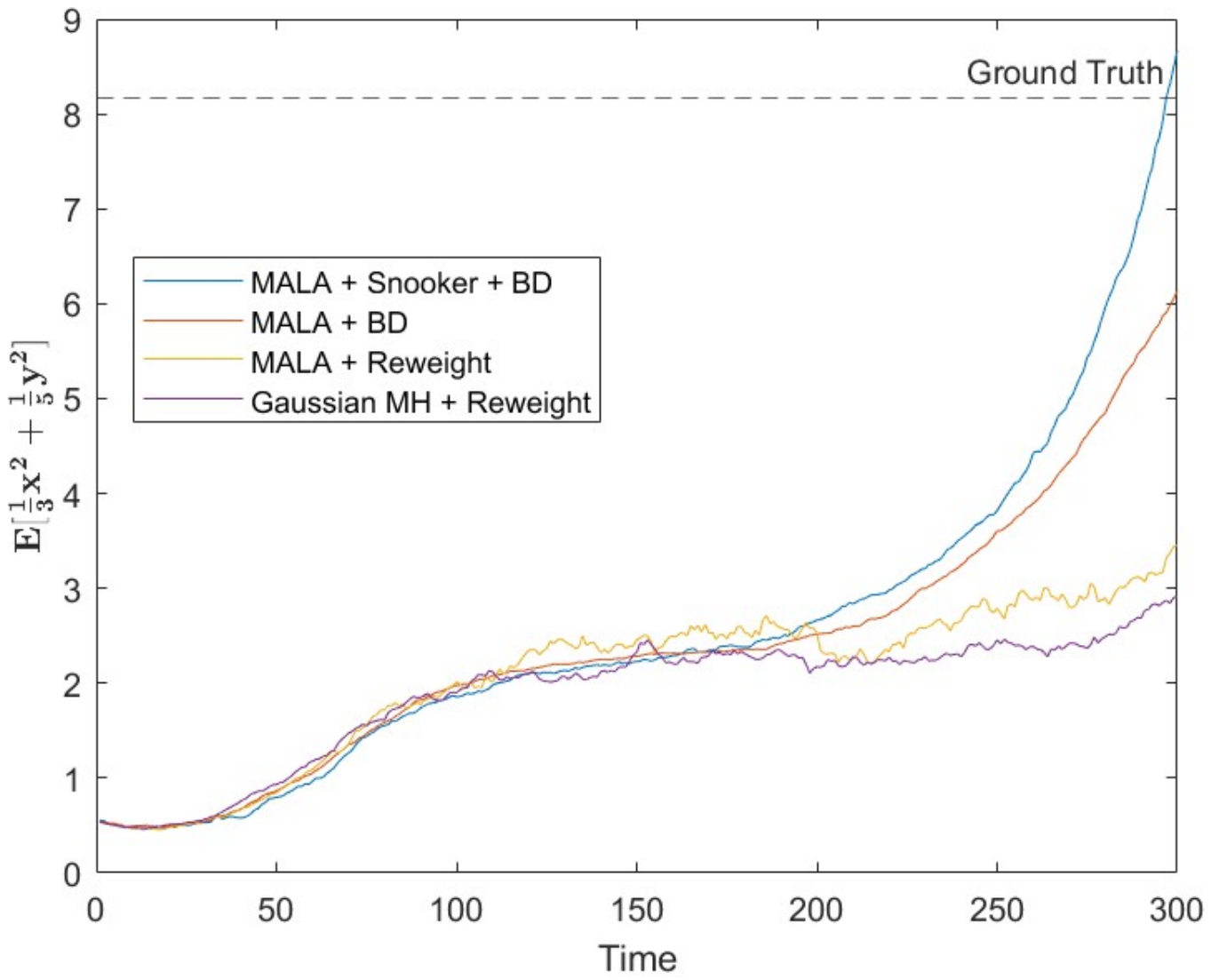}
\caption[]%
{{\small Estimation of $\mathbb{E}[\frac{1}{3}x^2 + \frac{1}{5}y^2]$}}    
\label{fig: 2D Gaussian Mixture Expectation quad small}
\end{subfigure}
\caption[]
{\small Evolving Estimations of $\mathbb{E}[f(x,y)]$ evaluated at weighted samples generated by different testing algorithms in \ref{example: 2D Gaussian Mixture} for $f_1(x,y) = y$ and $f_2(x,y) = \frac{1}{3}x^2 + \frac{1}{5}y^2$.} 
\label{fig: 2d gaussian mixture estimations of two functions}
\end{figure}

\begin{figure}[H]
\centering
\begin{subfigure}{.24\textwidth}
    \centering
    \includegraphics[width=.95\linewidth]{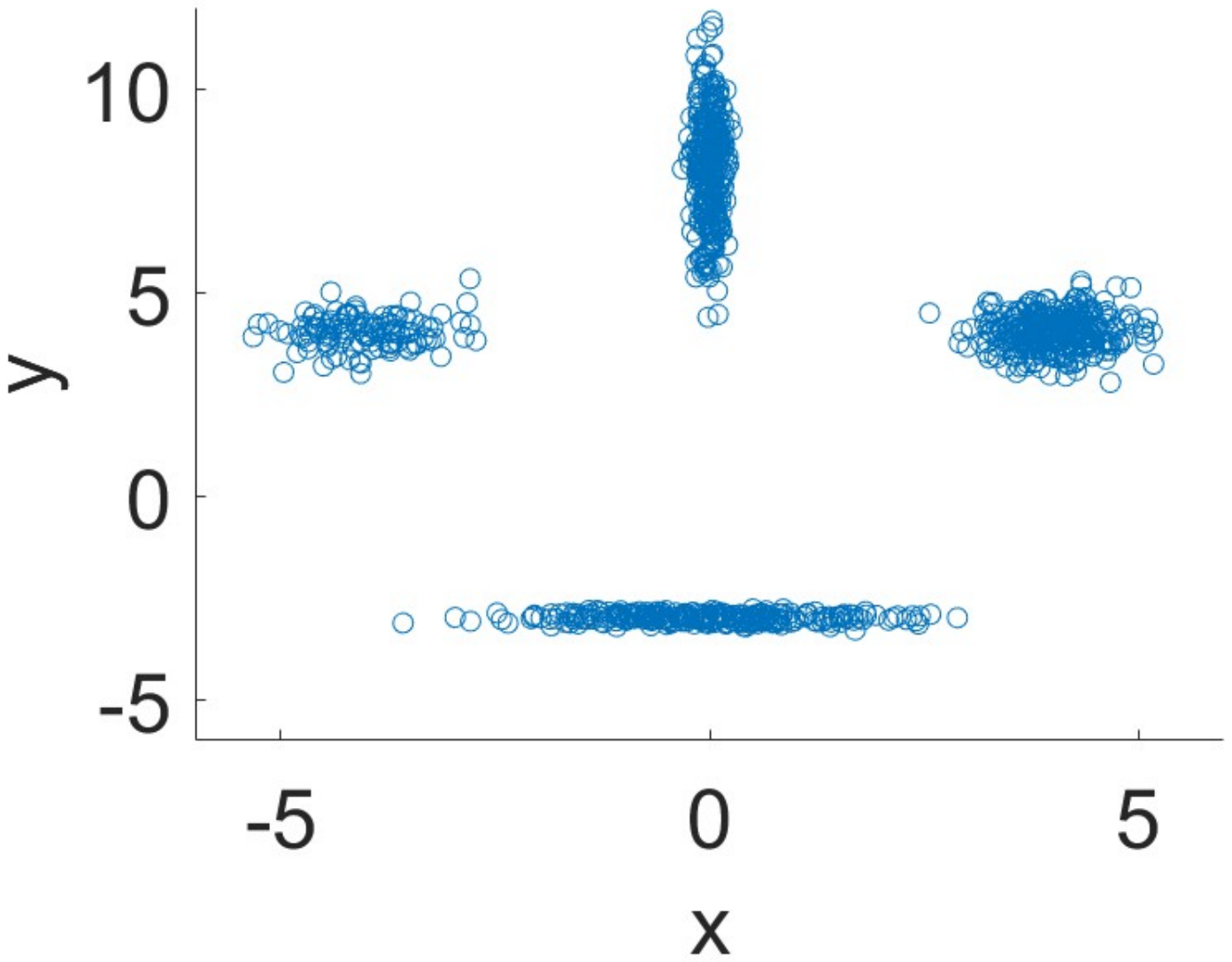} 
    \caption{}
    \label{fig: 2D Gaussian Mixture MALA Snooker BD small}
\end{subfigure}
\begin{subfigure}{.24\textwidth}
    \centering
    \includegraphics[width=.95\linewidth]{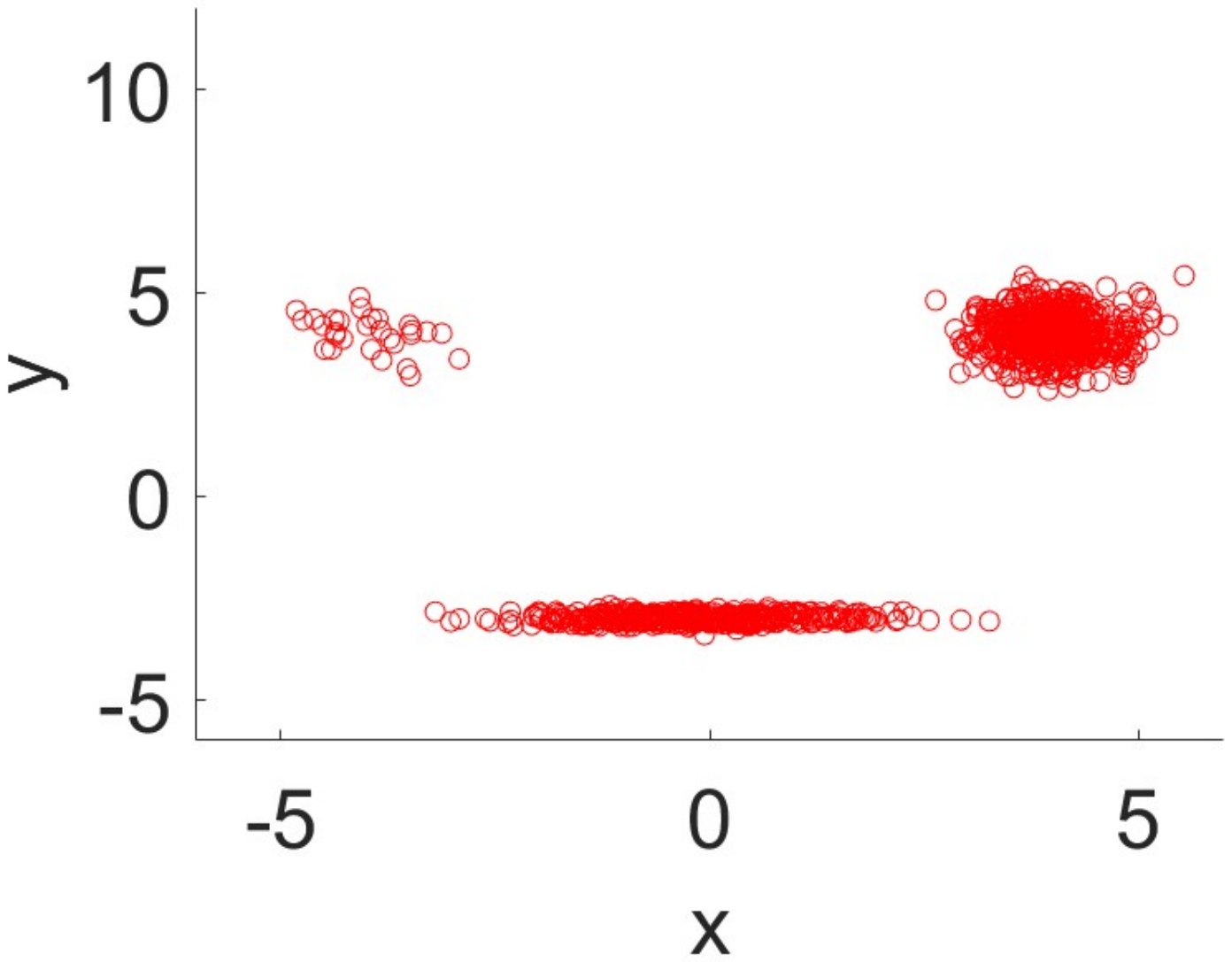}
    \caption{}
    \label{fig: 2D Gaussian Mixture MALA BD small}
\end{subfigure}
\begin{subfigure}{.24\textwidth}
    \centering
    \includegraphics[width=.95\linewidth]{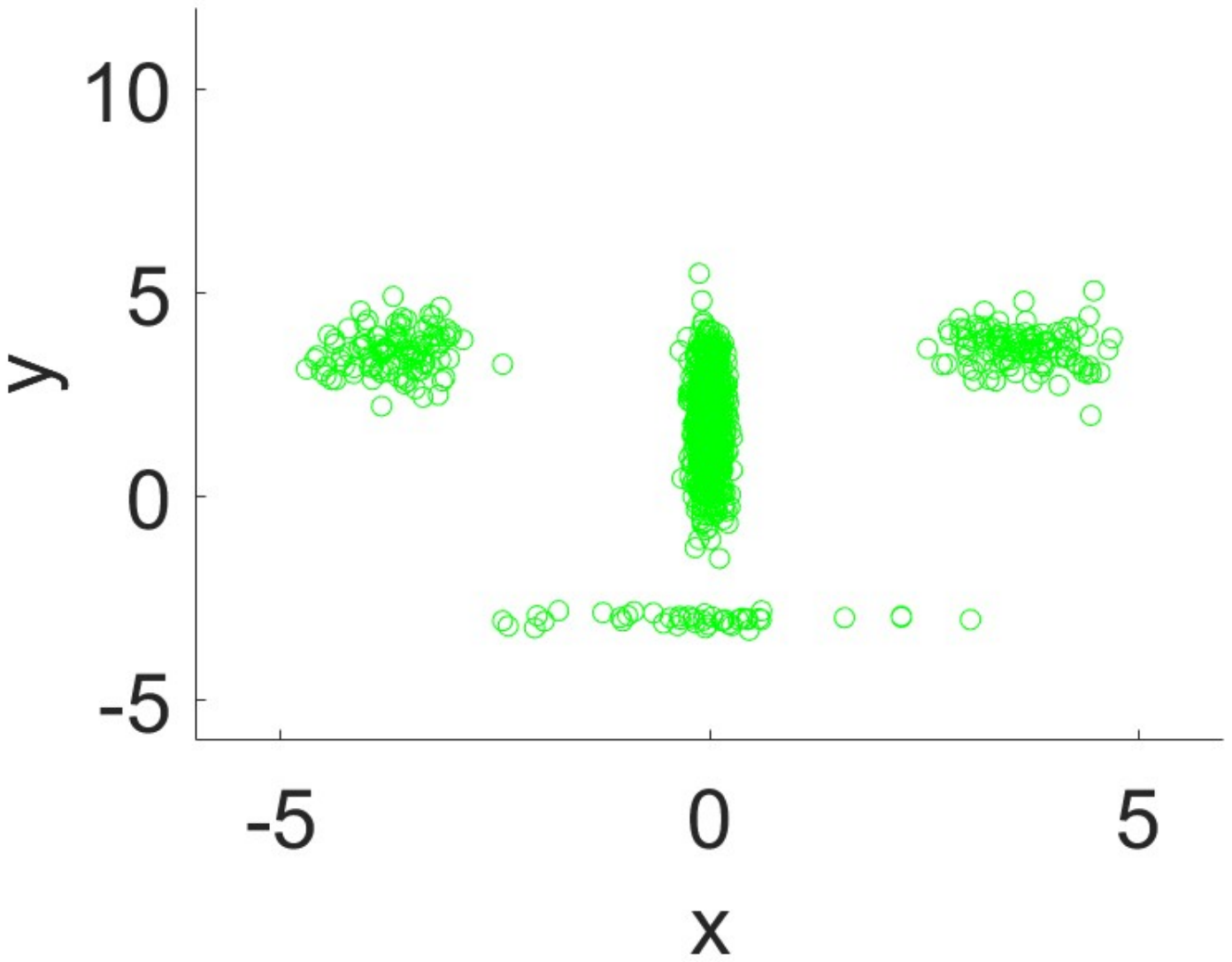}  
    \caption{}
    \label{fig: 2D Gaussian Mixture MALA Reweight small}
\end{subfigure}
\begin{subfigure}{.24\textwidth}
    \centering
    \includegraphics[width=.95\linewidth]{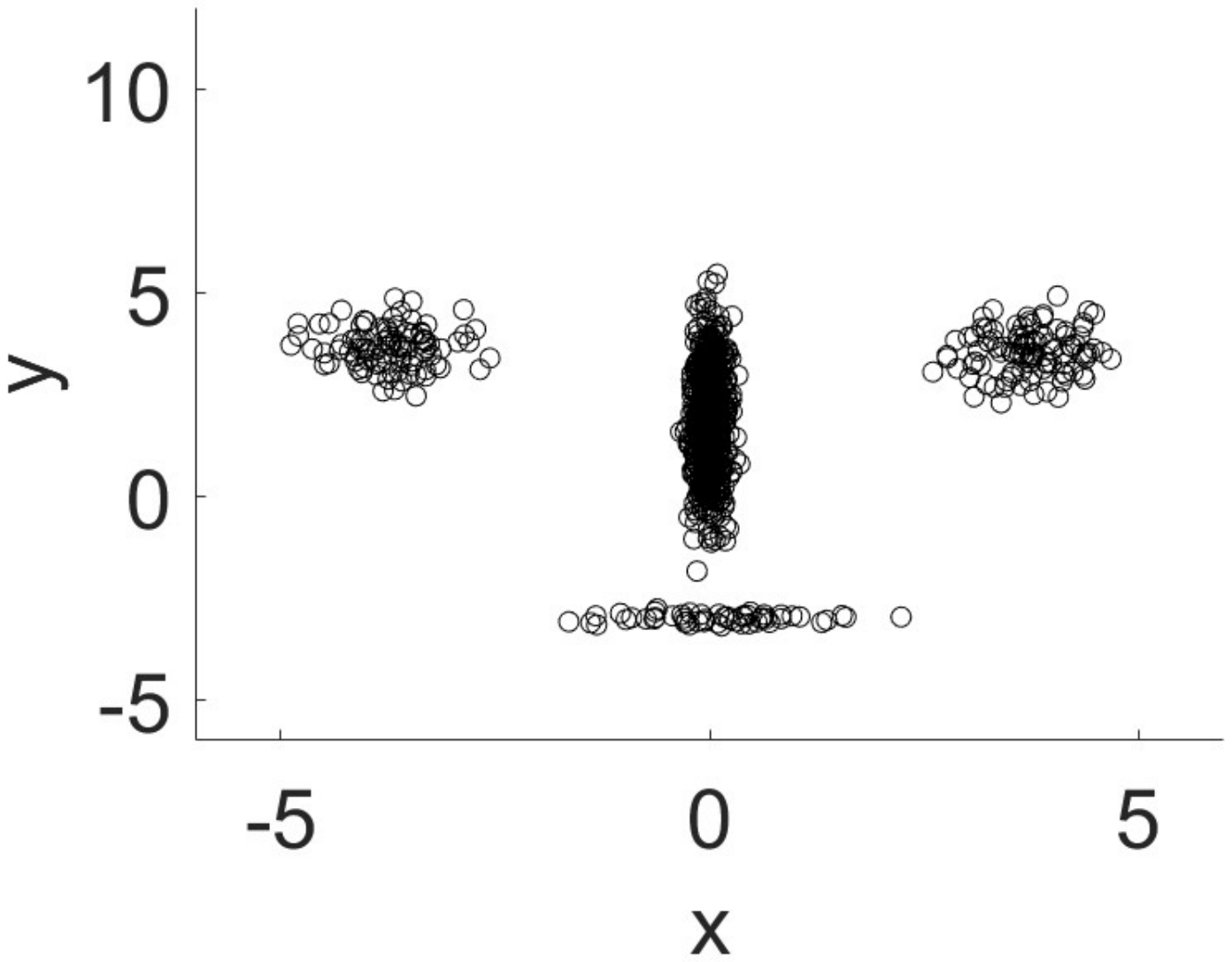}  
    \caption{}
    \label{fig: 2D Gaussian Mixture Gaussian MH Reweight small}
\end{subfigure}
\caption{Scatter plots of the empirical samples returned by different testing algorithms in \ref{example: 2D Gaussian Mixture}; (a) MALA + Snooker + BD; (b) MALA + BD; (c) MALA + Reweight; (d) Gaussian MH + Reweight.}
\label{fig: 2d gaussian mixture scattered plots}
\end{figure}

For two different test functions $f_1(x,y) = y$ and $f_2(x,y) = \frac{x^2}{3} + \frac{y^2}{5}$, we plot the estimated values of $\mathbb{E}[f_1]$ and $\mathbb{E}[f_2]$ given by weighted samples associated with different algorithms. From the two plots given in Figure (\ref{fig: 2d gaussian mixture estimations of two functions}) we can see that Algorithm \ref{alg: AIS + snooker} (\emph{i.e.}, ``MALA+Snooker+BD”) returns much more accurate estimations of the two expectation values than the other three algorithms. Furthermore, we also provide scattered plots of the empirical samples generated by the four algorithms in Figure (\ref{fig: 2d gaussian mixture scattered plots}), from which we can see that only Algorithm \ref{alg: AIS + snooker} succeeds in discovering the mode centered around $\mu_2$. From all the plots above, we conclude that the exploration part serves as an essential factor for discovering modes and generating samples of higher quality in this example. 

\subsubsection{Ginzburg-Landau Distributions}
\label{example: GL distribution}
The Ginzburg-Landau (GL) theory is a mathematical model used in the studies of superconductivity~\cite{hoffmann2012ginzburg, hohenberg2015introduction}. Here, we consider a continuous distribution coming from a simplified version of the Ginzburg-Landau model in $\mathbb{R}^d$, whose associated energy function $\mathcal{U}$ is defined as 
\begin{equation}
\label{eqn: continuous energy in GL}
\begin{aligned}
\mathcal{U}(x(\boldsymbol{r})) := \int_{\Omega}\Big(\frac{\lambda}{2}|\nabla_{\boldsymbol{r}} x(\boldsymbol{r})|^2+\frac{1}{\lambda}V(x(\boldsymbol{r}))\Big)d\boldsymbol{r}, 
\end{aligned}    
\end{equation}
where $\Omega$ is the domain, $x=x(\boldsymbol{r}):\Omega \rightarrow \mathbb{R}$ is a sufficiently smooth function with Dirichlet boundary condition $x(\boldsymbol{r})=0 \ (\forall \ \boldsymbol{r} \in \partial \Omega)$, $\lambda >0$ is some fixed parameter and $V(x)=\frac{1}{4}(1-x^2)^2$ is the potential function.

\begin{figure}[H]
\centering
\begin{subfigure}{.32\textwidth}
    \centering
    \includegraphics[width=.95\linewidth]{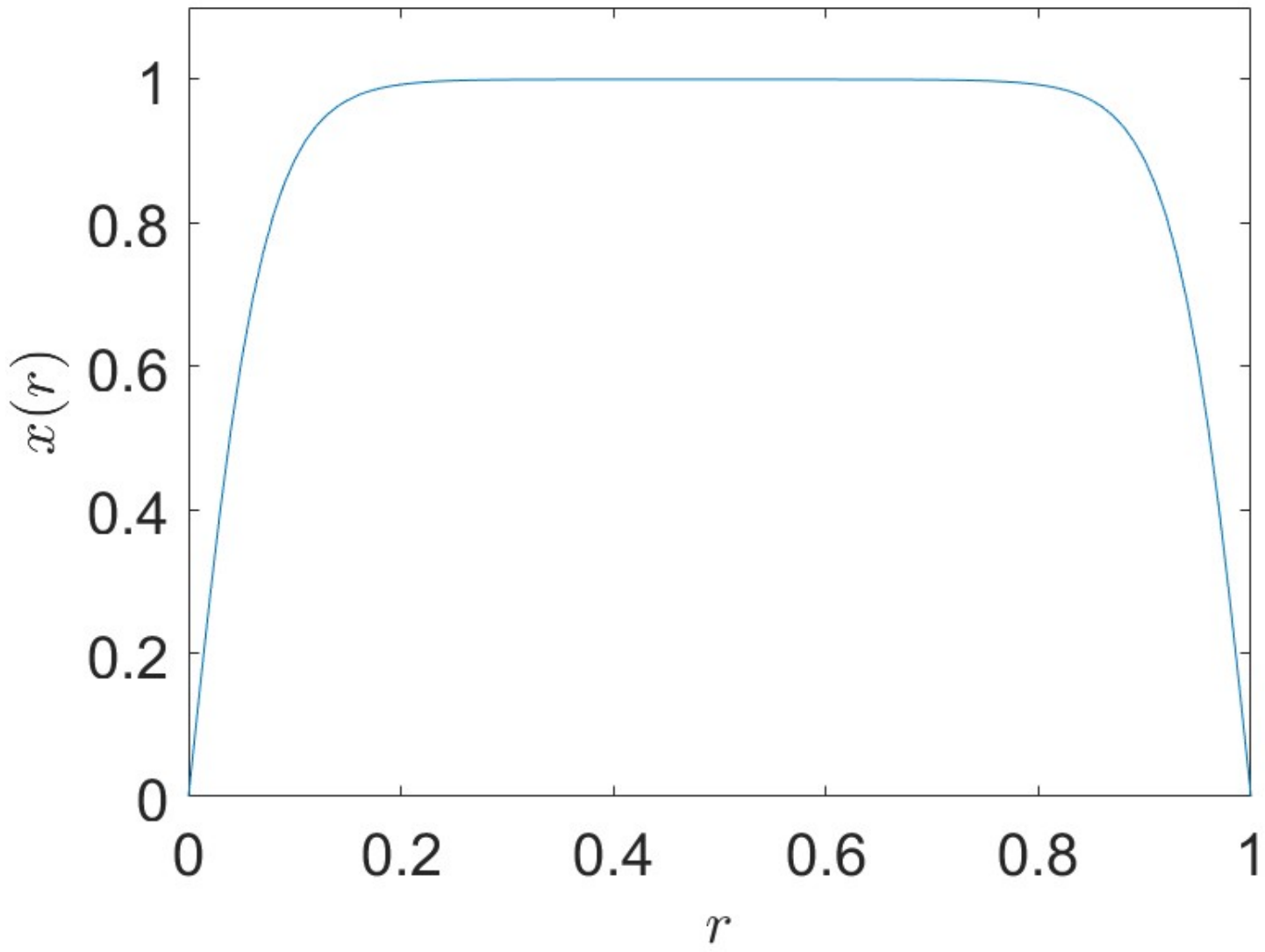} 
    \caption{Minimizer 1: $\boldsymbol{x}_{+}^{(1)}$}
    \label{fig: 16D Ginzburg Landau 1D minimizer1}
\end{subfigure}
\begin{subfigure}{.32\textwidth}
    \centering
    \includegraphics[width=.95\linewidth]{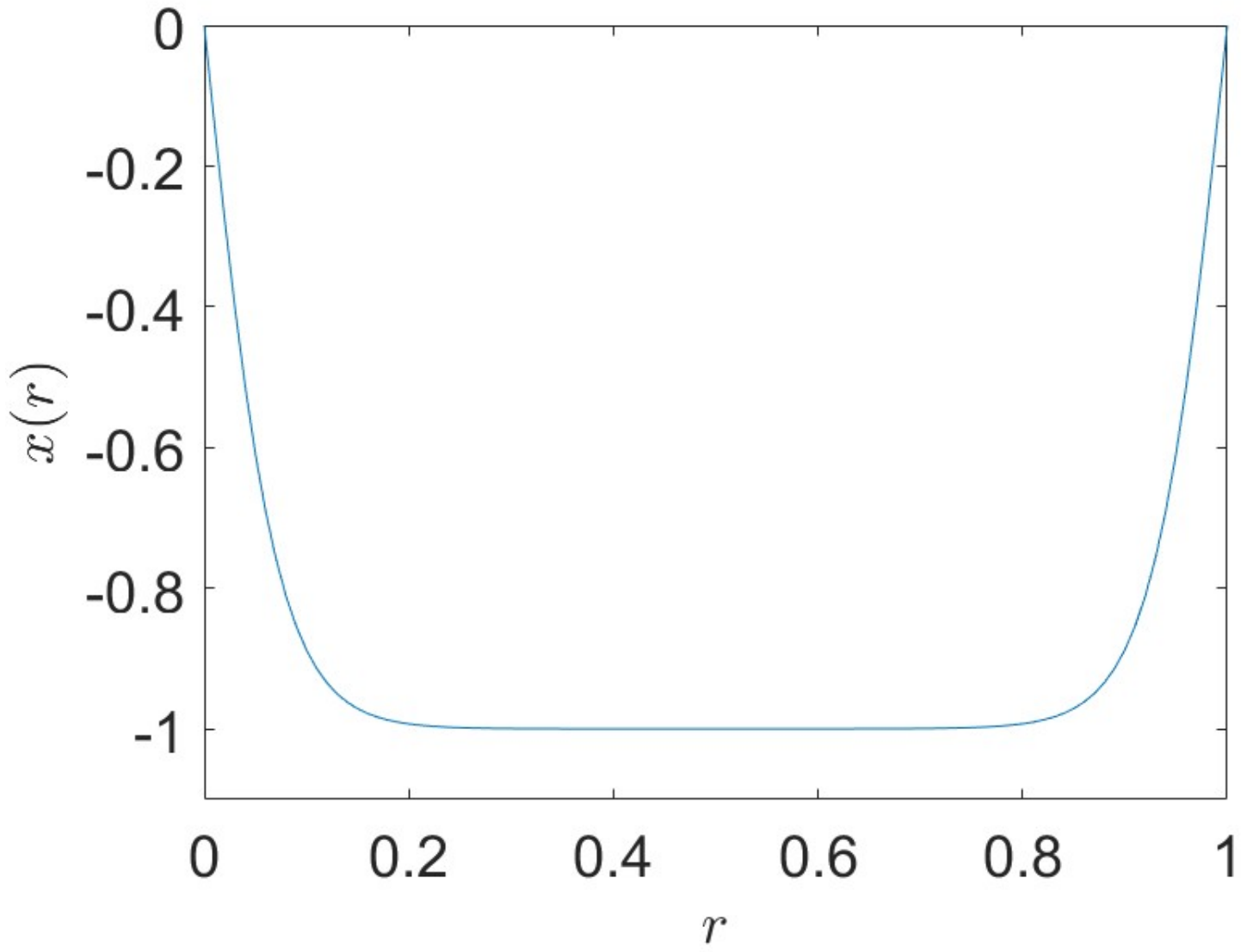}
    \caption{Minimizer 2: $\boldsymbol{x}_{-}^{(1)}$}
    \label{fig: 16D Ginzburg Landau 1D minimizer2}
\end{subfigure}
\begin{subfigure}{.32\textwidth}
    \centering
    \includegraphics[width=.95\linewidth]{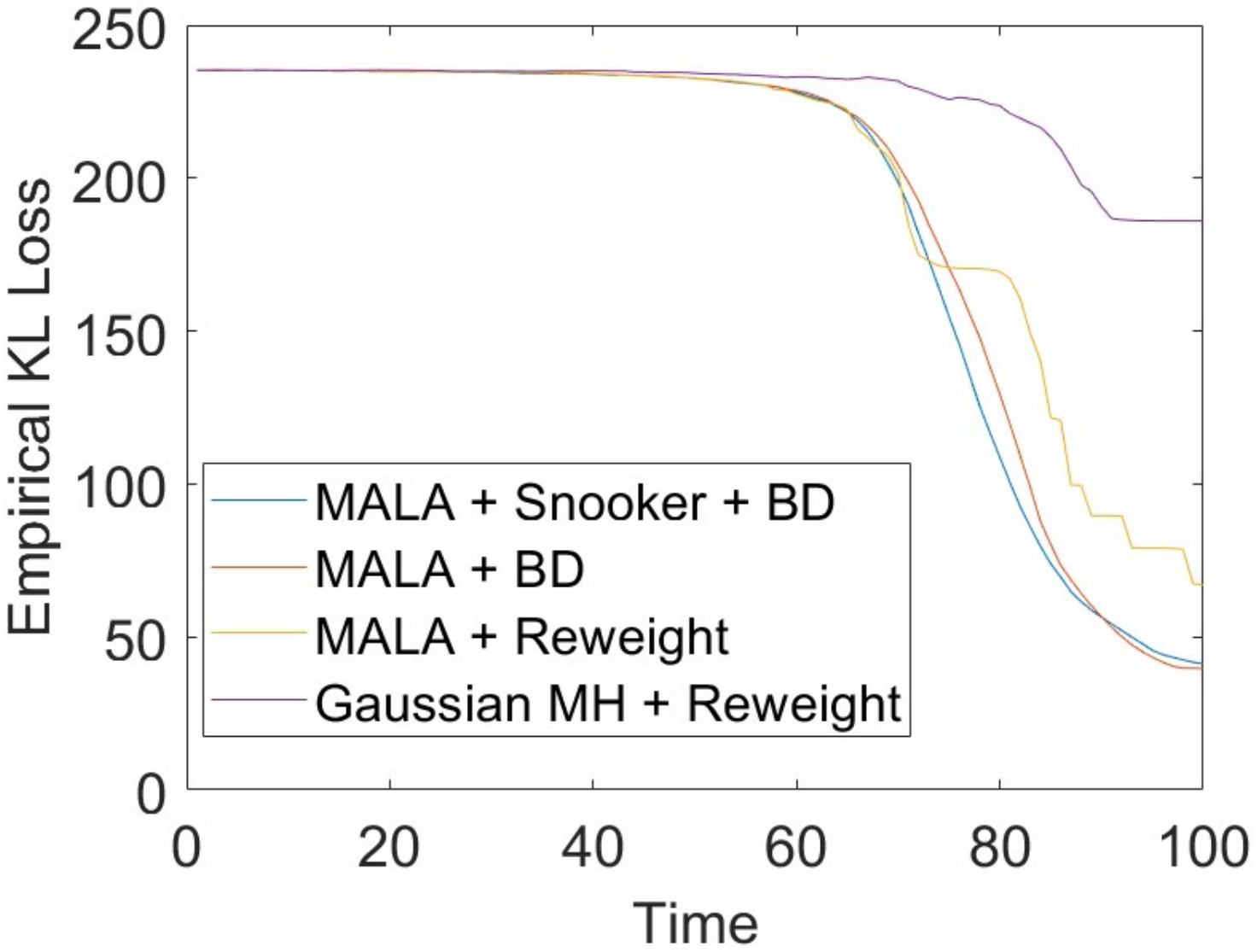} 
    \caption{Empirical KL Loss}
    \label{fig: 16D Ginzburg Landau 1D Empirical KL loss}
\end{subfigure}
\caption{(a) \& (b) Two minimizers $\boldsymbol{x}_{+}^{(1)}$ and $\boldsymbol{x}_{-}^{(1)}$ of the 1D Ginzburg-Landau Energy (\ref{eqn: continuous energy in GL}) with $\lambda = 0.05$; (c) Plots of $\mathcal{L}_{\text{KL}}$ evaluated at evolving weighted samples generated by different testing algorithms in the 1D case of \ref{example: GL distribution}.}
\label{fig: 16D GL 1D basics}
\end{figure}

For the first case when $\Omega = [0,L]$, the function $x=x(\boldsymbol{r})$ is discretized as $\boldsymbol{x} := (x_i)_{i=1}^{d}$ with boundary condition $x_0 = x_{d+1}=0$, which contains the values of the function $x=x(\boldsymbol{r})$ on the uniform grid points $\{ih\}_{i=0}^{d+1}$ with $h=\frac{L}{d+1}$. Then, the corresponding discrete approximation of (\ref{eqn: continuous energy in GL}) is given by 
\begin{equation}
\label{eqn: discrete energy in GL 1D}
\begin{aligned}
\mathcal{U}(x(r)) \approx U_1(\boldsymbol{x}) := \sum_{i=1}^{d+1}\Bigg(\frac{\lambda}{2}\Big|\frac{x_{i}-x_{i-1}}{h}\Big|^2 + \frac{1}{4\lambda}(1-x_{i}^2)^2\Bigg).
\end{aligned}    
\end{equation}

\begin{figure}[H]
\centering
\begin{subfigure}{.24\textwidth}
    \centering
    \includegraphics[width=.95\linewidth]{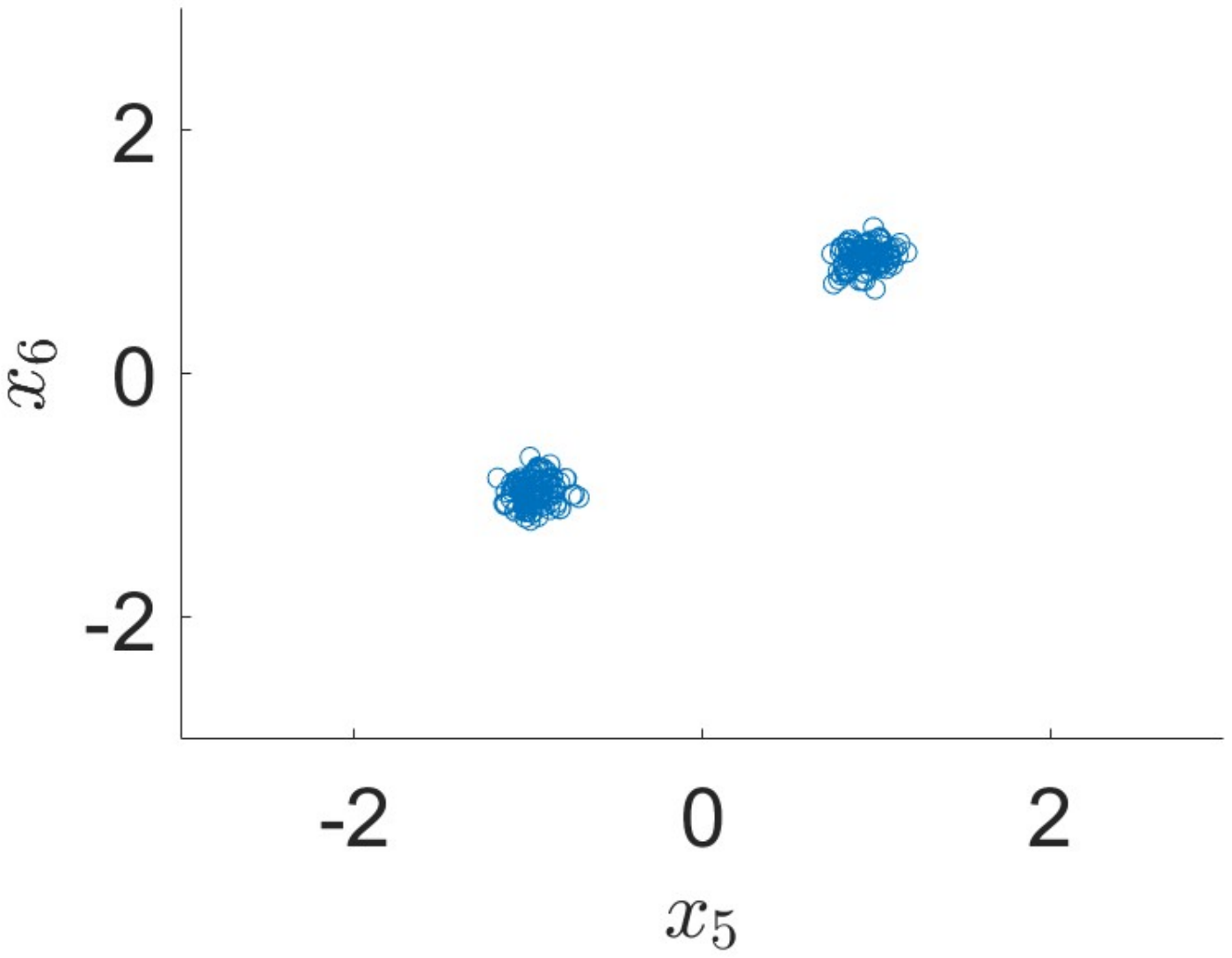} 
    \caption{}
    \label{fig: 16D GL 1D MALA Snooker BD}
\end{subfigure}
\begin{subfigure}{.24\textwidth}
    \centering
    \includegraphics[width=.95\linewidth]{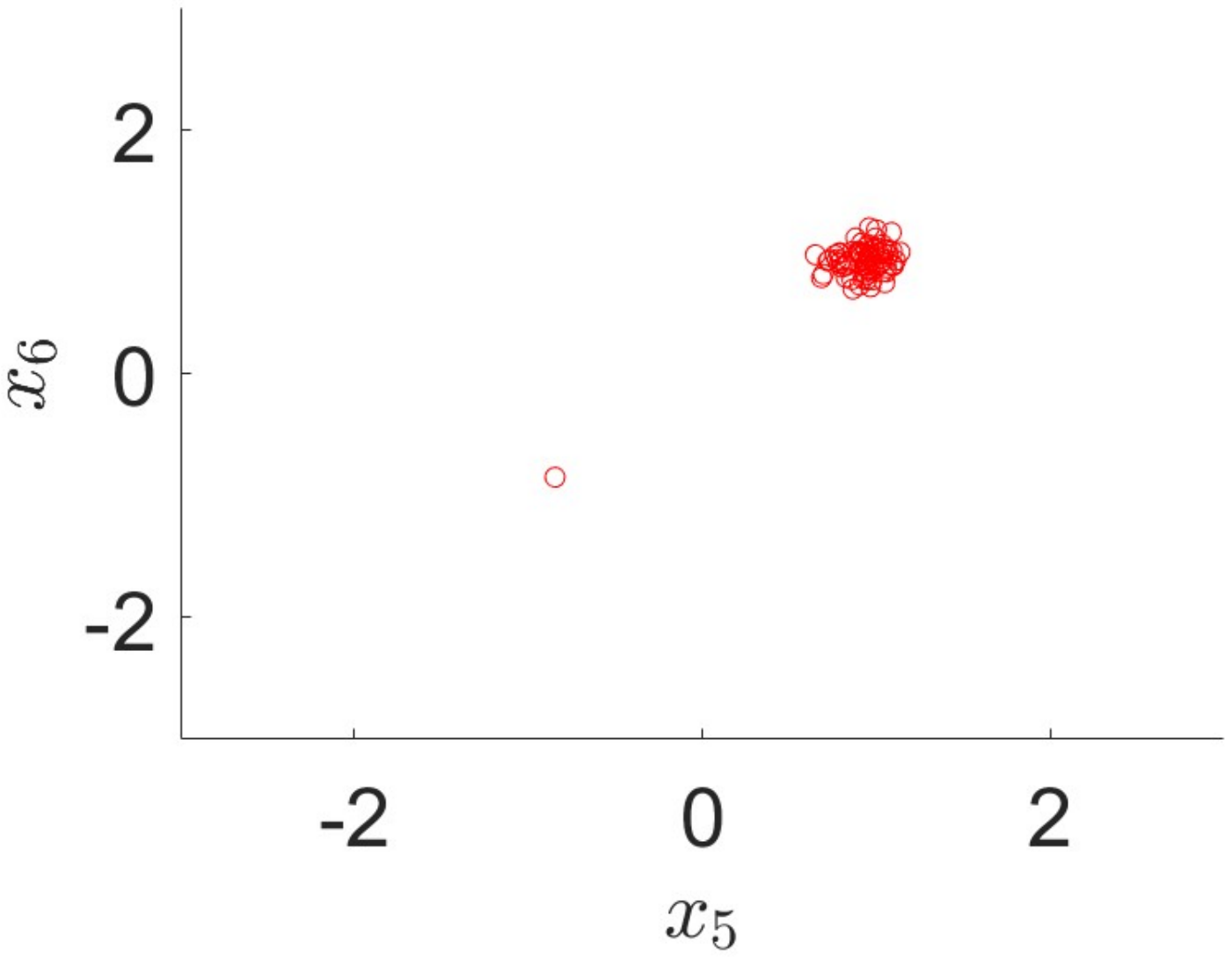}
    \caption{}
    \label{fig: 16D GL 1D MALA BD}
\end{subfigure}
\begin{subfigure}{.24\textwidth}
    \centering
    \includegraphics[width=.95\linewidth]{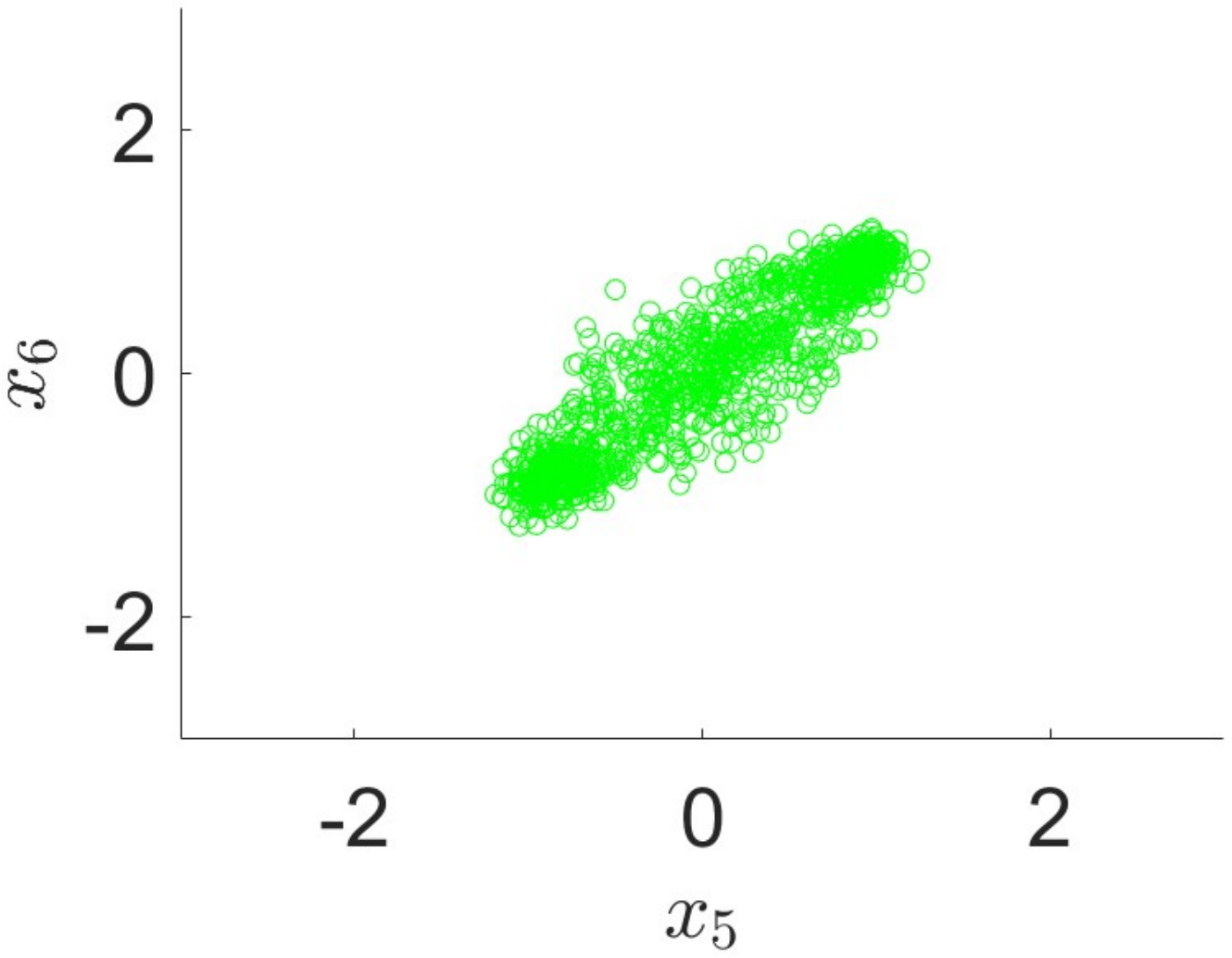}  
    \caption{}
    \label{fig: 16D GL 1D MALA Reweight}
\end{subfigure}
\begin{subfigure}{.24\textwidth}
    \centering
    \includegraphics[width=.95\linewidth]{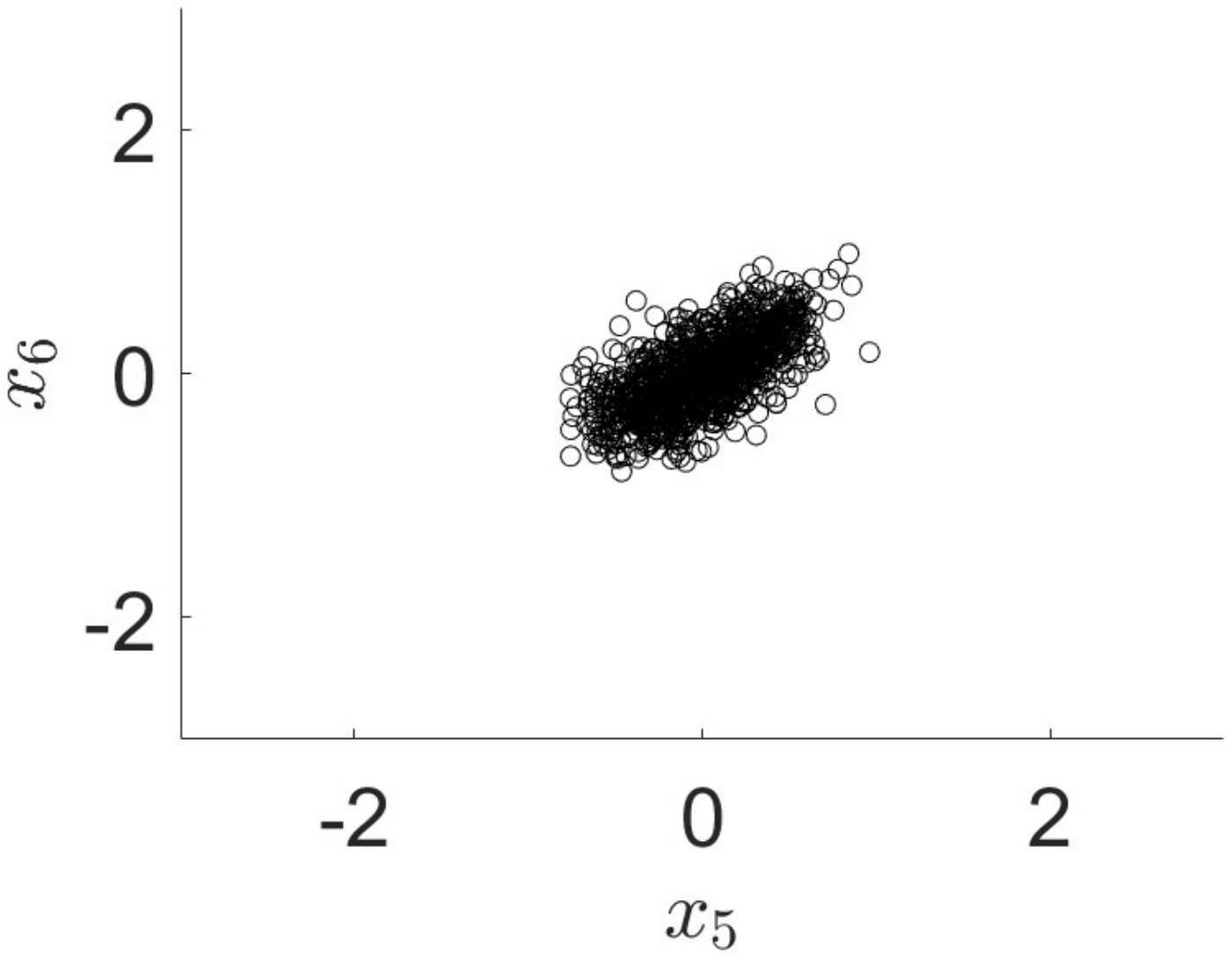}
    \caption{}
    \label{fig: 16D GL 1D Gaussian MH Reweight}
\end{subfigure}
\caption{Scatter plots of the marginal distribution at $(x_5, x_6)$ returned by different testing algorithms in the 1D case of \ref{example: GL distribution}; (a) MALA + Snooker + BD; (b) MALA + BD; (c) MALA + Reweight; (d) Gaussian MH + Reweight.}
\label{fig: 16D GL 1D scatter}
\end{figure}

The target distribution is set to be the Boltzmann distribution $p_1(\boldsymbol{x}) \propto e^{-\beta U_1(\boldsymbol{x})}$ associated with (\ref{eqn: discrete energy in GL 1D}), where $\beta = \frac{1}{T}$ is the inverse temperature parameter. For our numerical experiments on the 1D case, we choose $\lambda = 0.05, d = 16, \beta = 3$ and $L = 1$. From previous studies~\cite{weinan2004minimum}, we know that $p_1$ contains two local modes centered around the two local minimizers $\boldsymbol{x}_{+}^{(1)}$ and $\boldsymbol{x}_{-}^{(1)}$ of the 1D GL energy function (\ref{eqn: discrete energy in GL 1D}), which are plotted in Figure (\ref{fig: 16D Ginzburg Landau 1D minimizer1}) and (\ref{fig: 16D Ginzburg Landau 1D minimizer2}) respectively. In our implementation, we pick the number of samplers and time steps to be $N=1000$ and $L = 100$. Moreover, for the standard AIS whose transition kernel is given by an MH Algorithm, we pick its Gaussian proposal density to be $\mathcal{N}(\boldsymbol{0}, 0.01\boldsymbol{I}_{16})$. The starting distribution is selected to be the normal distribution $\mathcal{N}(\boldsymbol{0}, 0.01\boldsymbol{I}_{16})$ in $\mathbb{R}^{16}$. In order to compare the testing algorithms' abilities to discover modes of the underlying distribution, we also plot the marginal distribution of $(x_5, x_6)$ for the empirical samples returned by each algorithm, which are exhibited in Figure (\ref{fig: 16D GL 1D scatter}).

\begin{figure}[H]
\centering
\begin{subfigure}{.32\textwidth}
    \centering
    \includegraphics[width=.95\linewidth]{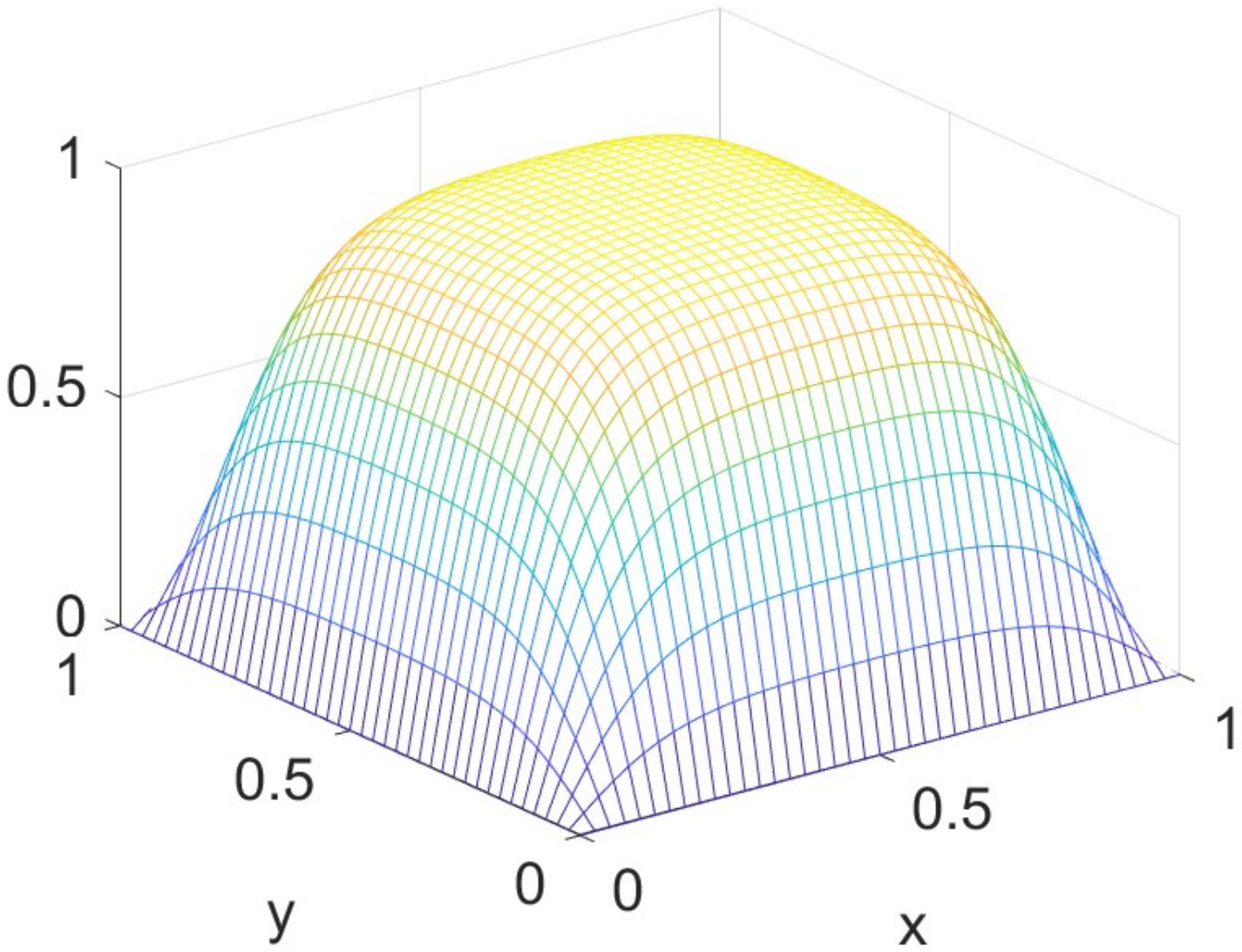} 
    \caption{Minimizer 1: $\boldsymbol{x}_{+}^{(2)}$}
    \label{fig: 16D Ginzburg Landau 2D minimizer1}
\end{subfigure}
\begin{subfigure}{.32\textwidth}
    \centering
    \includegraphics[width=.95\linewidth]{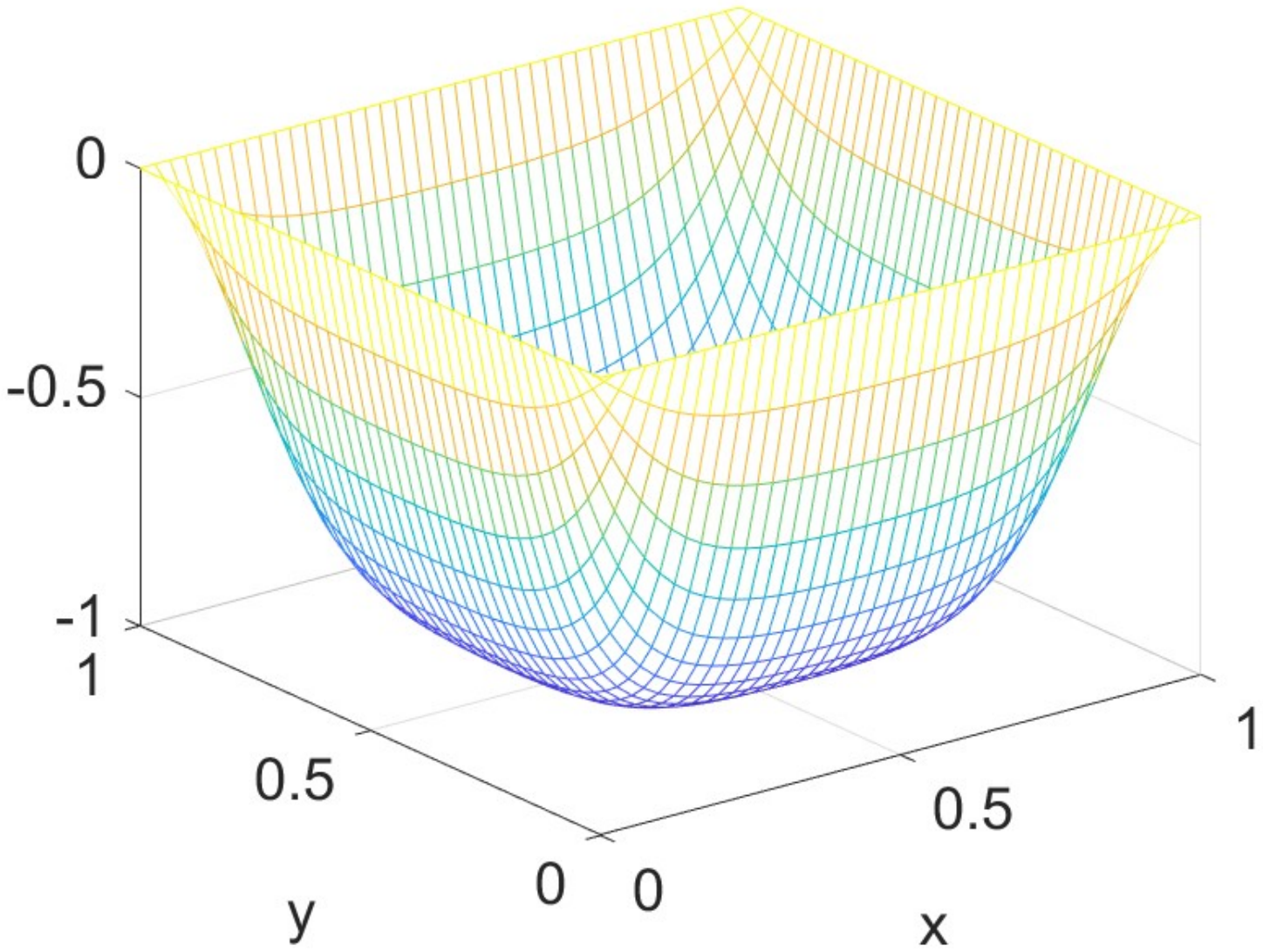}
    \caption{Minimizer 2: $\boldsymbol{x}_{-}^{(2)}$}
    \label{fig: 16D Ginzburg Landau 2D minimizer2}
\end{subfigure}
\begin{subfigure}{.32\textwidth}
    \centering
    \includegraphics[width=.95\linewidth]{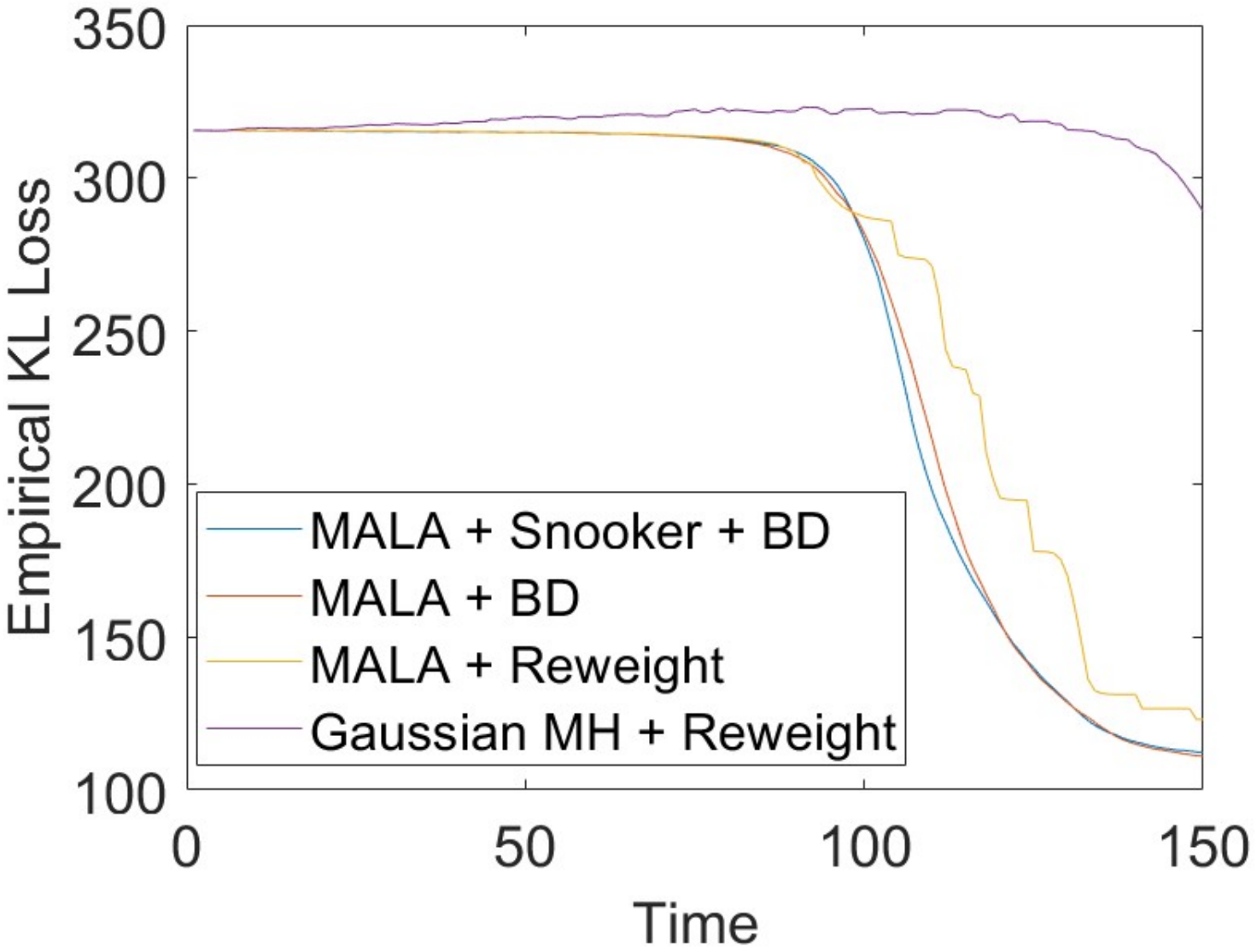}
    \caption{Empirical KL Loss}
    \label{fig: 16D Ginzburg Landau 2D Empirical KL loss}
\end{subfigure}
\caption{(a) \& (b) Two minimizers $\boldsymbol{x}_{+}^{(2)}$ and $\boldsymbol{x}_{-}^{(2)}$ of the 2D Ginzburg-Landau Energy (\ref{eqn: continuous energy in GL}) with $\lambda = 0.125$; (c) Plots of $\mathcal{L}_{\text{KL}}$ evaluated at evolving weighted samples generated by different testing algorithms in the 2D case of \ref{example: GL distribution}.}
\label{fig: 16D GL 2D basics}
\end{figure}

For the second case when $\Omega = [0, L]^2$, the function $x=x(\boldsymbol{r})$ is discretized as $\boldsymbol{x} := (x_{i,j})_{i,j=1}^{d}$ with boundary condition $x_{0,k} = x_{k,0} = x_{(d+1),l} = x_{l,(d+1)}=0 \ (\forall \ 0 \leq k,l \leq d+1)$, which contains the values of the function $x=x(\boldsymbol{r})$ on the uniform grid points $\{(ih, jh)\}_{i,j=0}^{d+1}$ with $h=\frac{L}{d+1}$. Then, the corresponding discrete approximation of (\ref{eqn: continuous energy in GL}) is given by 
\begin{equation}
\label{eqn: discrete energy in GL 2D}
\begin{aligned}
\mathcal{U}(x(r)) &\approx U_2(\boldsymbol{x}) := \sum_{i=1}^{d}\sum_{j=1}^{d}\Bigg(\frac{\lambda}{4}\Big|\frac{x_{i,j}-x_{(i-1),j}}{h}\Big|^2 + \frac{\lambda}{4}\Big|\frac{x_{i,j}-x_{(i+1),j}}{h}\Big|^2 \\
&+ \frac{\lambda}{4}\Big|\frac{x_{i,j}-x_{i,(j-1)}}{h}\Big|^2 + \frac{\lambda}{4}\Big|\frac{x_{i,j}-x_{i,(j+1)}}{h}\Big|^2 + \frac{1}{4\lambda}(1-x_{i,j}^2)^2\Bigg).
\end{aligned}    
\end{equation}

\begin{figure}[H]
\centering
\begin{subfigure}{.24\textwidth}
    \centering
    \includegraphics[width=.95\linewidth]{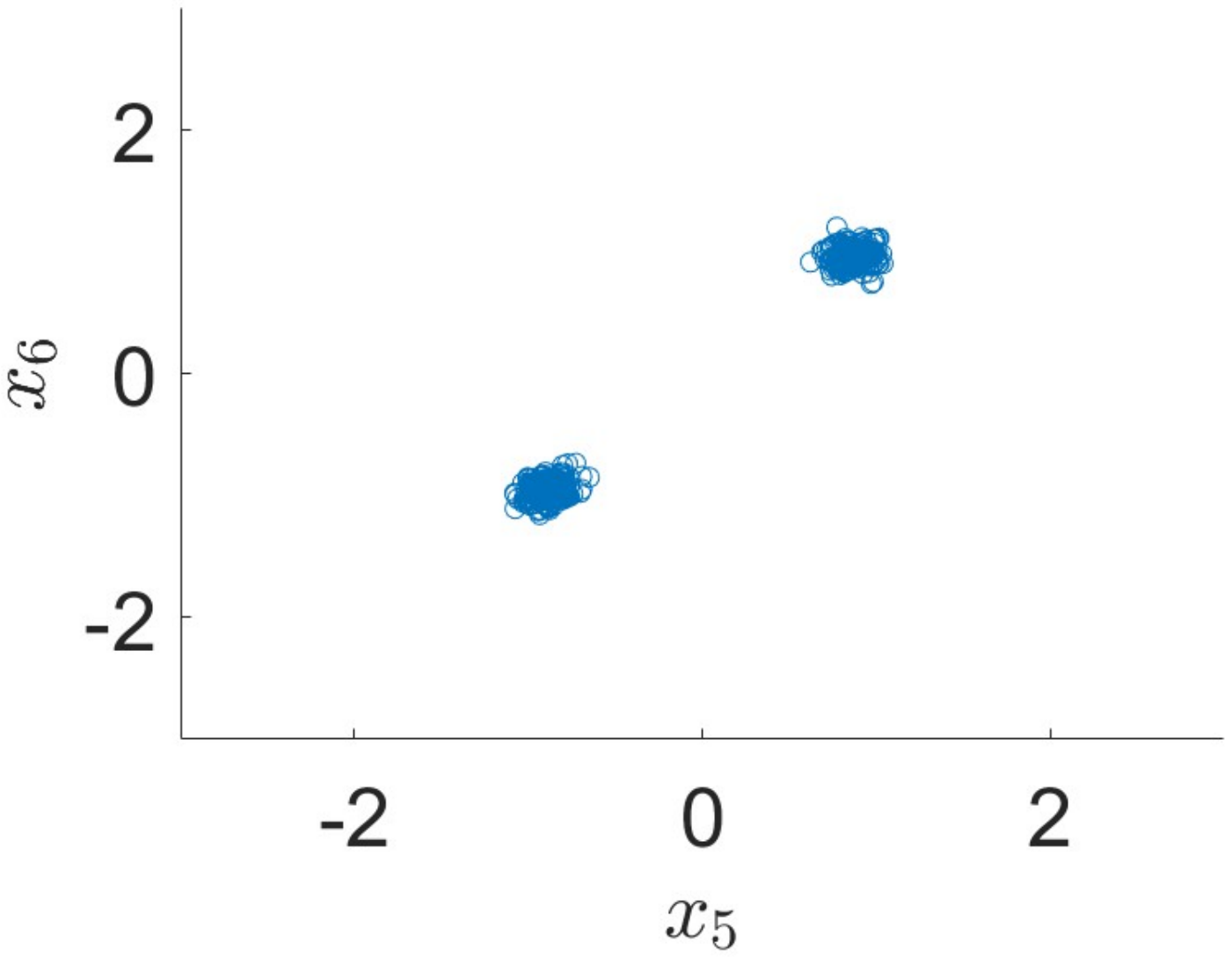} 
    \caption{}
    \label{fig: 16D GL 2D MALA Snooker BD}
\end{subfigure}
\begin{subfigure}{.24\textwidth}
    \centering
    \includegraphics[width=.95\linewidth]{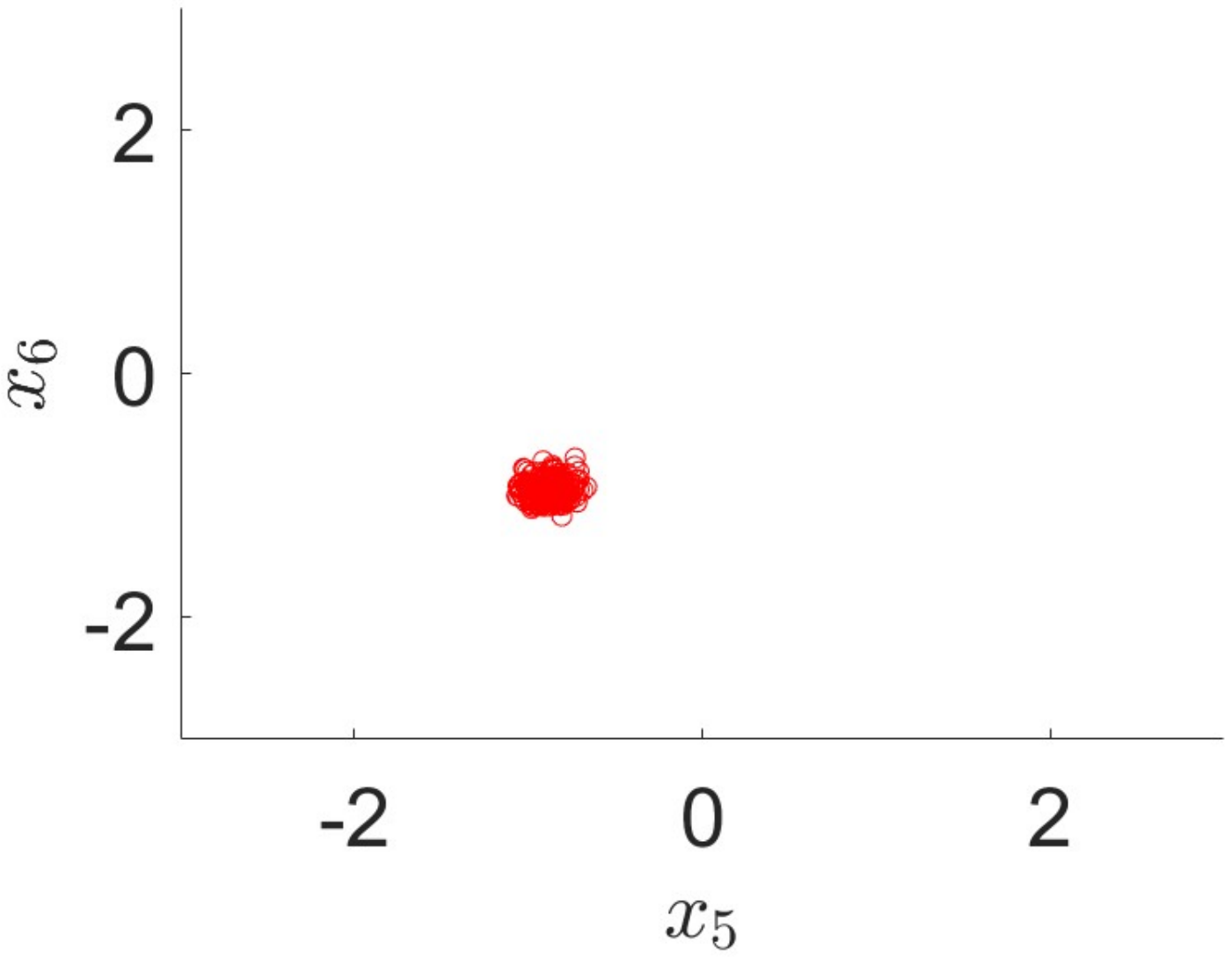}
    \caption{}
    \label{fig: 16D GL 2D MALA BD}
\end{subfigure}
\begin{subfigure}{.24\textwidth}
    \centering
    \includegraphics[width=.95\linewidth]{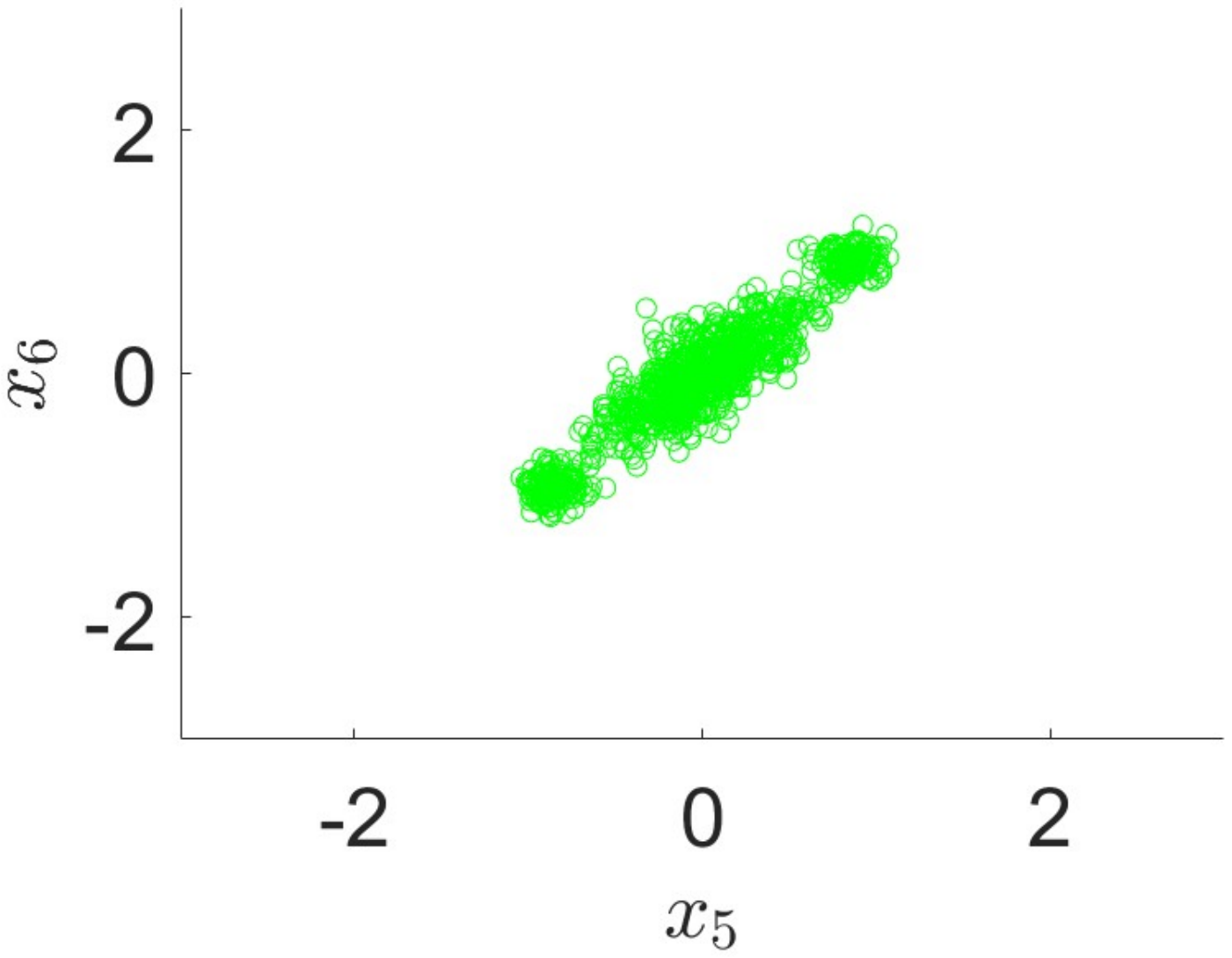}  
    \caption{}
    \label{fig: 16D GL 2D MALA Reweight}
\end{subfigure}
\begin{subfigure}{.24\textwidth}
    \centering
    \includegraphics[width=.95\linewidth]{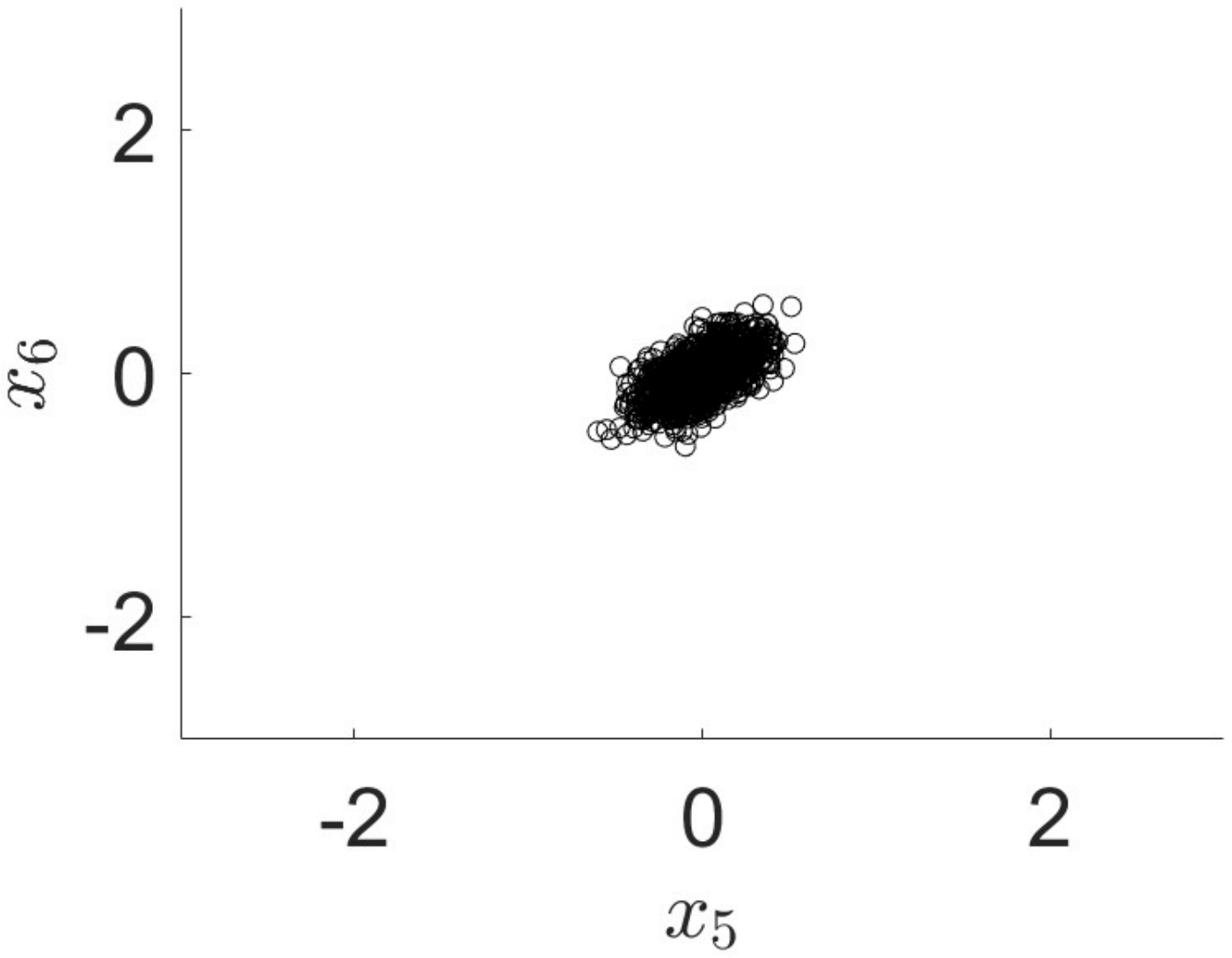}
    \caption{}
    \label{fig: 16D GL 2D Gaussian MH Reweight}
\end{subfigure}
\caption{Scatter plots of the marginal distribution at $(x_5, x_6)$ returned by different testing algorithms in the 2D case of \ref{example: GL distribution}; (a) MALA + Snooker + BD; (b) MALA + BD; (c) MALA + Reweight; (d) Gaussian MH + Reweight.}
\label{fig: 16D GL 2D scatter}
\end{figure}

In a similar manner to the 1D case, the target distribution for the 2D case is picked to be the Boltzmann distribution $p_2(\boldsymbol{x}) \propto e^{-\beta U_2(\boldsymbol{x})}$ associated with (\ref{eqn: discrete energy in GL 2D}), where $\beta = \frac{1}{T}$ and $T$ is the temperature. For our numerical experiments on the 2D case, we choose $\lambda = 0.125, d = 4, \beta = 10$ and $L = 1$. Analogously, $p_2$ contains two local modes centered around the two local minimizers $\boldsymbol{x}_{+}^{(2)}$ and $\boldsymbol{x}_{-}^{(2)}$ of the 2D GL energy function (\ref{eqn: discrete energy in GL 2D}), which are exhibited in Figure (\ref{fig: 16D Ginzburg Landau 2D minimizer1}) and (\ref{fig: 16D Ginzburg Landau 2D minimizer2}) respectively. In our implementation, we pick the number of samplers and time steps to be $N=1000$ and $L = 150$. For the standard AIS whose transition kernel is given by an MH Algorithm, we pick its Gaussian proposal density to be $\mathcal{N}(\boldsymbol{0}, 0.001\boldsymbol{I}_{16})$. The starting distribution is selected to be the normal distribution $\mathcal{N}(\boldsymbol{0}, 0.01\boldsymbol{I}_{16})$ in $\mathbb{R}^{16}$. Again, we examine the testing algorithms' capabilities of identifying the modes in the target distribution by plotting the marginal distribution of $(x_5, x_6)$ for the empirical samples returned by each algorithm in Figure (\ref{fig: 16D GL 2D scatter}).

On the one hand, from the two plots of the empirical KL loss functions above, we find that the two Interacting AIS algorithms (variants of Algorithm~\ref{alg: AIS + snooker}, \emph{i.e.}, ``MALA+BD” and ``MALA+Snooker+BD”) converge faster and generate samples of higher quality compared to the two standard AIS algorithms (variants of Algorithm~\ref{alg: AIS}, \emph{i.e.}, ``MALA+Reweight” and ``Gaussian MH + Reweight”) when measured under $\mathcal{L}_{\text{KL}}$. On the other hand, from the scattered plots of the marginal distributions, we also observe that the Algorithm \ref{alg: AIS + snooker} can discover the local modes in a more effective way than the other three algorithms. Both claims above justify that the exploration part can indeed help improve the efficiency of the AIS algorithm. 

\begin{remark}
Since marginal of the Ginzburg-Landau distributions above has only two modes, we also consider a synthetic target distribution with more modes in its marginal to further assess the performance of our proposed interacting AIS algorithms with exploration. Due to space constraints, we defer this additional experiment to the third section of the supplementary material.
\end{remark}

\subsubsection{Bayesian Neural Network (BNN)}
\label{example: Bayesian Neural Network}
For the third case of continuous distributions, we consider an experimental setting in Bayesian deep learning, which is similar to the experiments presented in~\cite{liang2024smc}. We evaluate the performance of the proposed method on Bayesian neural networks using the Boston Housing dataset~\footnote{https://www.kaggle.com/code/prasadperera/the-boston-housing-dataset}. For a fixed training set $\mathcal{X}_{\text{train}} := \{(x_i, y_i)\}_{i=1}^{N}$ with $x_i \in \mathbb{R}^{13}$ and $y_i \in \mathbb{R}$, we assume that the relation between the input and output can be modeled as $y_i = f_{\theta}(x_i) + \sigma_i$, where $\sigma_i \sim \mathcal{N}(0,1)$ for any $1 \leq i \leq N$ and $f_{\theta}: \mathbb{R}^{13} \rightarrow \mathbb{R}$ is a two-layer neural network of hidden dimension $16$. Specifically, $f_{\theta}$ can be expressed as $f_{\theta}(x) = W_2\eta(W_1x + b_1) + b_2$, where $W_{1} \in \mathbb{R}^{16 \times 13}, W_2 \in \mathbb{R}^{1 \times 16}, b_1 \in \mathbb{R}^{16}, b_{2} \in \mathbb{R}$ and $\eta$ is selected to be the ReLU activation function. We further pick the prior distribution of the neural network parameters $\theta = (W_1, W_2,b_1,b_2)$ to be a multivariate Gaussian distribution with covariance matrix $\sigma^2 \mI_{241}$ for $\sigma =3$, \emph{i.e.}, $\theta \sim \mathcal{N}(0,\sigma^2 \mI_{241}) =: p_{\text{prior}}(\theta)$. From Bayes' Theorem, we then have that the posterior distribution of $\theta$ conditioned on the training set can be expressed as 
\begin{equation}
\label{eqn: Bayes' Theorem BNN}
p_{\text{posterior}}(\theta \ | \ \mathcal{X}_{\text{train}}) \propto p_{\text{prior}}(\theta)p(\mathcal{X}_{\text{train}} \ | \ \theta) = p_{\text{prior}}(\theta)e^{-\frac{1}{2N}\sum_{i=1}^{N}(y_i - f_{\theta}(x_i))^2}
\end{equation}

\begin{figure}[H]
\centering
\begin{subfigure}[b]{0.49\textwidth}   
\centering 
\includegraphics[width=\textwidth]{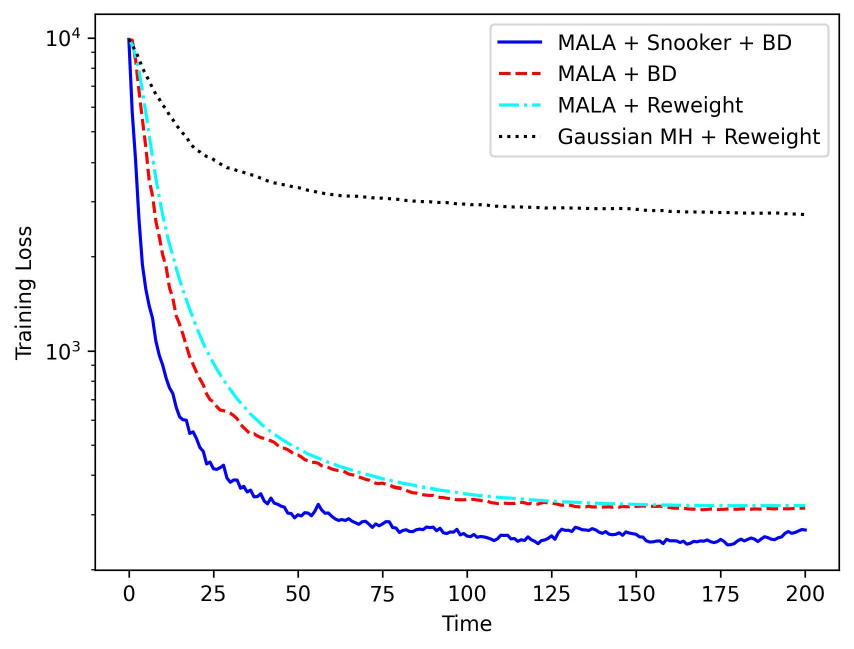}
\caption{BNN Training Loss}
\label{fig: BNN Empirical KL loss}
\end{subfigure}
\hfill
\begin{subfigure}[b]{0.49\textwidth}   
\centering
\includegraphics[width=\textwidth]{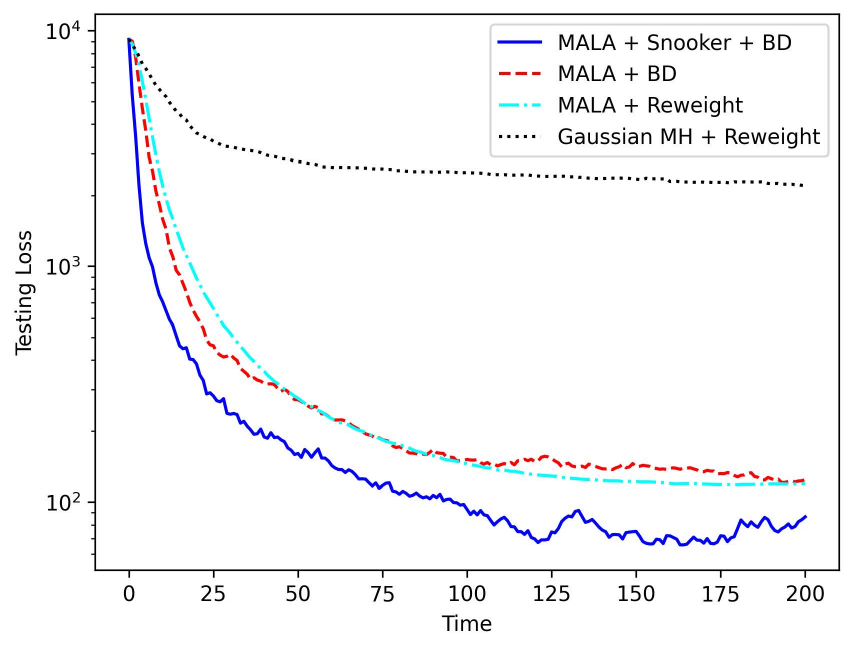} 
\caption{BNN Testing Loss}
\label{fig: BNN Validation loss} 
\end{subfigure}
\caption[]
{\small (a) Plots of $\mathcal{L}_{\text{KL}}$ for the posterior distribution~\eqref{eqn: Bayes' Theorem BNN} evaluated at evolving weighted samples generated by different testing algorithms; (b) Plots of the squared loss function evaluated at the testing set.} 
\label{fig: BNN}
\end{figure}

We then apply the two Interacting AIS algorithms (variants of Algorithm~\ref{alg: AIS + snooker}) and the two standard AIS algorithms (variants of Algorithm~\ref{alg: AIS}) to sample from the target posterior distribution~\eqref{eqn: Bayes' Theorem BNN}. The ensemble of sampled parameters are then evaluated on both the empirical KL loss function $\mathcal{L}_{KL}$ and the squared loss under the testing dataset. In our implementation of the four algorithms, we pick the number of samplers and time steps to be $N=200$ and $L = 200$. From the training and testing loss curves shown above, we observe that the Interacting AIS Algorithm~\ref{alg: AIS + snooker} (``MALA+Snooker+BD”) converges faster than the other three methods. This again highlights the effectiveness of the exploration component in improving the overall efficiency of the AIS algorithm.

\subsection{Discrete Distributions: Ising Models}
For the case of sampling discrete distributions, we focus on the Ising model, which is a simple but commonly used model in statistical physics. Specifically, we test Algorithm \ref{alg: AIS + genetic} on a few one-dimensional and two-dimensional Ising models of similar form as the ones used in~\cite{peng2023generative}. Similar to the cases of continuous distributions, we also compare Algorithm \ref{alg: AIS + genetic} with both interacting AIS without exploration and standard AIS to justify the essentiality of the exploration part. For the standard AIS, the transition kernel is picked to be the Glauber dynamics. For any empirical distribution $\hat{p}$ formed by the samples returned by a testing algorithm, we measure the quality of the samples via the $L_2$ loss function $\mathcal{L}_2(\hat{p}, p) = \|\hat{p}-p^\ast\|_2$, where $p^\ast(\cdot) \propto e^{-U(\cdot)}: \{-1,1\}^d \rightarrow [0,1]$ is the target distribution. For each Ising model, we perform three experiments to illustrate the effectiveness of Algorithm \ref{alg: AIS + genetic}. In the first experiment, we fix the sample size and run each algorithm multiple times to create a box plot of the $L_2$ loss function's value for each algorithm. For the second experiment, we investigate how the $L_2$ loss function depends on the sample size for each algorithm, where the standard Monte Carlo method is chosen to be the reference. In the third experiment, we set the sample size to be relatively large and plot the samples generated by Algorithm \ref{alg: AIS + genetic}.

\subsubsection{1D Ising Models}
\label{sec: 1D Ising}
Consider a generalized one-dimensional Ising model, where interaction exists between both the nearest and the second nearest neighbor pairs. The associated distribution $p(\cdot):\{-1,1\}^d \rightarrow [0,1]$ is defined as
\begin{equation}
\label{eqn: 1D Ising}
\begin{aligned}
p(\boldsymbol{x}) = p(x_1,x_2,\cdots,x_d) \propto \exp\Big(-\beta\sum_{i=1}^{d-1}J_1x_ix_{i+1}-\beta\sum_{i=1}^{d-2}J_2x_{i}x_{i+2}\Big).    
\end{aligned}    
\end{equation}

Note that $\beta =\frac{1}{T}$ in (\ref{eqn: 1D Ising}) is the inverse temperature parameter. In our experiments on the one-dimensional Ising models, we set $\beta =0.8$, $d= 20$ and the number of time steps to be $L=64$. The initial distribution is fixed to be the uniform distribution over $\{-1,1\}^{20}$. Moreover, for the experiments producing box plots, we fix the sample size to be $N = 512$. For the experiments investigating how the loss function $\mathcal{L}_2(\hat{p},p)$ depends on the sample size, we increase the sample size from $N=64$ to $N=1024$, doubling it at each round. Finally, when visualizing the empirical distribution formed by the samples generated by Algorithm \ref{alg: AIS + genetic}, we choose the sample size to be $N =65536$.

\begin{figure}[H]
\centering
\begin{subfigure}{.49\textwidth}
    \centering
    \includegraphics[width=1.0\linewidth]{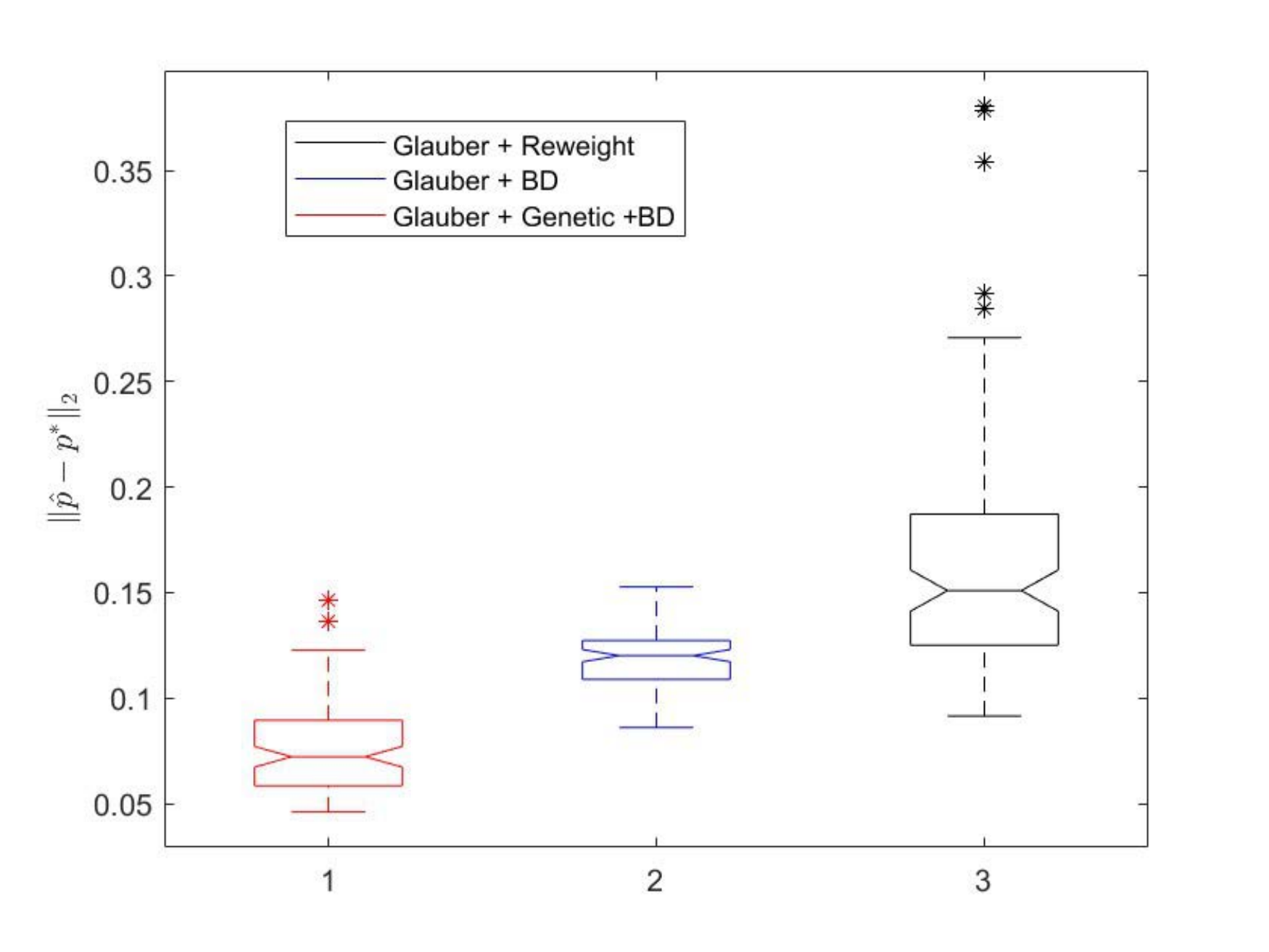} 
    \caption{Box plot with fixed sample size}    \label{20D_Ising_1D_F_boxplot_512_sample_size}
\end{subfigure}
\begin{subfigure}{.49\textwidth}
    \centering
    \includegraphics[width=1.0\linewidth]{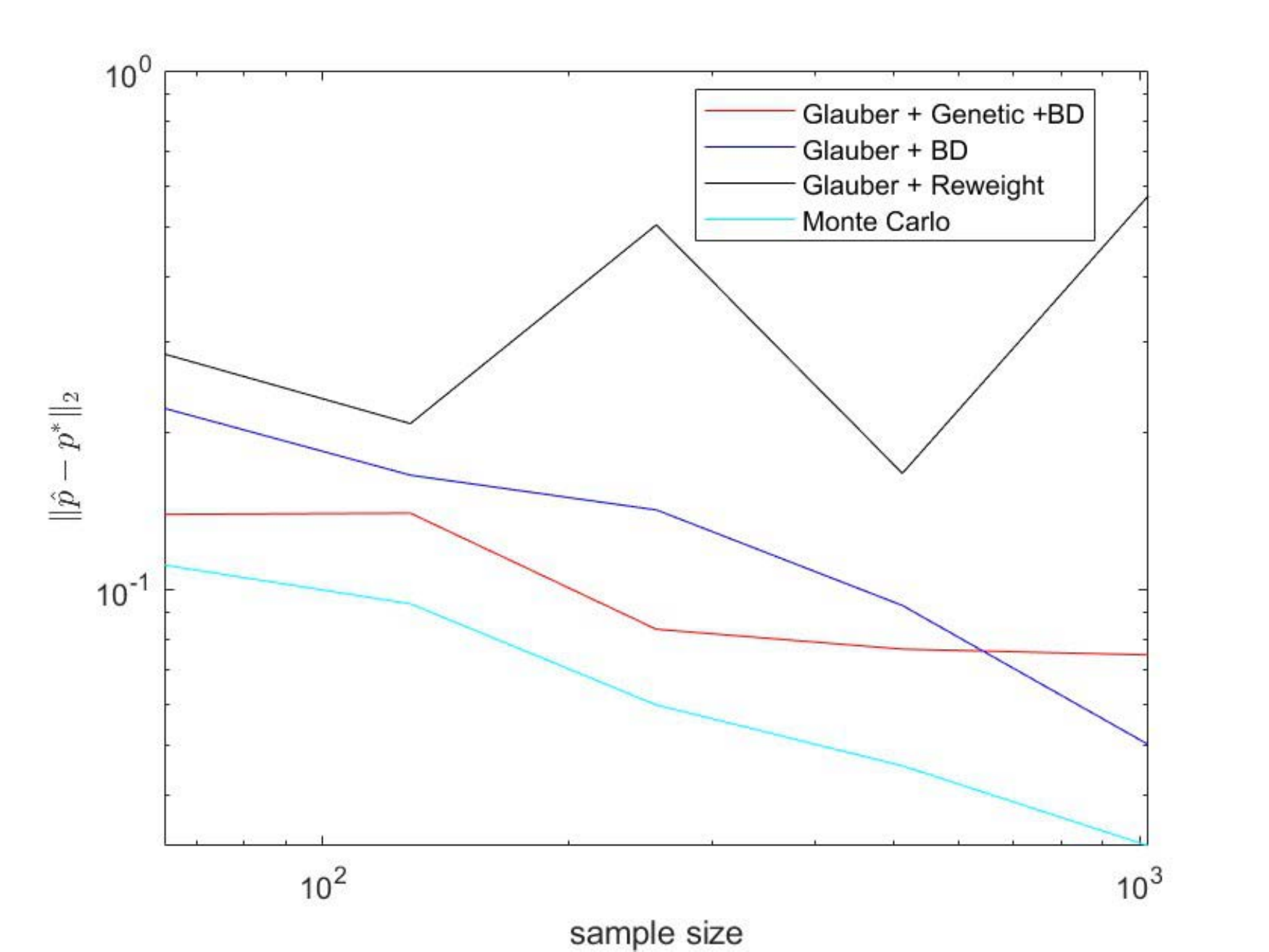}
    \caption{Error plot with varying sample sizes}
    \label{20D_Ising_1D_F_MC}
\end{subfigure}
\begin{subfigure}{.49\textwidth}
    \centering
    \includegraphics[width=1.0\linewidth]{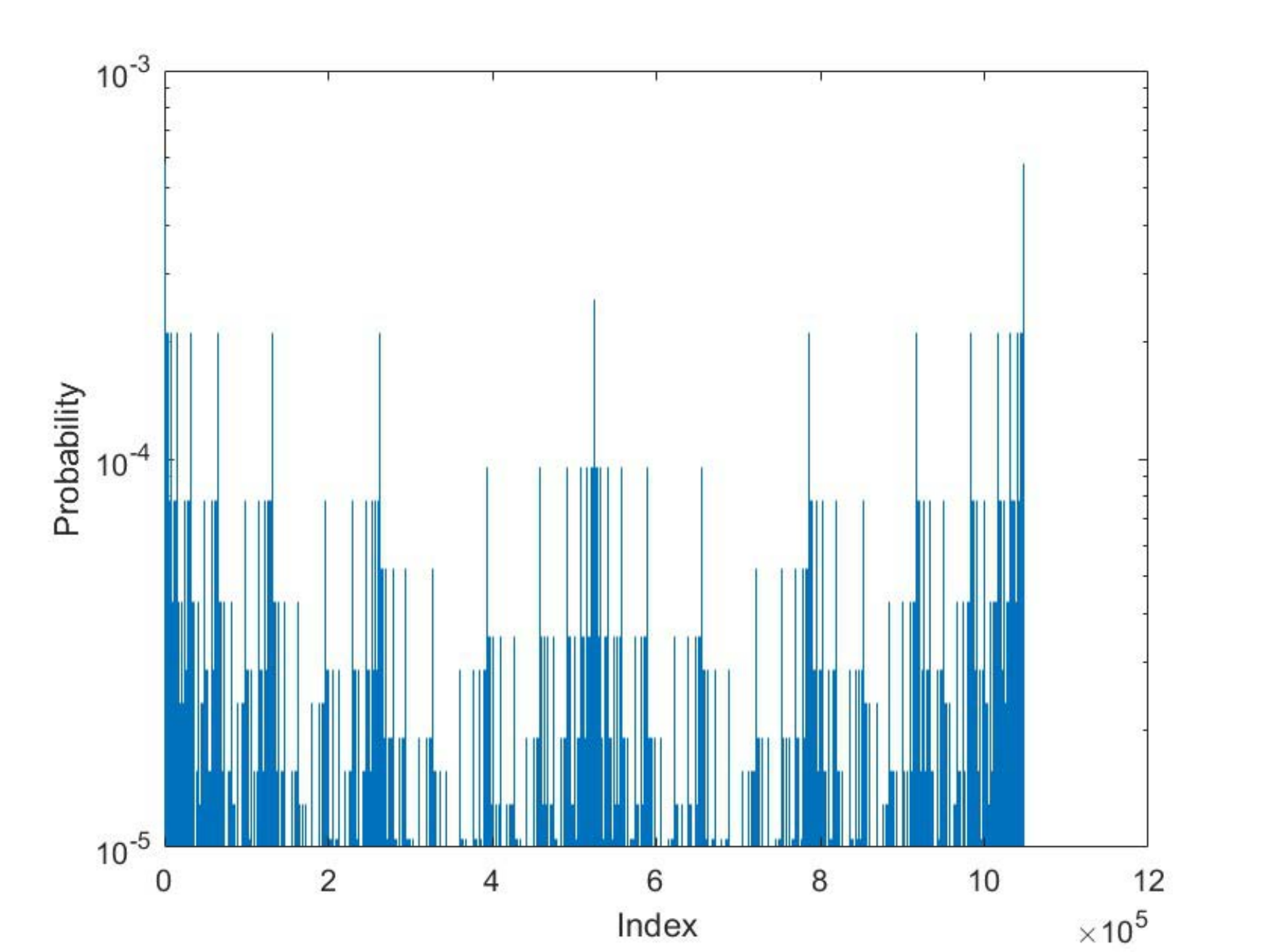} 
    \caption{Target Distribution}
    \label{20D_Ising_1D_F_ref_density}
\end{subfigure}
\begin{subfigure}{.49\textwidth}
    \centering
    \includegraphics[width=1.0\linewidth]{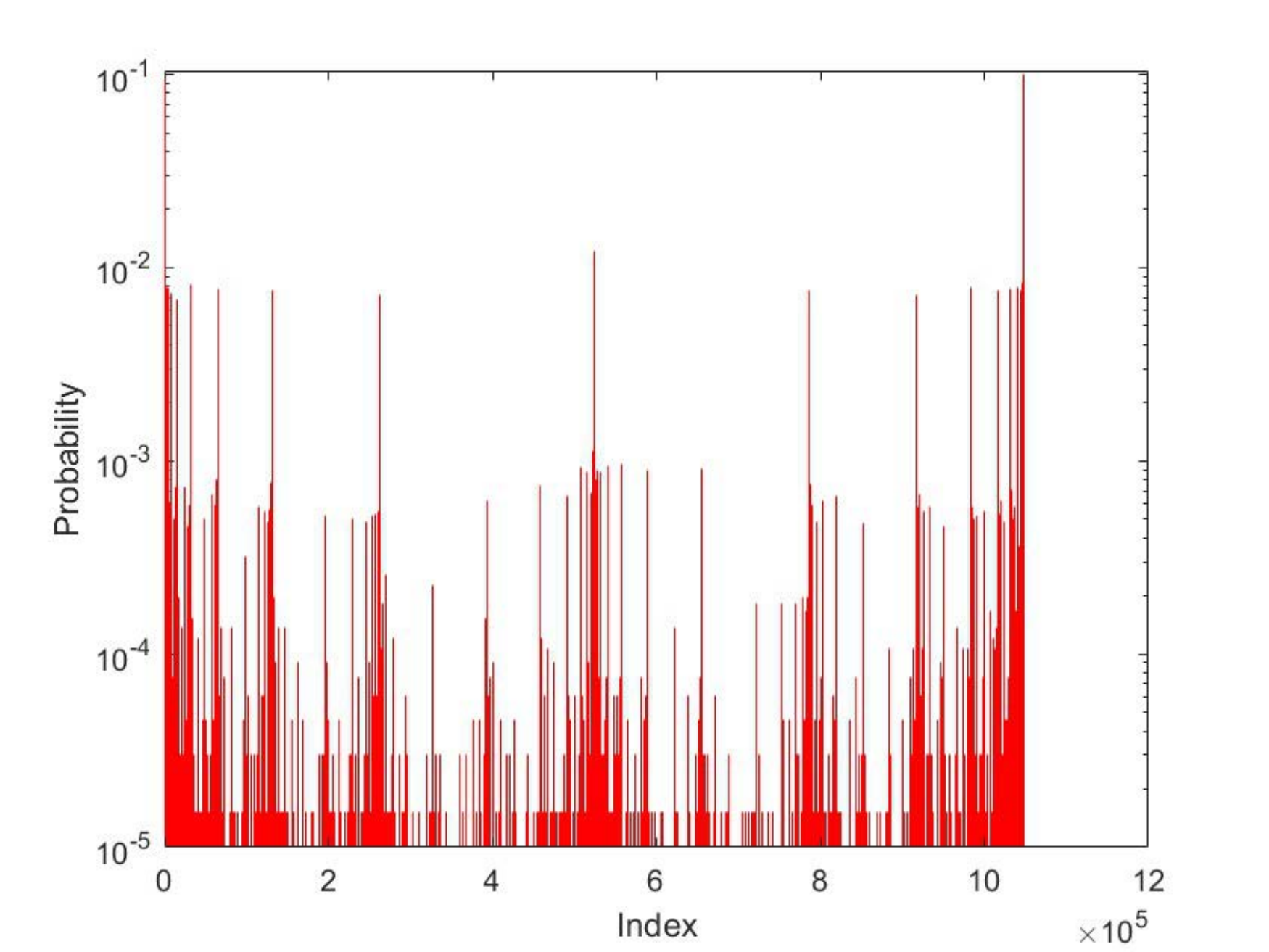}
    \caption{Generated Empirical Distribution}
    \label{20D_Ising_1D_F_empirical_density}
\end{subfigure}
\caption{One-dimensional ferromagnetic Ising model with $\beta = 0.8$ and $d=20$; (a) Box plot with fixed sample size $N =512$; (b) Error plot with varying sample sizes $N \in \{64, 128, 256, 512, 1024\}$; (c) Target distribution; (d) Empirical distribution formed by samples generated by Algorithm \ref{alg: AIS + genetic} (sample size $N = 65536$).}
\label{20D_Ising_1D_F_empirical}
\end{figure}

\begin{figure}[H]
\centering
\begin{subfigure}{.49\textwidth}
    \centering
    \includegraphics[width=1.0\linewidth]{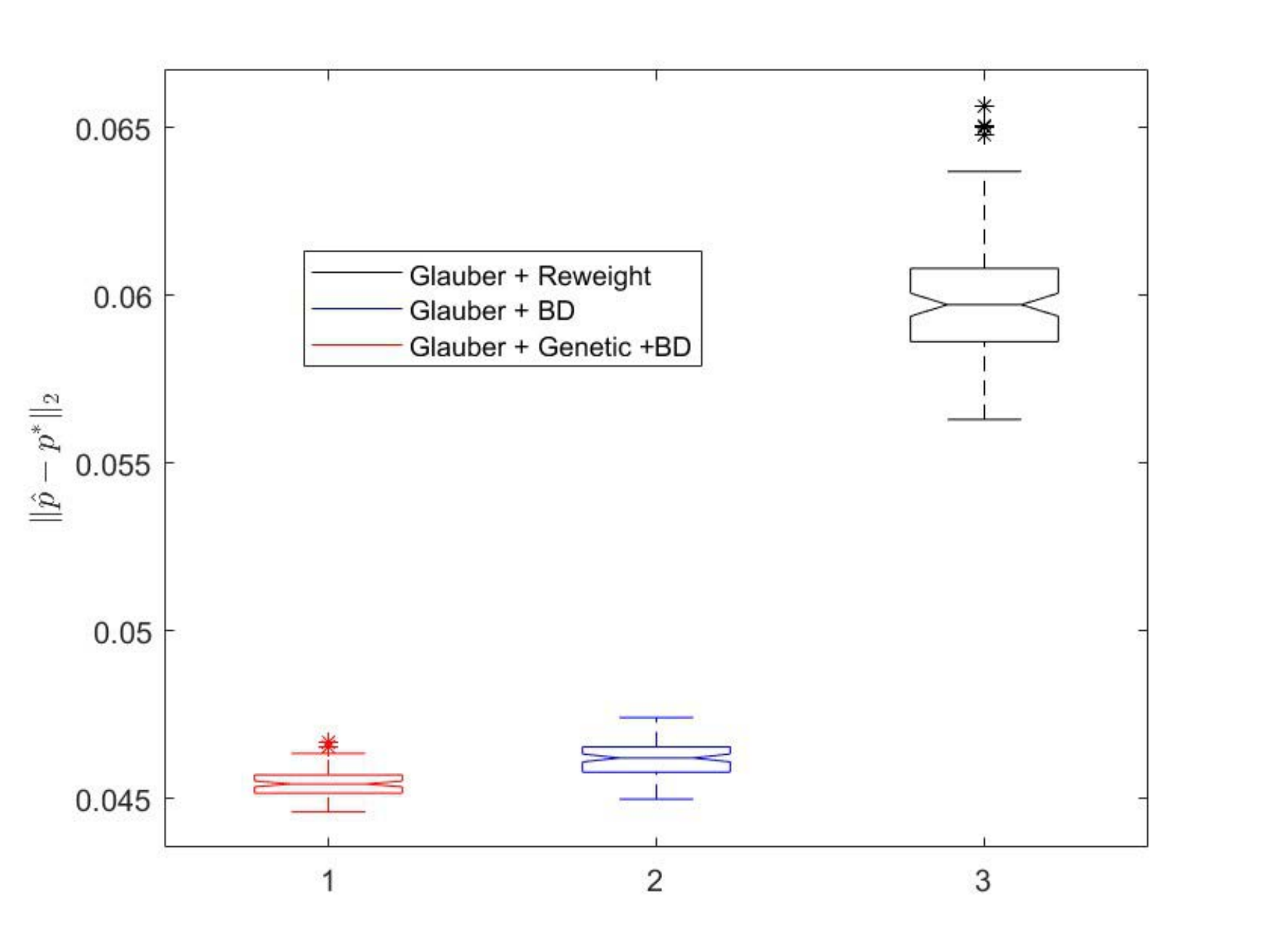} 
    \caption{Box plot with fixed sample size}   \label{20D_Ising_1D_Anti_F_boxplot_512_sample_size}
\end{subfigure}
\begin{subfigure}{.49\textwidth}
    \centering
    \includegraphics[width=1.0\linewidth]{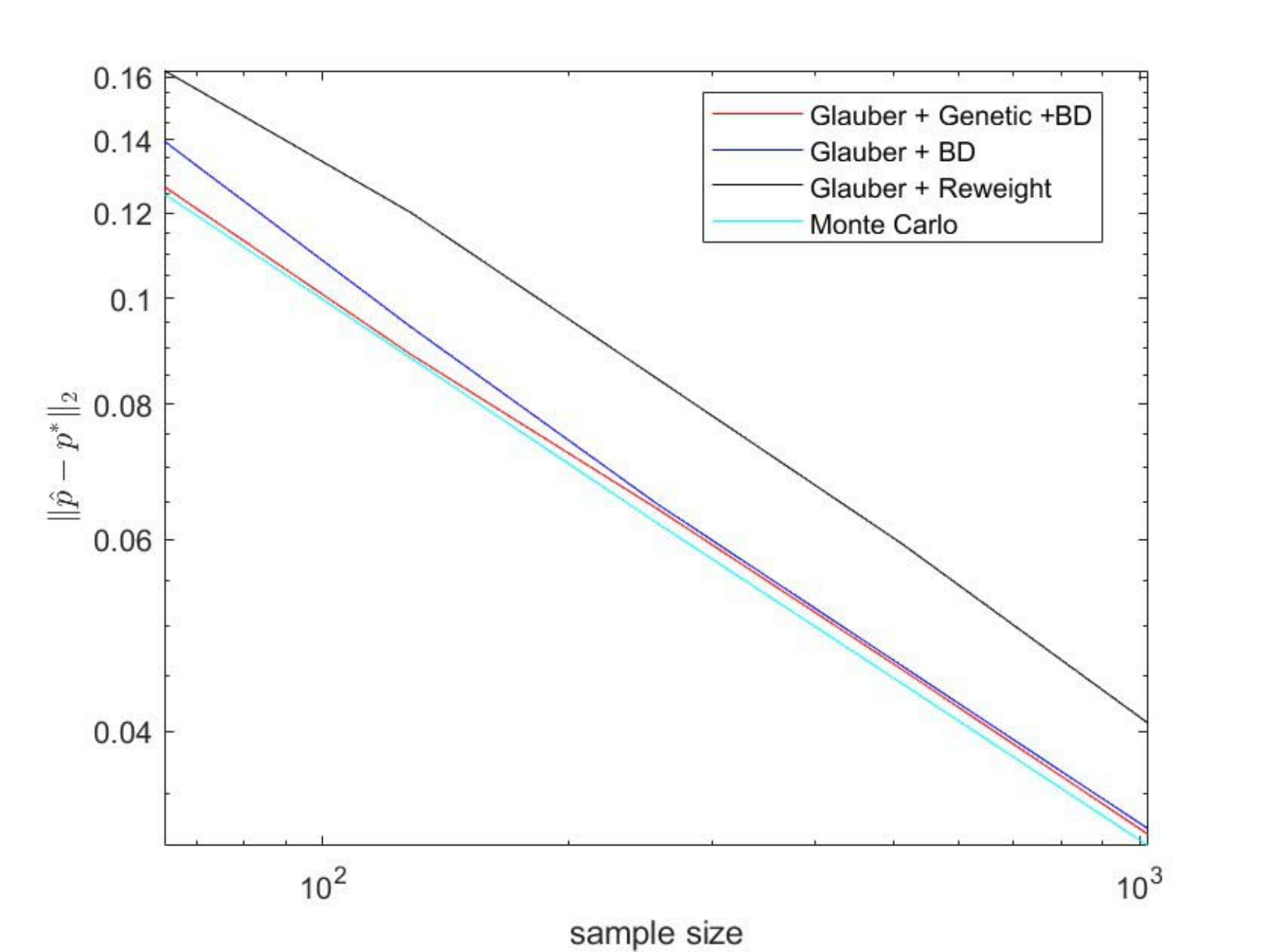}
    \caption{Error plot with varying sample sizes}
    \label{20D_Ising_1D_Anti_F_MC}
\end{subfigure}
\begin{subfigure}{.49\textwidth}
    \centering
    \includegraphics[width=1.0\linewidth]{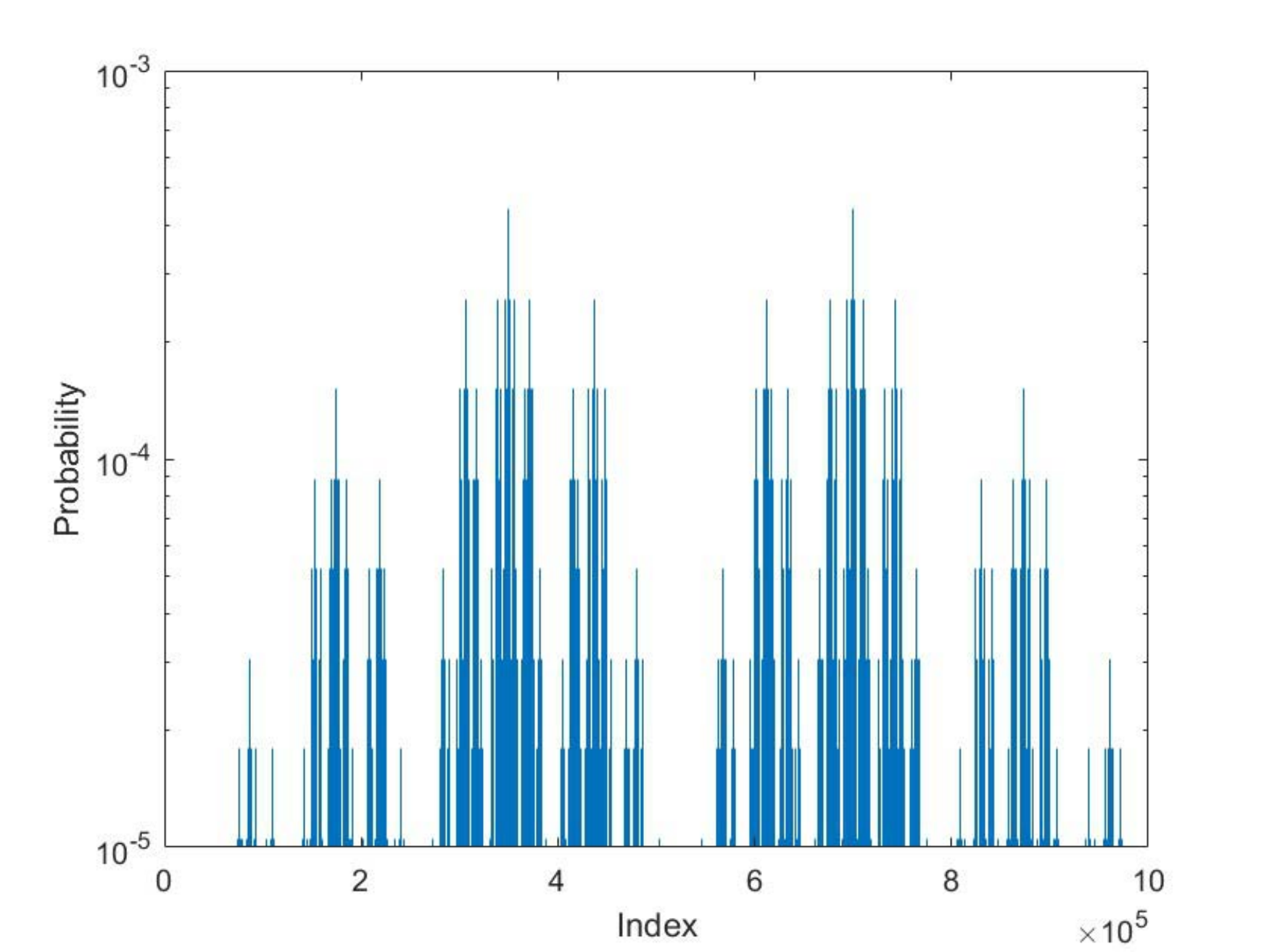} 
    \caption{Target Distribution}
    \label{20D_Ising_1D_Anti_F_ref_density}
\end{subfigure}
\begin{subfigure}{.49\textwidth}
    \centering
    \includegraphics[width=1.0\linewidth]{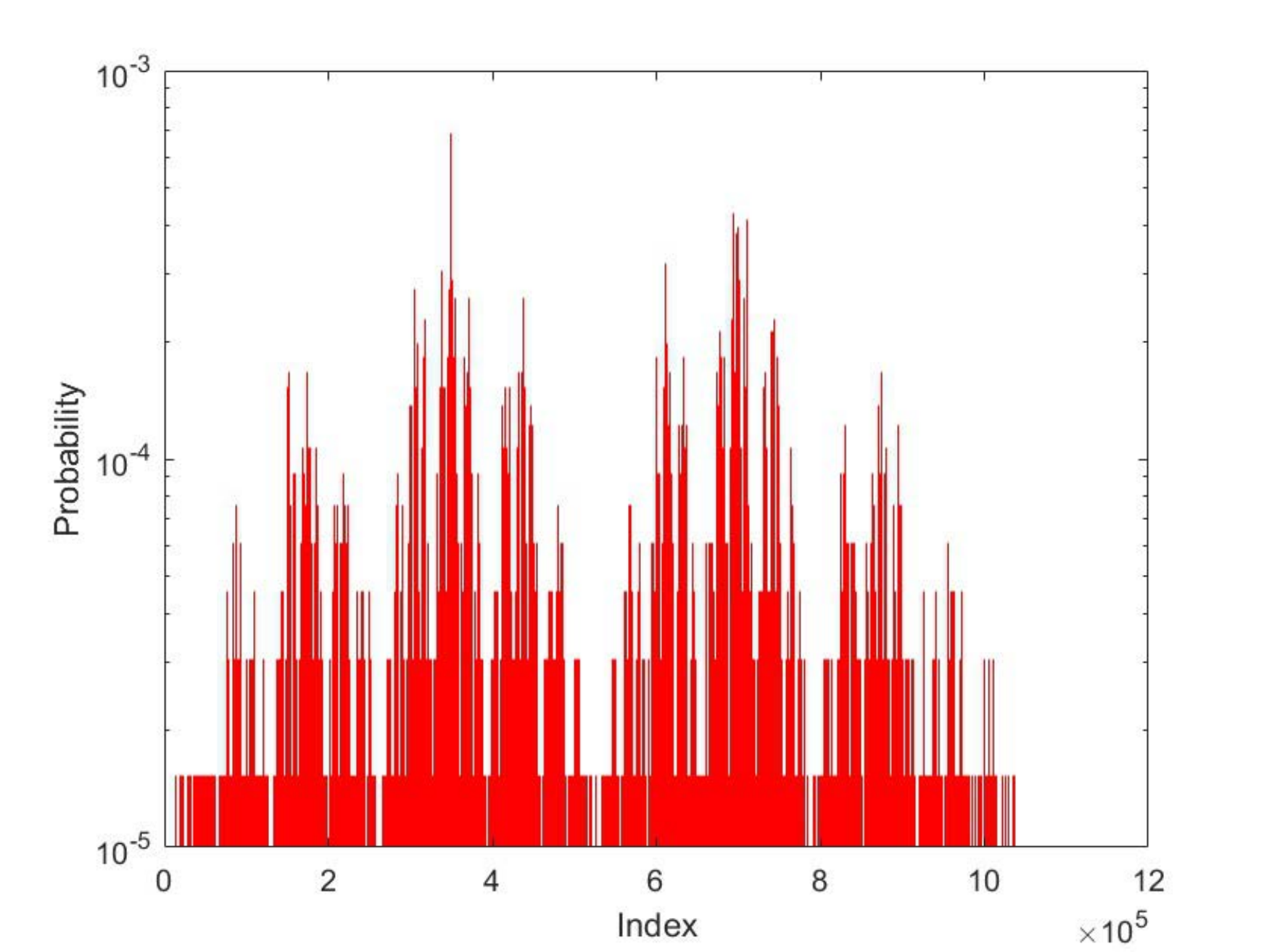}
    \caption{Generated Empirical Distribution}
    \label{20D_Ising_1D_Anti_F_empirical_density}
\end{subfigure}
\caption{One-dimensional antiferromagnetic Ising model with $\beta = 0.8$ and $d=20$; (a) Box plot with fixed sample size $N =512$; (b) Error plot with varying sample sizes $N \in \{64, 128, 256, 512, 1024\}$; (c) Target distribution; (d) Empirical distribution formed by samples generated by Algorithm \ref{alg: AIS + genetic} (sample size $N = 65536$).}
\label{20D_Ising_1D_Anti_F_empirical}
\end{figure}

The first model we test is the ferromagnetic one-dimensional Ising model, whose interaction coefficients in (\ref{eqn: 1D Ising}) are given by $J_1 = -1$ and $J_2 = -\frac{1}{3}$. The testing results are listed in Figure (\ref{20D_Ising_1D_F_empirical}), from which we can see that Algorithm \ref{alg: AIS + genetic} does outperform the ensemble-based AIS without exploration and the standard AIS on the first model. Specifically, the box plots reveal that Algorithm \ref{alg: AIS + genetic} returns more accurate samplers than the other two algorithms, while the error plots show that Algorithm \ref{alg: AIS + genetic} induces smaller error than the other two algorithms for varying sample sizes. Furthermore, when the sample size is set to be sufficiently large, we observe that the empirical distribution generated by Algorithm \ref{alg: AIS + genetic} is reasonably close to the target distributions. The second model we test is the antiferromagnetic Ising model, whose interaction coefficients in (\ref{eqn: 1D Ising}) are given by $J_1 = 1$ and $J_2 = \frac{1}{3}$. The plotted figures are given in Figure (\ref{20D_Ising_1D_Anti_F_empirical}), where we can infer that Algorithm \ref{alg: AIS + genetic} performs better than the other two algorithms on the second model by comparing the sub-figures in Figure (\ref{20D_Ising_1D_Anti_F_empirical}) in a similar way.

\begin{figure}[H]
\centering
\begin{subfigure}{.49\textwidth}
    \centering
    \includegraphics[width=1.0\linewidth]{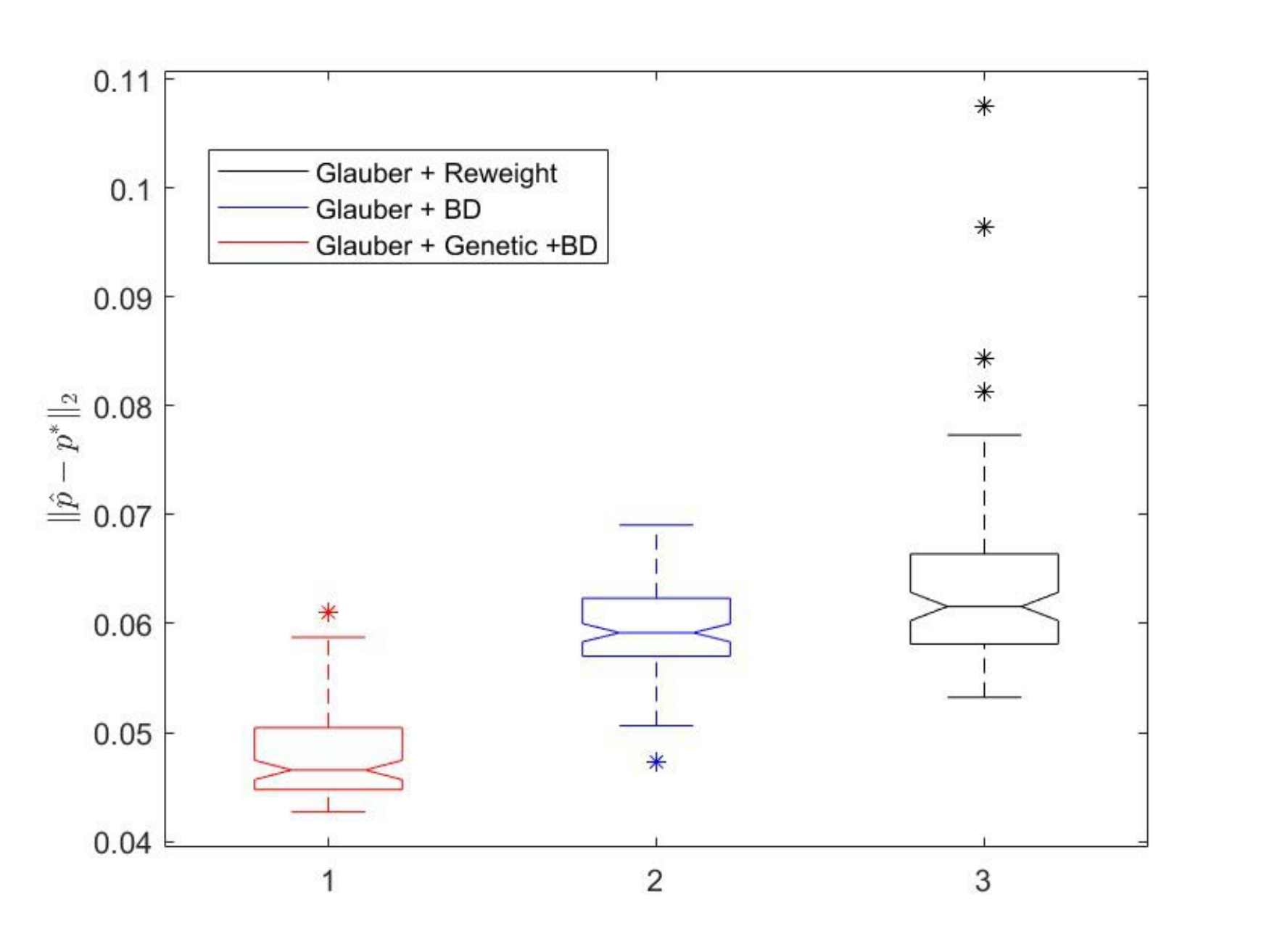} 
    \caption{Box plot with fixed sample size}
    \label{16D_Ising_2D_F_boxplot_512_sample_size}
\end{subfigure}
\begin{subfigure}{.49\textwidth}
    \centering
    \includegraphics[width=1.0\linewidth]{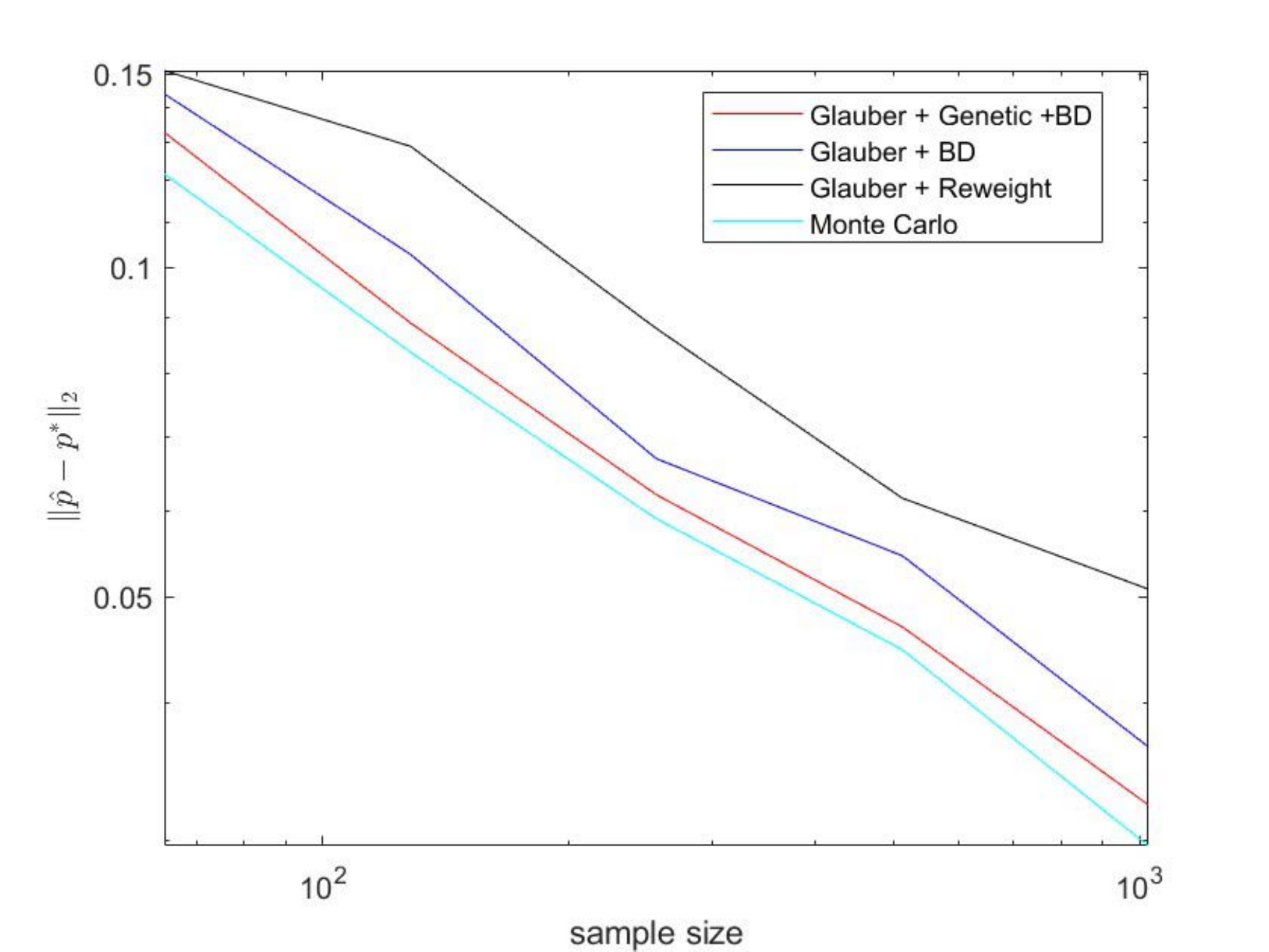}
    \caption{Error plot with varying sample sizes}
    \label{16D_Ising_2D_F_MC}
\end{subfigure}
\begin{subfigure}{.49\textwidth}
    \centering
    \includegraphics[width=1.0\linewidth]{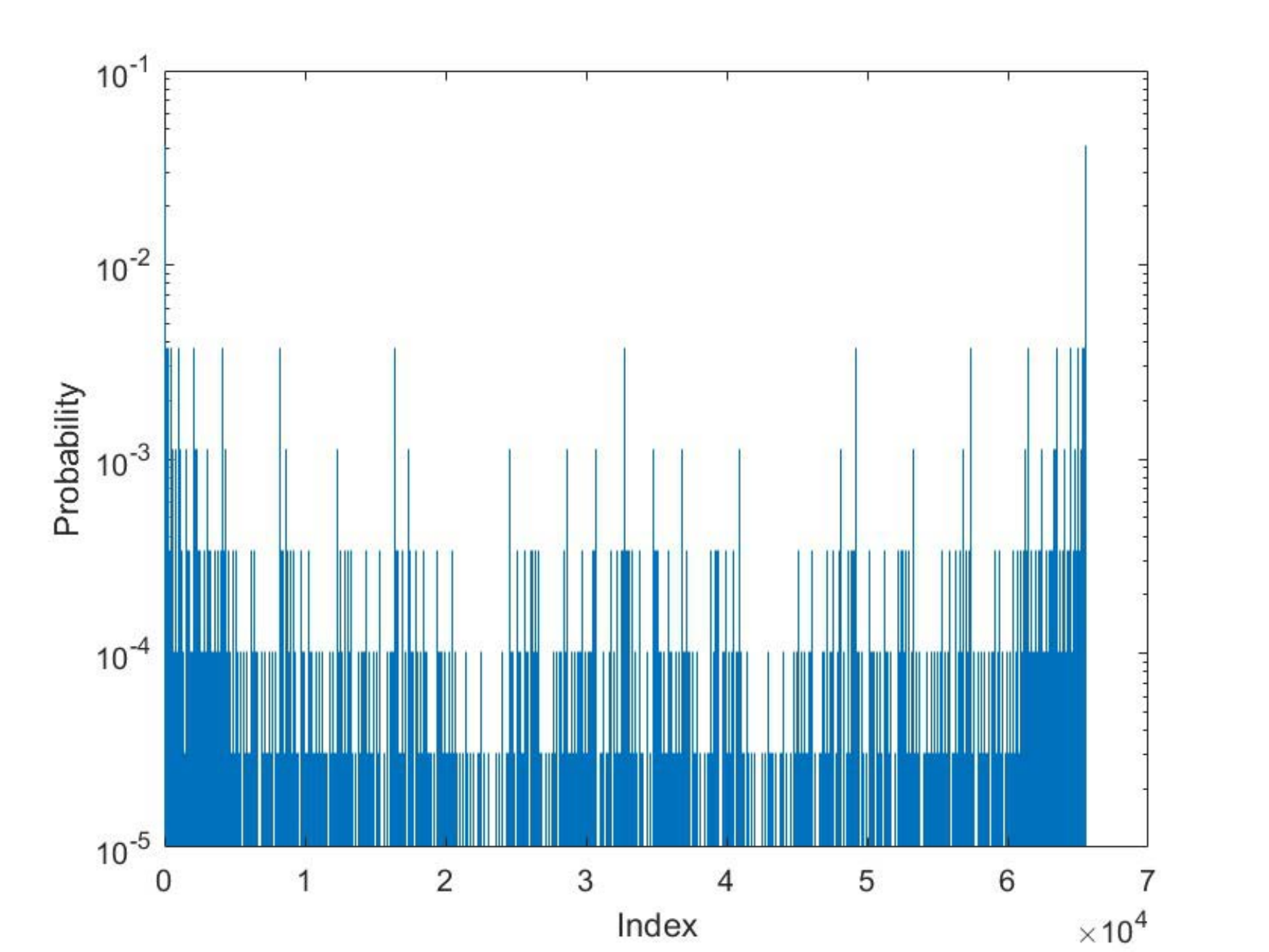} 
    \caption{Target Distribution}
    \label{16D_Ising_2D_F_ref_density}
\end{subfigure}
\begin{subfigure}{.49\textwidth}
    \centering
    \includegraphics[width=1.0\linewidth]{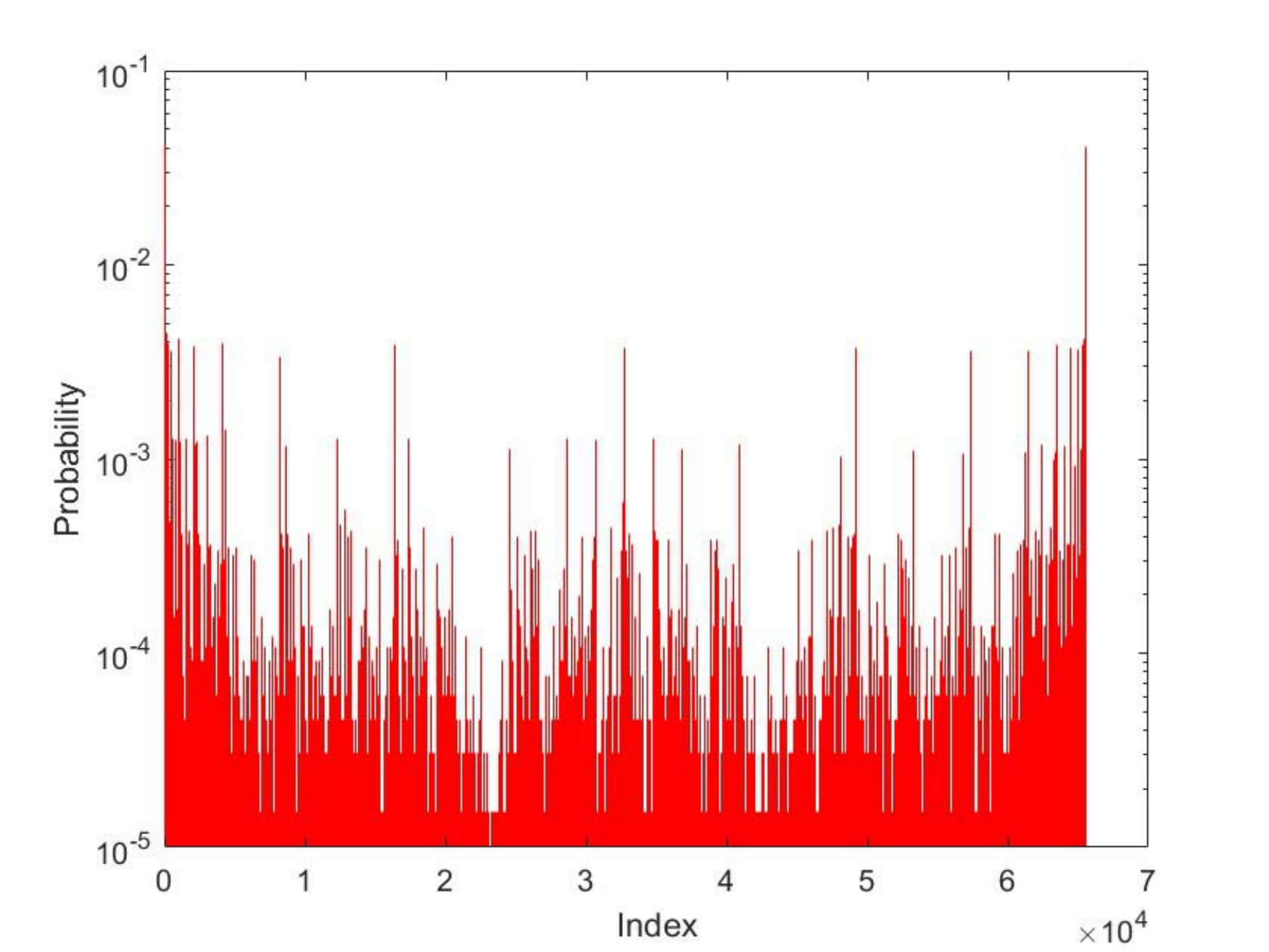}
    \caption{Generated Empirical Distribution}
    \label{16D_Ising_2D_F_empirical_density}
\end{subfigure}
\caption{Two-dimensional ferromagnetic Ising model with $\beta = 0.3$ and $d=4$; (a) Box plot with fixed sample size $N =512$; (b) Error plot with varying sample sizes $N \in \{64, 128, 256, 512, 1024\}$; (c) Target distribution; (d) Empirical distribution formed by samples generated by Algorithm \ref{alg: AIS + genetic} (sample size $N = 65536$).}
\label{16D_Ising_2D_F_empirical}
\end{figure}

\subsubsection{2D Ising Models}
\begin{figure}[H]
\centering
\begin{subfigure}{.49\textwidth}
    \centering
    \includegraphics[width=1.0\linewidth]{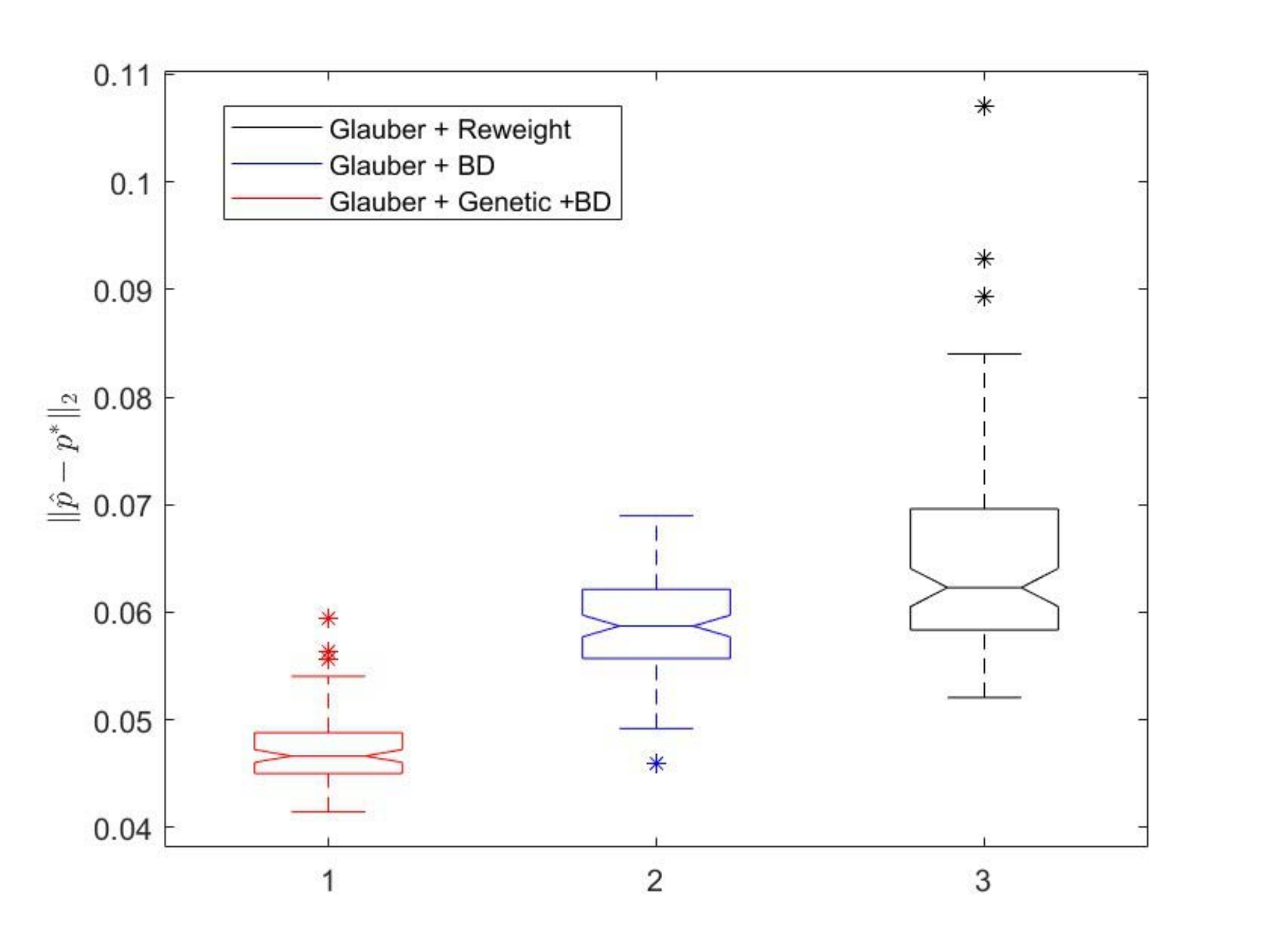} 
    \caption{Box plot with fixed sample size}
    \label{16D_Ising_2D_Anti_F_boxplot_512_sample_size}
\end{subfigure}
\begin{subfigure}{.49\textwidth}
    \centering
    \includegraphics[width=1.0\linewidth]{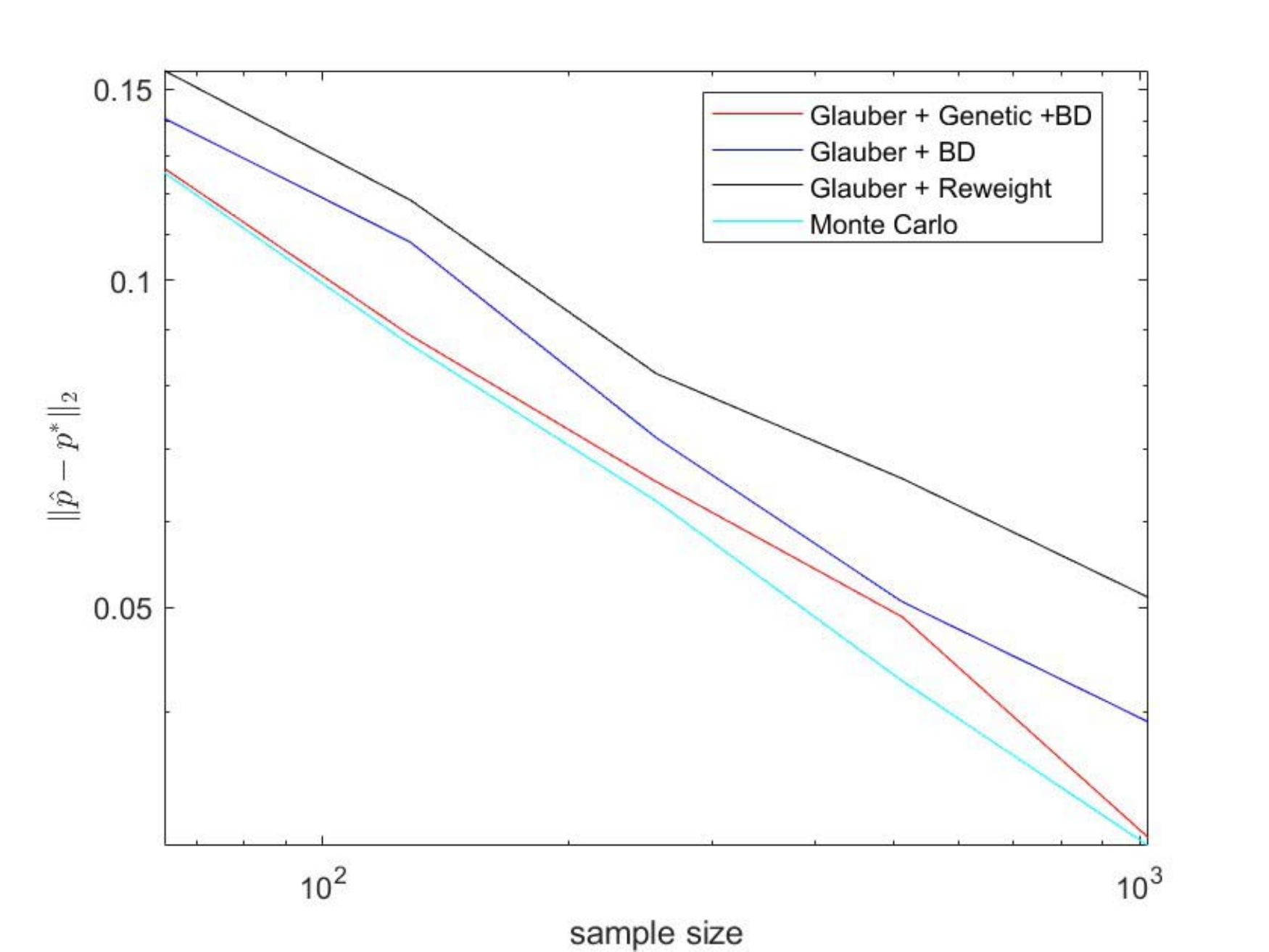}
    \caption{Error plot with varying sample sizes}
    \label{16D_Ising_2D_Anti_F_MC}
\end{subfigure}
\begin{subfigure}{.49\textwidth}
    \centering
    \includegraphics[width=1.0\linewidth]{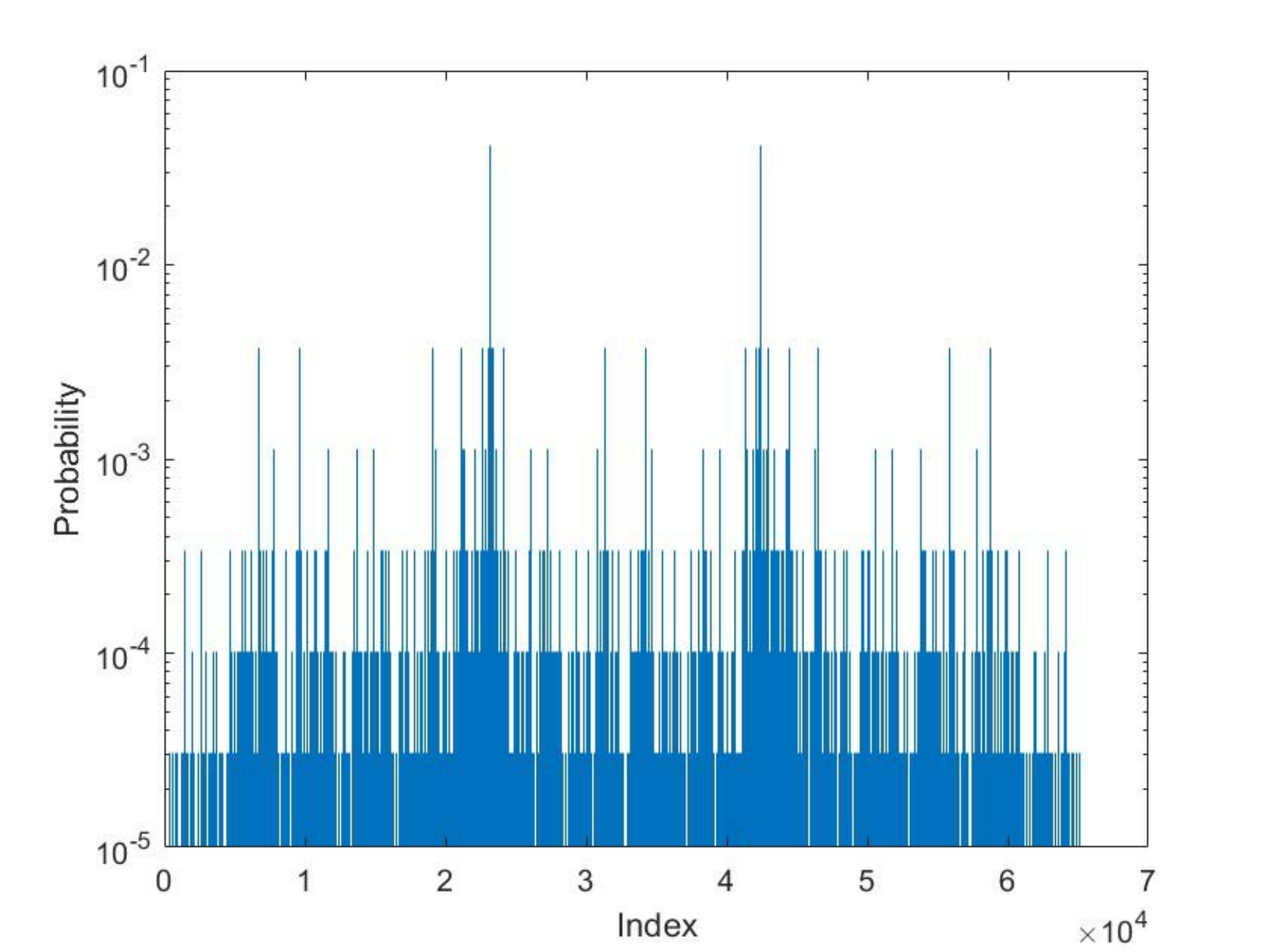} 
    \caption{Target Distribution}
    \label{16D_Ising_2D_Anti_F_ref_density}
\end{subfigure}
\begin{subfigure}{.49\textwidth}
    \centering
    \includegraphics[width=1.0\linewidth]{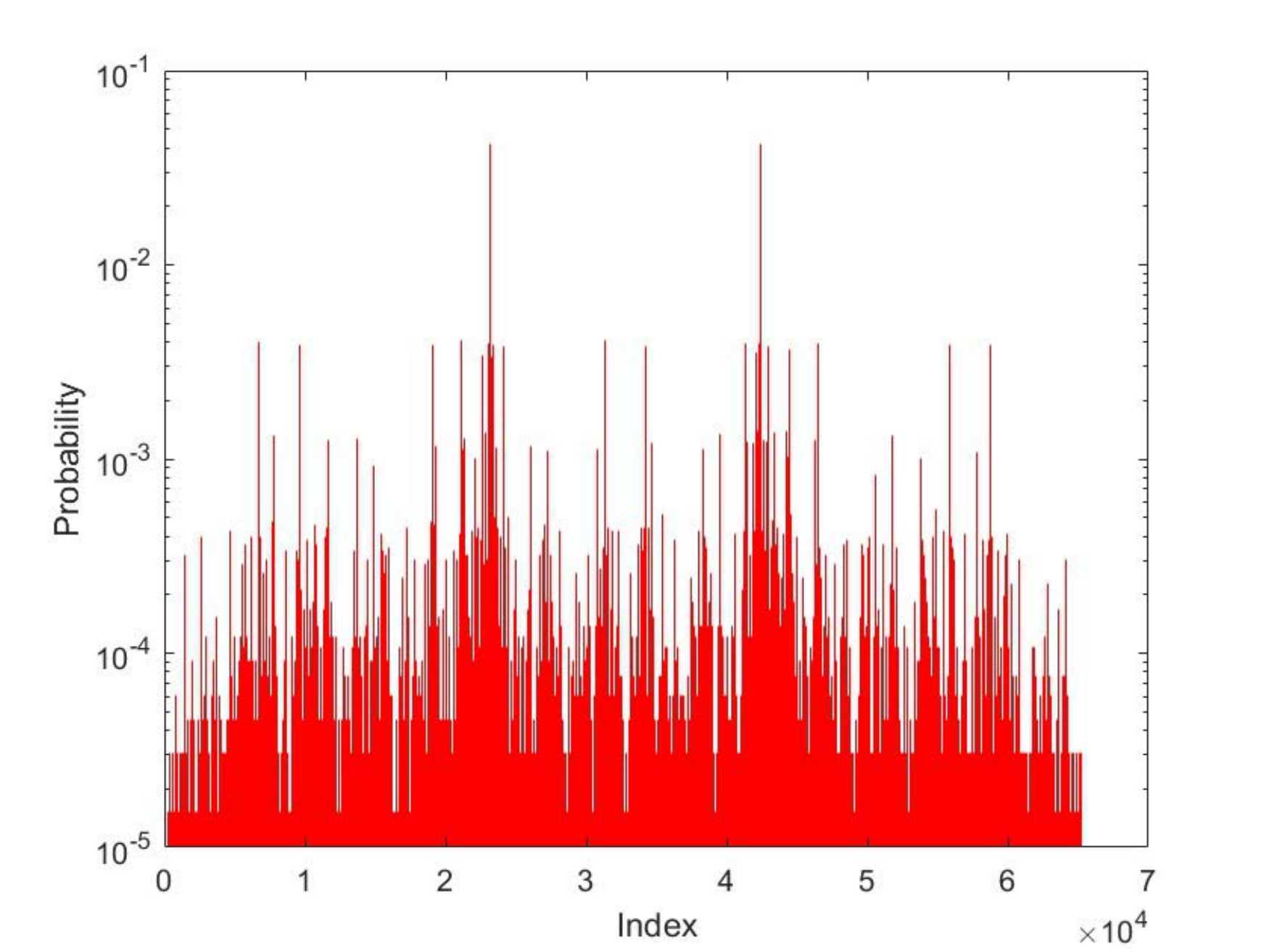}
    \caption{Generated Empirical Distribution}
    \label{16D_Ising_2D_Anti_F_empirical_density}
\end{subfigure}
\caption{Two-dimensional antiferromagnetic Ising model with $\beta = 0.3$ and $d=4$; (a) Box plot with fixed sample size $N =512$; (b) Error plot with varying sample sizes $N \in \{64, 128, 256, 512, 1024\}$; (c) Target distribution; (d) Empirical distribution formed by samples generated by Algorithm \ref{alg: AIS + genetic} (sample size $N = 65536$).}
\label{16D_Ising_2D_Anti_F_empirical}
\end{figure}

We further consider the two-dimensional Ising model with the periodic boundary condition, where interaction exists between only the nearest neighbor pairs. The corresponding distribution $p(\cdot):\{-1,1\}^{d^2} \rightarrow [0,1]$ can be written as
\begin{equation}
\label{eqn: 2D Ising}
\begin{aligned}
p(\boldsymbol{x}) = p(\{x_{i,j}\}_{i,j=1}^{d}) \propto \exp\Big(-\beta\sum_{i=1}^{d}\sum_{j=1}^{d}J(x_{i,j}x_{i+1,j} + x_{i,j}x_{i,j+1})\Big),    
\end{aligned}    
\end{equation}
where $\beta = \frac{1}{T}$ is the inverse of the temperature and $x_{i,d+1} = x_{i,1}, x_{d+1,j} = x_{1,j}$ for any $1 \leq i,j \leq d$. For our experiments on the two-dimensional Ising models, we pick $\beta =0.3, d= 4$ and the number of time steps to be $L=64$. The initial distribution is selected to be the uniform distribution over $\{-1,1\}^{16}$. Again, we fix the sample size to be $N = 512$ for the experiments producing box plots. For the experiments studying how the loss function $\mathcal{L}_2(\hat{p},p)$ scales with respect to the sample size, we set the range of the sample sizes to be $N \in \{64, 128, 256, 512, 1024\}$. Finally, we set the sample size to be $N =65536$ in our visualization of the empirical distribution formed by the samples generated by Algorithm \ref{alg: AIS + genetic}. Similar to subsection \ref{sec: 1D Ising} above, we test the three algorithms on both the ferromagnetic and antiferromagnetic two-dimensional Ising models, whose interaction coefficients in (\ref{eqn: 2D Ising}) are given by $J=-1$ and $J=1$ respectively. For the ferromagnetic model, whose testing results are listed in Figure (\ref{16D_Ising_2D_F_empirical}), we again observe that the interacting AIS with exploration performs better than the other two algorithms. Regarding the antiferromagnetic model, a comparison between the sub-figures given in Figure (\ref{16D_Ising_2D_Anti_F_empirical}) leads to the same conclusion. 

\section{Conclusion and Discussion}
\label{sec:conclusions}
In this paper, we considered the task of sampling multimodal distributions and proposed an improved version of AIS by combining it with population-based MCMC kernels. Our method leverages the interaction between different particles in the ensemble, which allows us to find undiscovered modes more effectively. Regarding future work, one potential direction is to investigate how the interaction-based methods can be combined with other continuation methods like Simulated Tempering and Tempered Transitions. Another interesting project is to study how the interaction-based AIS may be used to sample distributions with approximate symmetries under the framework proposed in~\cite{ying2023multimodal}. 
Moreover, studying the theoretical properties of the mean-field PDE derived in the supplementary materials under appropriate assumptions on the acceptance ratio function may also be of interest. Furthermore, one might also investigate how the interaction-based exploration part used in this paper can be combined with samplers based on pretrained diffusion models, which are proposed in existing work like~\cite{chen2025solving,ren2025driftlite,wu2025reverse,phillips2024particle}.

\section*{Acknowledgments}
Haoxuan Chen would like to thank Dr. Yifan Chen for many helpful discussions on sampling-related problems and Dr. Xun Tang, Dr. Nan Sheng for proofreading an early draft of this manuscript.  

\renewcommand{\theequation}{A.\arabic{equation}}
\setcounter{equation}{0}
\section*{Appendix A: Derivation of the Weights in Annealed Importance Sampling (AIS)}
\label{section: appendix one}
In the first subsection, we provide a formal proof for the weight expression stated in the middle of subsection~\ref{subsec: finite time} of the main text.  We begin by recalling the notations used in the main text. Specifically, consider the task of sampling from some fixed continuous distribution $p \propto e^{-U}$ on $\mathbb{R}^d$. Under the continuous-time limit, the intermediate distributions used in AIS are given by $p_t \propto e^{-U_t}$, where the interpolating potential $U_t := (1-c(t))U_0 + c(t)U$ for any $t \in [0,1]$, where $c(0) = 0$ and $c(1)=1$. 

Let $\dot{x} = -\nabla_{x}U_t(x) + \sqrt{2}\dot{W}$ denote the dynamics governing the evolution of the particle in AIS, where $W$ denotes the standard Brownian motion in $\mathbb{R}^d$. We then state the following lemma, which derives the expression for the weight $w$ associated with the particle in AIS. 

\begin{lemma}
Consider the following expression of the weight used in AIS, which is derived via path integrals in subsection~\ref{subsec: finite time} of the main text:
\begin{equation}
w = U_0(x(0)) - U(x(1)) + \frac{1}{4}\int_{0}^{1}\Big((\dot{x} + \nabla_xU_t(x))^2-(\dot{x} - \nabla_xU_t(x))^2\Big)\dif t.    
\end{equation}
We have that the weight expression above also simplifies to $w = -\int_{0}^{1}\frac{\partial}{\partial t}U_t(x)\dif t$. 
\end{lemma}
\begin{proof}
We note that the simplified identity above directly follows from integration by parts along with the given assumption that $c(0)=0$ and $c(1)=1$. Specifically, 
\begin{align*}
w &= U_0(x(0)) - U(x(1)) +\int_{0}^{1}\langle \dot{x}, \nabla_xU_t(x) \rangle \dif t\\
 &= U_0(x(0)) - U(x(1)) + \int_{0}^{1}(1-c(t))\langle \dot{x}, \nabla U_0(x) \rangle \dif t + \int_{0}^{1}c(t) \langle \dot{x}, \nabla U(x) \rangle \dif t\\
 &=U_0(x(0)) - U(x(1)) + \int_{0}^{1}(1-c(t))\cdot \frac{\dif}{\dif t}U_0(x(t))\dif t + \int_{0}^{1}c(t) \cdot \frac{\dif}{\dif t}U(x(t)) \dif t\\
 &= U_0(x(0)) - U(x(1)) + \int_{0}^{1}c'(t)U_0(x(t)) \dif t + (1-c(1))U_0(x(1)) \\
 &- (1-c(0))U_0(x(0))- \int_{0}^{1}c'(t) U(x(t))\dif t + c(1)U(x(1)) -c(0)U(x(0))\\
 &= -\int_{0}^{1}c'(t)(U(x) - U_0(x))\dif t
= -\int_{0}^{1}\frac{\partial}{\partial t}U_t(x)\dif t.
\end{align*}
\end{proof}

\renewcommand{\theequation}{B.\arabic{equation}}
\setcounter{equation}{0}
\section*{Appendix B: Mean-Field Limit and Convergence Properties of Interacting Annealed Importance Sampling for Continuous Distributions}
\label{section: appendix two}
In the second subsection, we provide a mean-field analysis of the Interacting Annealed Importance Sampling (IAIS) Algorithm with exploration proposed in subsection~\ref{subsec: interacting AIS for cont distribution} of the main text under the continuous-time limit. 

Below we use the same set of notations covered in the first subsection above. Moreover, for the snooker algorithm described in Algorithm~\ref{alg:snooker} of subsection~\ref{subsubsec: snooker algo}, we use the stretch move proposed in \cite{goodman2010ensemble} to sample from the density $\rho(r)$ given in equation~\eqref{eqn: density in snooker} of Algorithm~\ref{alg:snooker}. Specifically, let $g:\mathbb{R} \rightarrow [0,\infty)$ denote some prescribed density function satisfying $\text{supp}(g) = \left[a^{-1}, a\right]$ for some $a \in (1,\infty)$ and the symmetry condition, i.e., $g\left(z^{-1}\right) = zg(z)$ for any $z \in \left[a^{-1}, a\right]$. Furthermore, for any time $t \in [0,1]$ and collinear points $u,v,w \in \mathbb{R}^d$, where $w = \lambda u + (1-\lambda)v$ for some $\lambda \in \mathbb{R}$, we represent the acceptance ratio of the move $u \rightarrow w$ given in equation~\eqref{eqn: snooker accept ratio} by some function $A_t(x,y,\lambda)$ defined as follows:
\begin{equation*}
\begin{aligned}
A_t(u,v,\lambda) := \min\Bigg\{1,\frac{\|w-v\|^{d-1} \cdot p_t(w)}{\|u-v\|^{d-1} \cdot p_t(u)}\Bigg\} = \min\Bigg\{1,|\lambda|^{d-1}\frac{e^{-U_t(\lambda u + (1-\lambda)v)}}{e^{-U_t(u)}}\Bigg\}.    
\end{aligned}    
\end{equation*}

We further represent the empirical measure formed by the ensemble $\boldsymbol{X}^{(t)} := \{x^{(t)}_i\}_{i=1}^{N}$ as $\mu_{N}^{(t)} := \frac{1}{N}\sum_{i=1}^{N}\delta_{x^{(t)}_i}$ for any time $t \in [0,1]$. Under the continuous-time limit of the discretized IAIS Algorithm presented in the main text, we have that the evolution of the particles in $\boldsymbol{X}^{(t)}$ consists of three parts as $t$ ranges from $0$ to $1$:
\begin{itemize}
    \item Each particle $x^{(t)}_i$ diffuses independently according to the time-dependent Langevin dynamics $\dot{x} = -\nabla_{x}U_t(x) + \sqrt{2}\dot{W}$.
    \item Each particle $x^{(t)}_i$ is updated via the stretch move governed by the acceptance ratio function $A_t$ described above. 
    \item For any $1 \leq i \leq N$, each particle $x^{(t)}_i$ possesses an independent exponential clock with the following instantaneous birth-death rate
\end{itemize}
    \begin{equation}
    \begin{aligned}
    \beta^{(t)}_i := c'(t)\left(U\left(x^{(t)}_i\right) - U_0\left(x^{(t)}_i\right)\right) - \frac{1}{N}\sum_{j=1}^{N}c'(t)\left(U\left(x^{(t)}_j\right) - U_0\left(x^{(t)}_j\right)\right).    
    \end{aligned}    
    \end{equation}
With all notations introduced above, we now present the partial differential equation (PDE) that governs the evolution of $\mu^{(t)}_N$ under the limit $N \rightarrow \infty$. 
\begin{proposition}
\label{prop: PDE of the evolving density}
Assume that the initial distribution $\mu^{(0)}_N$ converges to some $\rho_0$ in law under the large particle limit $N \rightarrow \infty$. Then for any $t \in [0,1]$, $\mu^{(t)}_N$ converges to some $\rho_t$ in law as $N \rightarrow \infty$, where $\rho_t(x) = \rho(t,x)$ satisfies the following PDE:
\begin{equation*}
\begin{aligned}
&\frac{\partial}{\partial t}\rho(t,x) = \nabla_{x} \cdot \Bigg(\Big((1-c(t))\nabla_{x}U_0(x) +c(t)\nabla_x U(x) \Big)\rho(t,x)\Bigg)+ \Delta_{x}\rho(t,x)\\
&+\int_{\frac{1}{a}}^{a}|\lambda|^{d-2}g(\lambda^{-1})\Bigg(\int_{\mathbb{R}^d}A_t(\lambda x +(1-\lambda)y,y,\lambda^{-1})\rho(t,y) \cdot \\
&\rho(t,\lambda x +(1-\lambda)y)dy\Bigg)d\lambda-\Bigg(\int_{\frac{1}{a}}^{a}g(\lambda)\Big(\int_{\mathbb{R}^d}A_t(x,y,\lambda)\rho(t,y)dy\Big)d\lambda\Bigg)\rho(t,x)  \\
&-\Bigg(c'(t)\Big(U(x) - U_0(x)\Big) - \int_{\mathbb{R}^d}c'(t)\Big(U(x) - U_0(x)\Big)\rho(t,x)dx \Bigg)\rho(t,x),
\end{aligned}
\end{equation*}    
with initial condition $\rho_0(x) = \rho(0,x)$.
\end{proposition}
\begin{remark}
For the RHS of the PDE above, the first and second terms correspond to the time-dependent Langevin dynamics, the third and fourth terms come from the stretch move, while the last term stands for the birth-death process.
\end{remark}

\begin{proof}
We note that the proof mainly uses the theory of measure-valued Markov processes~\cite{dawson1993measure}, which is similar to that of~\cite{lu2019accelerating, rotskoff2019neuron}. We begin by introducing a few notations used in the proof. For any smooth functional $\Psi \in \mathcal{B}(\mathcal{P}(\mathbb{R}^d), \mathbb{R})$ and any measure $\mu \in \mathcal{P}(\mathbb{R}^d)$, the functional derivative of $\Psi$ evaluated at $\mu$ is defined to be a function $D_{\mu}\Psi:\mathbb{R}^d \rightarrow \mathbb{R}$ such that the following identity holds for any signed measure $\nu$ satisfying $\int_{\mathbb{R}^d}\nu(x)\dif x =0$, \emph{i.e.},
\begin{equation}
\label{eqn: defn of functional derivative}
\begin{aligned}
\lim_{\epsilon \rightarrow 0}\frac{\Psi(\mu+\epsilon \nu) - \Psi(\mu)}{\epsilon} = \int_{\mathbb{R}^d}D_{\mu}\Psi(x)\nu(x)\dif x.  
\end{aligned}
\end{equation}

Moreover, for any time $t \in [0,1]$, empirical measure $\mu_N \in \mathcal{P}_N(\mathbb{R}^d)$ and smooth functional $\Psi \in \mathcal{B}(\mathcal{P}(\mathbb{R}^d), \mathbb{R})$, we define the time-dependent infinitesimal generator $\mathcal{L}_{N,t}$ associated with the measure-valued markov process $\mu^{(t)}_N$ as below:
\begin{equation}
\label{eqn: infinitesimal generator}
\begin{aligned}
(\mathcal{L}_{N,t} \Psi)(\mu_N) := \lim_{\Delta t \rightarrow 0^{+}}\frac{\mathbb{E}[\Psi(\mu_N^{(t+\Delta t)}) \ | \ \mu_{N}^{(t)} = \mu_N] - \Psi(\mu_N)}{\Delta t}.    
\end{aligned}
\end{equation}

To evaluate the infinitesimal generator defined above, we need to explicitly depict the change of the empirical measure caused by the stretch move and the birth-death dynamics. On the one hand, after a stretch move going from $(x,y)$ to $(z,y)$ happens at time $t$, we define the changed empirical measure to be 
\begin{equation}
\label{eqn: measure after stretch move}
\begin{aligned}
\mu^{(t)}_{N,S}\{(x,y) \rightarrow (z,y)\} := \mu^{(t)}_N + \frac{1}{N}(\delta_{z} - \delta_{x}).   
\end{aligned}   
\end{equation}

On the other hand, after a swap between two particles $x$ and $x'$ happens at time $t$ under the birth-death dynamics, we denote the changed empirical measure by 
\begin{equation}
\label{eqn: measure after birth-death}
\begin{aligned}
\mu^{(t)}_{N, BD}\{x \leftrightarrow x'\} := \mu^{(t)}_N -\frac{1}{N}\text{sign}\left(\beta^{(t)}(x,\mu^{(t)}_N)\right)(\delta_x - \delta_{x'}), 
\end{aligned}    
\end{equation}
where the birth-death rate function $\beta^{(t)}(\cdot, \cdot): \mathbb{R}^d \times \mathcal{P}(\mathbb{R}^d) \rightarrow \mathbb{R}$ above is defined as 
\begin{equation}
\label{eqn: rate function}
\begin{aligned}
\beta^{(t)}(x, \mu) &:= \frac{\partial}{\partial t}U_t(x) -\int_{\mathbb{R}^d}\frac{\partial}{\partial t}U_t(y)\mu(y)\dif y\\
&= c'(t)\Big(U(x) - U_0(x)\Big) - \int_{\mathbb{R}^d}c'(t)\Big(U(y) - U_0(y)\Big)\mu(y)\dif y.
\end{aligned}
\end{equation}

Recall that each particle $x^{(t)}_i$ is also driven by the Langevin dynamics before the stretch move and the birth-death dynamics. Combining such fact with (\ref{eqn: measure after stretch move}) and (\ref{eqn: measure after birth-death}) defined above allows us to express the generator $\mathcal{L}_{N,t}$ defined in (\ref{eqn: infinitesimal generator}) above as
\begin{equation}
\label{eqn: initial derivation of infinitesimal generator}
\begin{aligned}
&(\mathcal{L}_{N,t} \Psi)(\mu_N) \\
&= \frac{1}{N}\sum_{i=1}^{N}\int_{\mathbb{R}^d}\Big(-\langle\nabla_{x}D_{\mu_N}\Psi(x), \nabla_{x}U_t(x)\rangle + \Delta_{x}D_{\mu_N}\Psi(x)\Big)\delta_{x^{(t)}_i}(x)\dif x\\
&+ \sum_{i=1}^{N}\int_{\mathbb{R}^d}\Bigg(\frac{1}{N}\sum_{j=1}^{N}\int_{\mathbb{R}^d}\Bigg(\int_{\frac{1}{a}}^{a}\Big(\Psi\left(\mu^{(t)}_{N,S}\{(x,y) \rightarrow (\lambda x + (1-\lambda)y,y)\}\right)\\
&- \Psi\left(\mu^{(t)}_N\right)\Big)\cdot g(\lambda)A_t(x,y,\lambda)\dif\lambda\Bigg)\delta_{x^{(t)}_j}(y)\dif y\Bigg)\delta_{x^{(t)}_i}(x)\dif x\\ 
&+\sum_{i=1}^{N}\int_{\mathbb{R}^d}\Bigg(\frac{1}{N}\sum_{j=1}^{N}\int_{\mathbb{R}^d}\Big(\Psi\left(\mu^{(t)}_{N, BD}\{x \leftrightarrow x'\}\right) - \Psi\left(\mu^{(t)}_N\right)\Big) \\
&\cdot\delta_{x^{(t)}_j}(x')\dif x'\Bigg)\left|\beta^{(t)}(x, \mu^{(t)}_N)\right|\delta_{x^{(t)}_i}(x)\dif x, 
\end{aligned}    
\end{equation}
where $\mu_N = \mu^{(t)}_N = \frac{1}{N}\sum_{i=1}^{N}\delta_{x^{(t)}_i}$. Using the definition of functional derivatives in (\ref{eqn: defn of functional derivative}) along with the definition of changed measures in (\ref{eqn: measure after stretch move}) and (\ref{eqn: measure after birth-death}), one may further simplify the second term in (\ref{eqn: initial derivation of infinitesimal generator}), which corresponds to the stretch move, as follows:
\begin{equation*}
\label{eqn: simplified stretch move term}
\begin{aligned}
&\sum_{i=1}^{N}\int_{\mathbb{R}^d}\Bigg(\frac{1}{N}\sum_{j=1}^{N}\int_{\mathbb{R}^d}\Bigg(\int_{\frac{1}{a}}^{a}\left(\frac{1}{N}\int_{\mathbb{R}^d}\left(D_{\mu_N}\Psi\right)(z) \left(\delta_{\lambda x+(1-\lambda)y}(z) - \delta_{x}(z)\right)\dif z\right)\\
&\cdot g(\lambda)A_t(x,y,\lambda)\dif\lambda\Bigg)\delta_{x^{(t)}_j}(y)\dif y\Bigg)\delta_{x^{(t)}_i}(x)\dif x\\
&= \int_{\mathbb{R}^d}\left(\int_{\mathbb{R}^d}\int_{\frac{1}{a}}^{a}\left(D_{\mu_N}\Psi\right)(\lambda x + (1-\lambda)y)g(\lambda)A_t(x,y,\lambda)\mu_N(y)\dif\lambda \dif y\right)\mu_{N}(x)\dif x\\
&- \int_{\mathbb{R}^d}\left(\int_{\mathbb{R}^d}\int_{\frac{1}{a}}^{a}(D_{\mu_N}\Psi)(x)g(\lambda)A_t(x,y,\lambda)\mu_N(y)\dif\lambda \dif y\right)\mu_{N}(x)\dif x.
\end{aligned}
\end{equation*}

Similarly, the third term in (\ref{eqn: initial derivation of infinitesimal generator}), which corresponds to the birth-death dynamics, can be further rewritten as 
\begin{equation}
\label{eqn: simplified birth-death term}
\begin{aligned}
&\sum_{i=1}^{N}\int_{\mathbb{R}^d}\Bigg(\frac{1}{N}\sum_{j=1}^{N}\int_{\mathbb{R}^d}\Big(-\frac{1}{N}\int_{\mathbb{R}^d}(D_{\mu_N}\Psi)(z)\cdot \text{sign}(\beta^{(t)}(x, \mu_N)) \\
&\cdot (\delta_{x}(z) - \delta_{x'}(z))\dif z\Big)\cdot\delta_{x^{(t)}_j}\left(x'\right)\dif x'\Bigg)\left|\beta^{(t)}(x, \mu_N)\right|\delta_{x^{(t)}_i}(x)\dif x\\
&=-\int_{\mathbb{R}^d}(D_{\mu_N}\Psi)(x)\beta^{(t)}(x,\mu_N)\mu_N(x)\dif x +\Big(\int_{\mathbb{R}^d}(D_{\mu_N}\Psi)(x')\mu_N(x')\dif x'\Big)\\
&\cdot\Big(\int_{\mathbb{R}^d}\beta^{(t)}(x,\mu_N)\mu_N(x)\dif x\Big). 
\end{aligned}
\end{equation}

Substituting the two expressions derived above into (\ref{eqn: initial derivation of infinitesimal generator}) then gives us a simplified form of $\mathcal{L}_{N,t}$:
\begin{equation*}
\label{eqn: simplified infinitesimal generator}
\begin{aligned}
&(\mathcal{L}_{N,t} \Psi)(\mu_N) = \int_{\mathbb{R}^d}\Big(-\langle\nabla_{x}(D_{\mu_N}\Psi)(x), \nabla_{x}U_t(x)\rangle + \Delta_{x}(D_{\mu_N}\Psi)(x)\Big)\mu_N(x)\dif x\\
&+ \int_{\mathbb{R}^d}\Bigg(\int_{\mathbb{R}^d}\int_{\frac{1}{a}}^{a}(D_{\mu_N}\Psi)(\lambda x + (1-\lambda)y)g(\lambda)\cdot A_t(x,y,\lambda)\mu_N(y)\dif\lambda \dif y\Bigg)\mu_{N}(x)\dif x\\
&- \int_{\mathbb{R}^d}\Bigg(\int_{\mathbb{R}^d}\int_{\frac{1}{a}}^{a}(D_{\mu_N}\Psi)(x)g(\lambda)A_t(x,y,\lambda)\mu_N(y)\dif\lambda \dif y\Bigg)\mu_{N}(x)\dif x\\ 
&-\int_{\mathbb{R}^d}(D_{\mu_N}\Psi)(x)\beta^{(t)}(x,\mu_N)\mu_N(x)\dif x +\Big(\int_{\mathbb{R}^d}(D_{\mu_N}\Psi)(x')\mu_N(x')\dif x'\Big)\\
&\cdot\Big(\int_{\mathbb{R}^d}\beta^{(t)}(x,\mu_N)\mu_N(x)\dif x\Big). 
\end{aligned}    
\end{equation*}

Using the generator $\mathcal{L}_{N,t}$ derived above, we may write out the backward Kolmogorov equation of $\Psi\left(\mu^{(t)}_N\right)$ as follows: 
\begin{equation}
\label{eqn: backward eqn of empirical}
\begin{aligned}
\frac{\partial}{\partial t}\Psi\left(\mu^{(t)}_N\right) = (\mathcal{L}_{N,t} \Psi)\left(\mu^{(t)}_N\right), \ \Psi\left(\mu^{(t)}_N\right)|_{t=0} = \Psi\left(\mu^{(0)}_N\right).
\end{aligned}
\end{equation}

Now, we may consider the equation above under the mean-field limit $N \rightarrow \infty$. Given that $\mu^{(0)}_N$ converges to some $\rho_0$ in law as $N \rightarrow \infty$, we have that $\mu^{(t)}_N$ converges in law to some $\rho_t(\cdot) = \rho(t,\cdot)$ as $N \rightarrow \infty$ for any $t \in [0,1]$, where $\Psi(\rho_t)$ solves the following mean-field backward Kolmogorov equation for any fixed $\Psi \in \mathcal{B}(\mathcal{P}(\mathbb{R}^d), \mathbb{R})$:
\begin{equation}
\label{eqn: limiting backward eqn}
\begin{aligned}
\frac{\partial}{\partial t}\Psi(\rho_t) = (\mathcal{L}_{t} \Psi)(\rho_t), \ \Psi(\rho_t)|_{t=0} = \Psi(\rho_0).    
\end{aligned}
\end{equation}

Above the limiting operator $\mathcal{L}_t = \lim_{N \rightarrow \infty}\mathcal{L}_{N,t}$ is obtained by replacing the empirical measure $\mu_N$ with a generic measure $\rho$, i.e., 
\begin{equation}
\label{eqn: initial limiting generator}
\begin{aligned}
&(\mathcal{L}_{t} \Psi)(\rho) = \int_{\mathbb{R}^d}\Big(-\langle\nabla_{x}(D_{\rho}\Psi)(x), \nabla_{x}U_t(x)\rangle + \Delta_{x}(D_{\rho}\Psi)(x)\Big)\rho(x)\dif x\\
&+ \int_{\mathbb{R}^d}\Bigg(\int_{\mathbb{R}^d}\int_{\frac{1}{a}}^{a}(D_{\rho}\Psi)(\lambda x + (1-\lambda)y)g(\lambda)A_t(x,y,\lambda)\rho(y)\dif\lambda \dif y\Bigg)\rho(x)\dif x\\
&- \int_{\mathbb{R}^d}\Bigg(\int_{\mathbb{R}^d}\int_{\frac{1}{a}}^{a}(D_{\rho}\Psi)(x)g(\lambda)A_t(x,y,\lambda)\rho(y)\dif\lambda \dif y\Bigg)\rho(x)\dif x\\ 
&-\int_{\mathbb{R}^d}(D_{\rho}\Psi)(x)\beta^{(t)}(x,\rho)\rho(x)\dif x +\Big(\int_{\mathbb{R}^d}(D_{\rho}\Psi)(x')\rho(x')\dif x'\Big)\\
&\cdot\Big(\int_{\mathbb{R}^d}\beta^{(t)}(x,\rho)\rho(x)\dif x\Big), 
\end{aligned}    
\end{equation}
for any $\Psi \in \mathcal{B}(\mathcal{P}(\mathbb{R}^d), \mathbb{R})$ and $\rho \in \mathcal{P}(\mathbb{R}^d)$. Now it suffices to show that $\Psi(\rho_t)$ evolves according to equation (\ref{eqn: limiting backward eqn}) when $\rho_t$ solves the target PDE given in Proposition~\ref{prop: PDE of the evolving density}, which requires us to simplify further the expression given in (\ref{eqn: initial limiting generator}). Firstly, we apply integration by parts to rewrite the first term in (\ref{eqn: initial limiting generator}) as below:
\begin{equation}
\label{eqn: rewrite Langevin}
\begin{aligned}
&\int_{\mathbb{R}^d}\Big(-\langle\nabla_{x}D_{\rho}\Psi(x), \nabla_{x}U_t(x)\rangle + \Delta_{x}(D_{\rho}\Psi)(x)\Big)\rho(x)dx\\
&= \int_{\mathbb{R}^d}(D_{\rho}\Psi)(x) \Big(\nabla_{x} \cdot(\nabla_{x}U_t(x)\rho(x)) + \Delta_x\rho(x)\Big)dx.
\end{aligned}    
\end{equation}

Secondly, we consider using a change of variables to rewrite the second term in (\ref{eqn: initial limiting generator}). Note that for fixed $y \in \mathbb{R}^d$, the Jacobian of the transform $(x, \lambda) \rightarrow (z, \alpha)$ defined via $z=\lambda x+(1-\lambda)y$ and $\alpha = \lambda^{-1}$ satisfies $\Big|\det\Big(\frac{\partial(x, \lambda)}{\partial(z,\alpha)}\Big)\Big| = |\alpha^d||\alpha^{-2}| = |\alpha|^{d-2}$. Then we may use $(z,\alpha)$ to rewrite the second term in (\ref{eqn: initial limiting generator}) in the following way:
\begin{equation}
\label{eqn: rewrite stretch move}
\begin{aligned}
&\int_{\mathbb{R}^d}\Big(\int_{\mathbb{R}^d}\int_{\frac{1}{a}}^{a}(D_{\rho}\Psi)(\lambda x + (1-\lambda)y)g(\lambda)A_t(x,y,\lambda)\rho(x)\dif\lambda \dif x\Big)\rho(y)\dif y\\
&= \int_{\mathbb{R}^d}\Big(\int_{\mathbb{R}^d}\int_{\frac{1}{a}}^{a}(D_{\rho}\Psi)(z)g\left(\alpha^{-1}\right)A_t\left(\alpha z+(1-\alpha)y,y,\alpha^{-1}\right)\\
&\cdot \rho(\alpha z+(1-\alpha)y)|\alpha|^{d-2}\dif\alpha \dif z\Big)\rho(y)\dif y\\
&=\int_{\mathbb{R}^d}(D_{\rho}\Psi)(z)\Bigg(\int_{\frac{1}{a}}^{a}|\alpha|^{d-2}g\left(\alpha^{-1}\right)\Big(\int_{\mathbb{R}^d}A_t\left(\alpha z+(1-\alpha)y,y,\alpha^{-1}\right)\\
&\cdot \rho(y)\rho(\alpha z+(1-\alpha)y)\dif y\Big)\dif\alpha\Bigg)\dif z 
\end{aligned}    
\end{equation}

Thirdly, plugging in the expression of the rate function given in (\ref{eqn: rate function}) allows us to rewrite the fourth and fifth terms in (\ref{eqn: initial limiting generator}) as follows:
\begin{equation}
\label{eqn: rewrite birth-death}
\begin{aligned}
&-\int_{\mathbb{R}^d}(D_{\rho}\Psi)(x)\beta^{(t)}(x,\rho)\rho(x)\dif x +\Big(\int_{\mathbb{R}^d}(D_{\rho}\Psi)(x')\rho(x')\dif x'\Big)\\
&\cdot\left(\int_{\mathbb{R}^d}\left(\frac{\partial}{\partial t}U_t(x) -\int_{\mathbb{R}^d}\frac{\partial}{\partial t}U_t(y)\rho(y)\dif y\right)\rho(x)\dif x\right)\\
&= -\int_{\mathbb{R}^d}(D_{\rho}\Psi)(x)\beta^{(t)}(x,\rho)\rho(x)\dif x\\
&= -\int_{\mathbb{R}^d}(D_{\rho}\Psi)(x)\left(\frac{\partial}{\partial t}U_t(x) -\int_{\mathbb{R}^d}\frac{\partial}{\partial t}U_t(y)\rho(y)\dif y\right)\rho(x)\dif x
\end{aligned}
\end{equation}

Finally, by substituting (\ref{eqn: rewrite Langevin}), (\ref{eqn: rewrite stretch move}) and (\ref{eqn: rewrite birth-death}) and switching the variables in (\ref{eqn: rewrite stretch move}), we obtain that
\begin{equation}
\label{eqn: final version limiting generator}
\begin{aligned}
&(\mathcal{L}_t\Psi)(\rho) = \int_{\mathbb{R}^d}(D_{\rho}\Psi)(x)\Bigg(\Big(\nabla_{x} \cdot(\nabla_{x}U_t(x)\rho(x)) + \Delta_x\rho(x)\Big)\\
&+\int_{\frac{1}{a}}^{a}|\lambda|^{d-2}g(\lambda^{-1})\Big(\int_{\mathbb{R}^d}A_t\left(\lambda x +(1-\lambda)y,y,\lambda^{-1}\right)\rho(y)\\
&\rho(\lambda x +(1-\lambda)y)\dif y\Big)\dif\lambda -\Big(\int_{\frac{1}{a}}^{a}g(\lambda)\left(\int_{\mathbb{R}^d}A_t(x,y,\lambda)\rho(y)\dif y\right)\dif\lambda\Big)\rho(x) \\
&-\left(\frac{\partial}{\partial t}U_t(x) -\int_{\mathbb{R}^d}\frac{\partial}{\partial t}U_t(y)\rho(y)\dif y\right)\rho(x)\Bigg)\dif x.
\end{aligned}    
\end{equation}

Comparing (\ref{eqn: final version limiting generator}) with the RHS of the target PDE and plugging in $U_t(x) = (1-c(t))U_0(x) + c(t)U(x)$ indicate that the PDE governing the evolution of $\Psi(\rho_t)$ is indeed (\ref{eqn: limiting backward eqn}) when $\rho_t$ solves the target PDE, as desired. 
\end{proof}

\section*{Appendix C: Extra Numerical Test (Product of Multiple Double Well Distributions)}
\label{example: many well distribution}
\begin{figure}[H]
\centering
\begin{subfigure}[b]{0.475\textwidth}   
\centering 
\includegraphics[width=\textwidth]{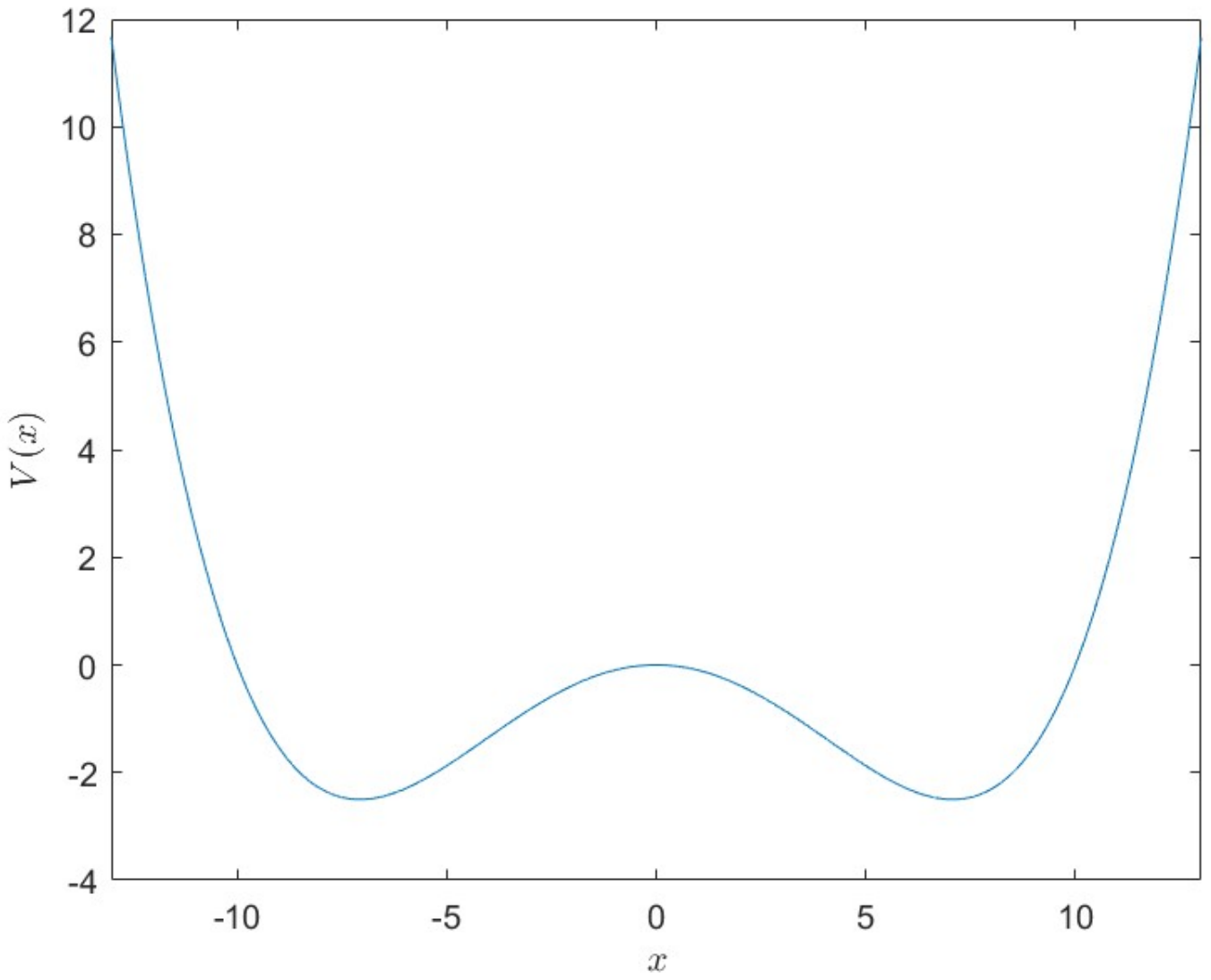}
\caption[]%
{{\small Double-Well Potential Energy $V(x)$}}    
\label{fig: 20D Many Well potential plot small}
\end{subfigure}
\hfill
\begin{subfigure}[b]{0.475\textwidth}   
\centering
\includegraphics[width=\textwidth]{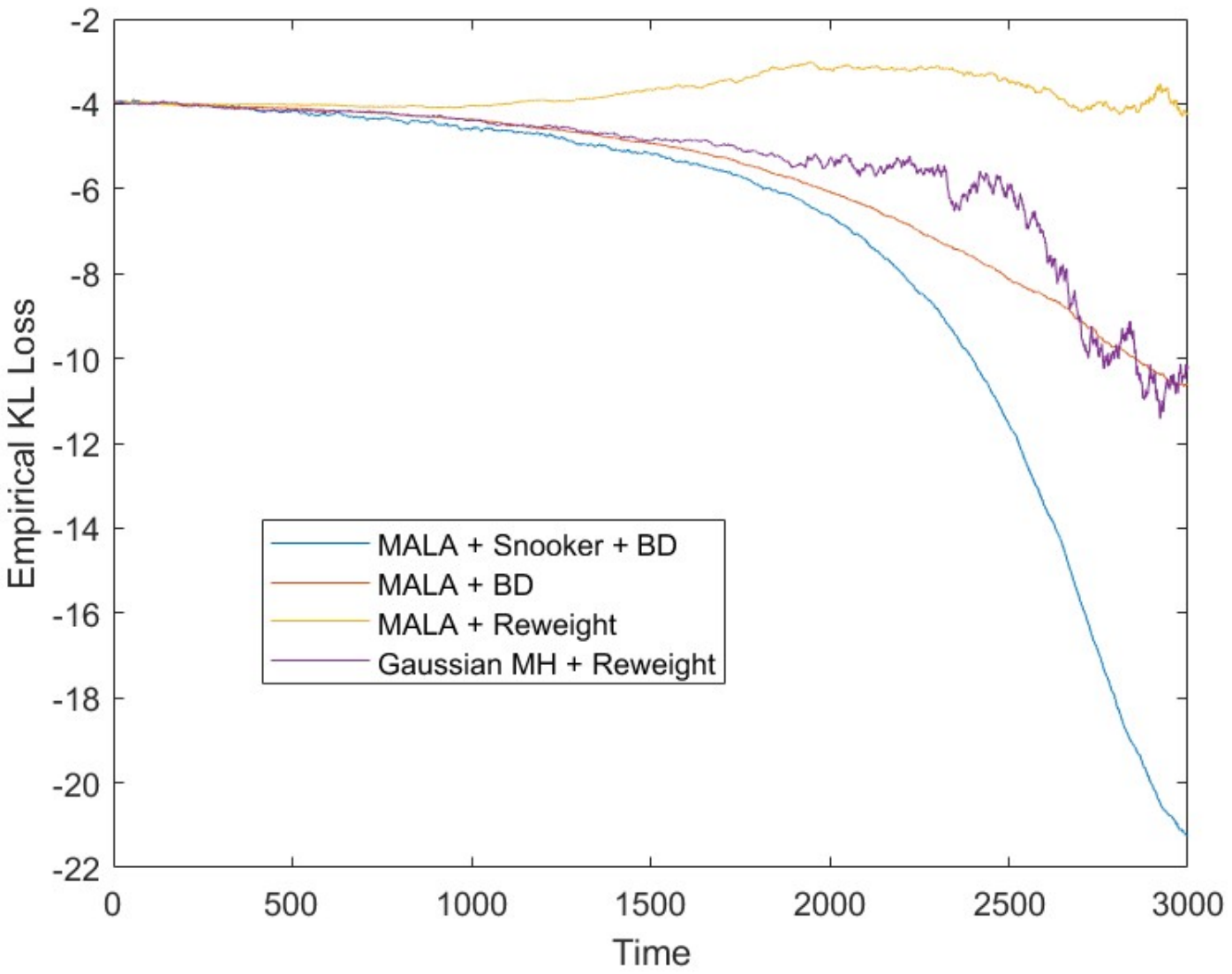} 
\caption{Empirical KL Loss}
\label{fig: 20D Many Well Empirical KL loss} 
\end{subfigure}
\caption[]
{\small (a) Plot of the potential function $V(x)$ with $\beta = 0.001$; (b) Plots of $\mathcal{L}_{\text{KL}}$ evaluated at evolving weighted samples generated by different testing algorithms.} 
\label{fig: 20D Many Well basics}
\end{figure}

In the last subsection, we will present an additional numerical test on continuous distributions to show that IAIS with exploration is capable of searching modes in multimodal distributions. Specifically, we choose the target density to be a product of multiple double-well densities in $\mathbb{R}$, which is of similar form as one of the testing cases used in \cite{midgley2022flow}. Specifically, we first consider a double-well distribution $q:\mathbb{R} \rightarrow [0,1]$ defined as $q(x) \propto \exp(-\beta(x^4-100x^2))$, where $\beta = \frac{1}{T}$ is the inverse of the temperature associated with this model. From the expression of $q(\cdot)$ one can see that it has two modes centered at $x_{+} = 5\sqrt{2}$ and $x_{-} = -5\sqrt{2}$, respectively. This can also be inferred from the plot of the associated potential function $V(x):= \beta(x^4-100x^2)$ in Figure (\ref{fig: 20D Many Well potential plot small}). The target density function in $\mathbb{R}^{20}$ is further defined as $p(\boldsymbol{x}) = p(x_1,x_2,\cdots,x_{20}) \propto \Big(\prod_{j=1}^{10}q(x_j)\Big) \cdot \Big(\prod_{j=11}^{20}\frac{1}{\sqrt{2\pi}}e^{-\frac{1}{2}x_j^2}\Big)$, \emph{i.e.}, the first $10$ coordinates are distributed as $q(\cdot)$ while the last $10$ coordinates are distributed as standard normal random variables. Since each $q(\cdot)$ contains $2$ local modes, a direct computation yields that the target density $p(\cdot)$ has $2^{10} =1024$ local modes. 

\begin{figure}[H]
\centering
\begin{subfigure}{.24\textwidth}
    \centering
    \includegraphics[width=.95\linewidth]{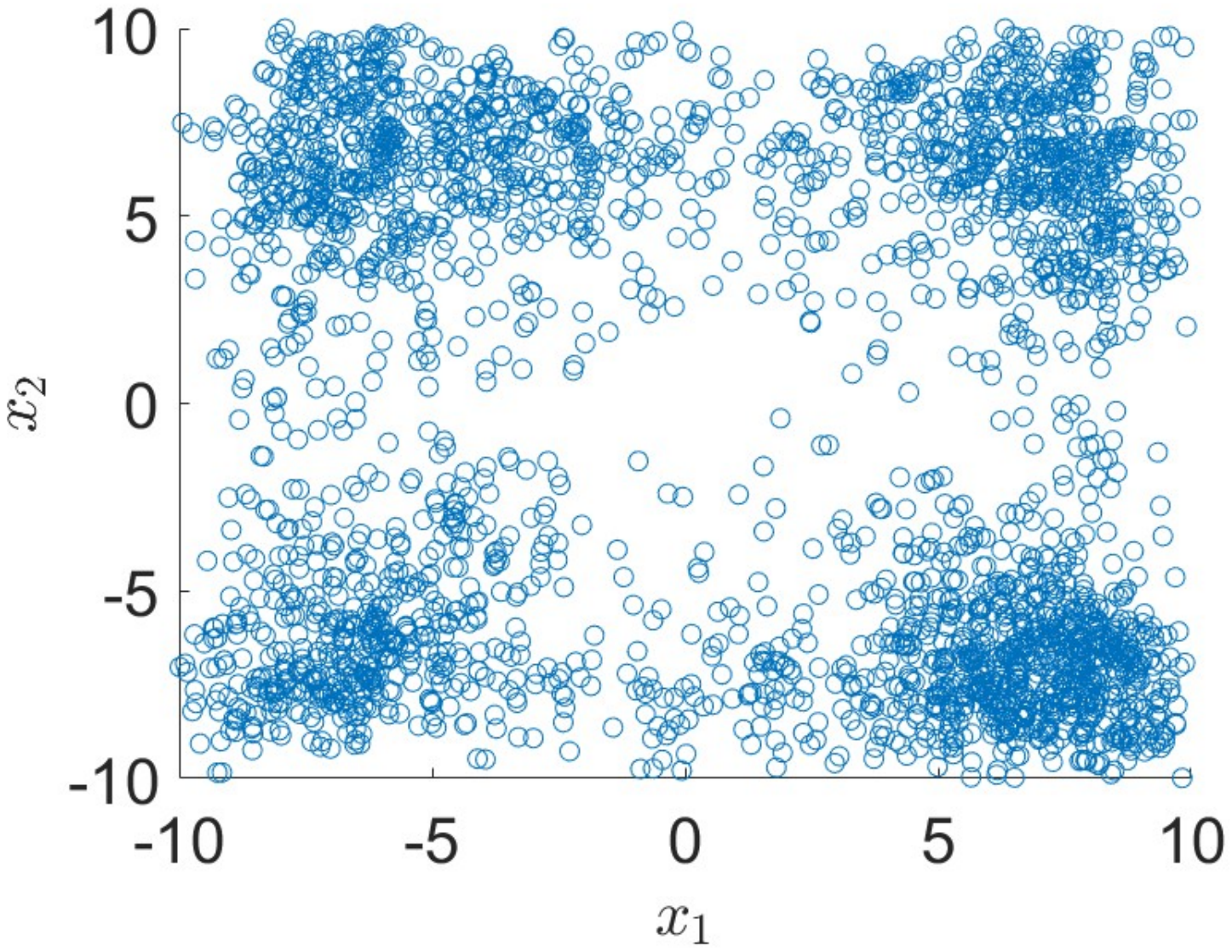} 
    \caption{}
    \label{20D_Many_Well_MALA_Snooker_BD}
\end{subfigure}
\begin{subfigure}{.24\textwidth}
    \centering
    \includegraphics[width=.95\linewidth]{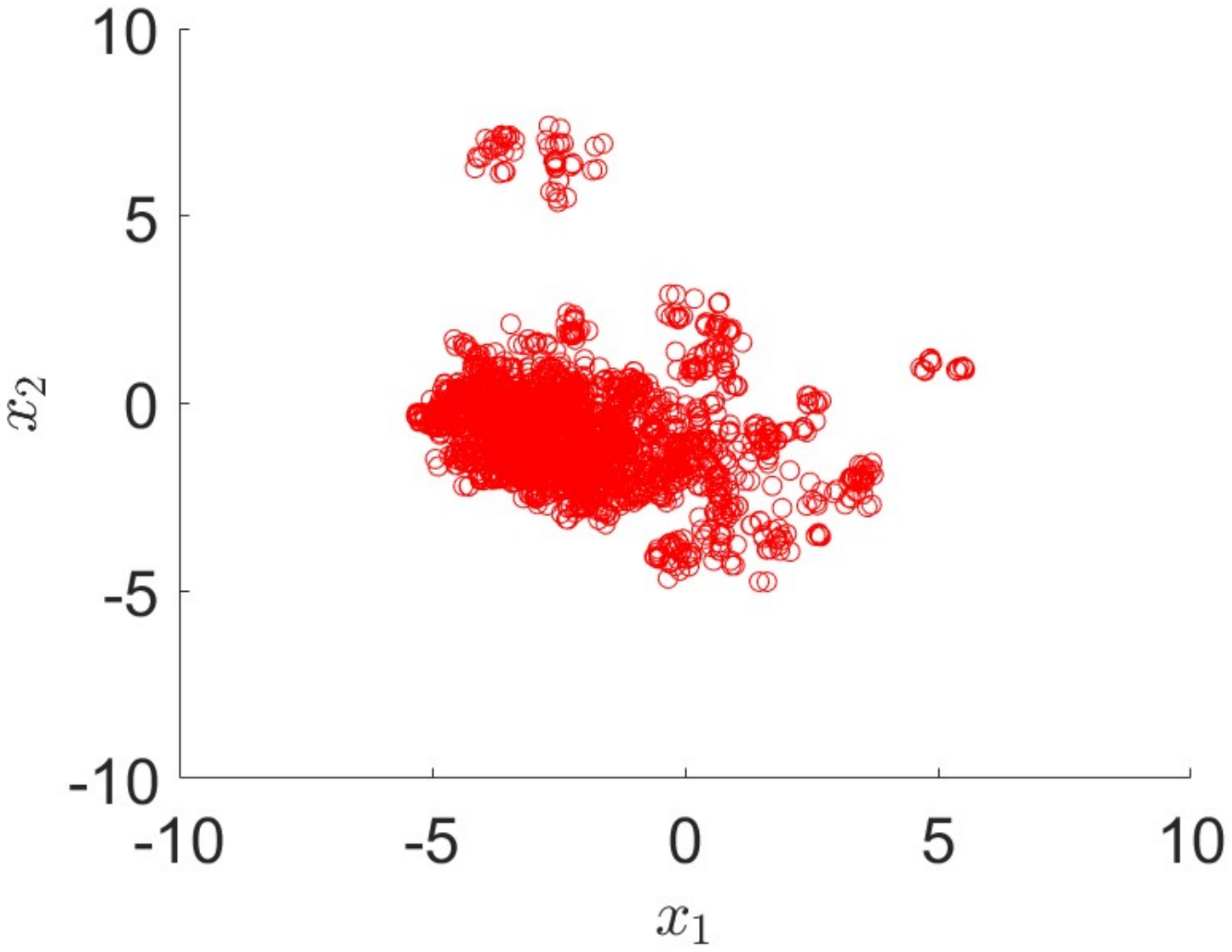}
    \caption{}
    \label{20D_Many_Well_MALA_BD}
\end{subfigure}
\begin{subfigure}{.24\textwidth}
    \centering
    \includegraphics[width=.95\linewidth]{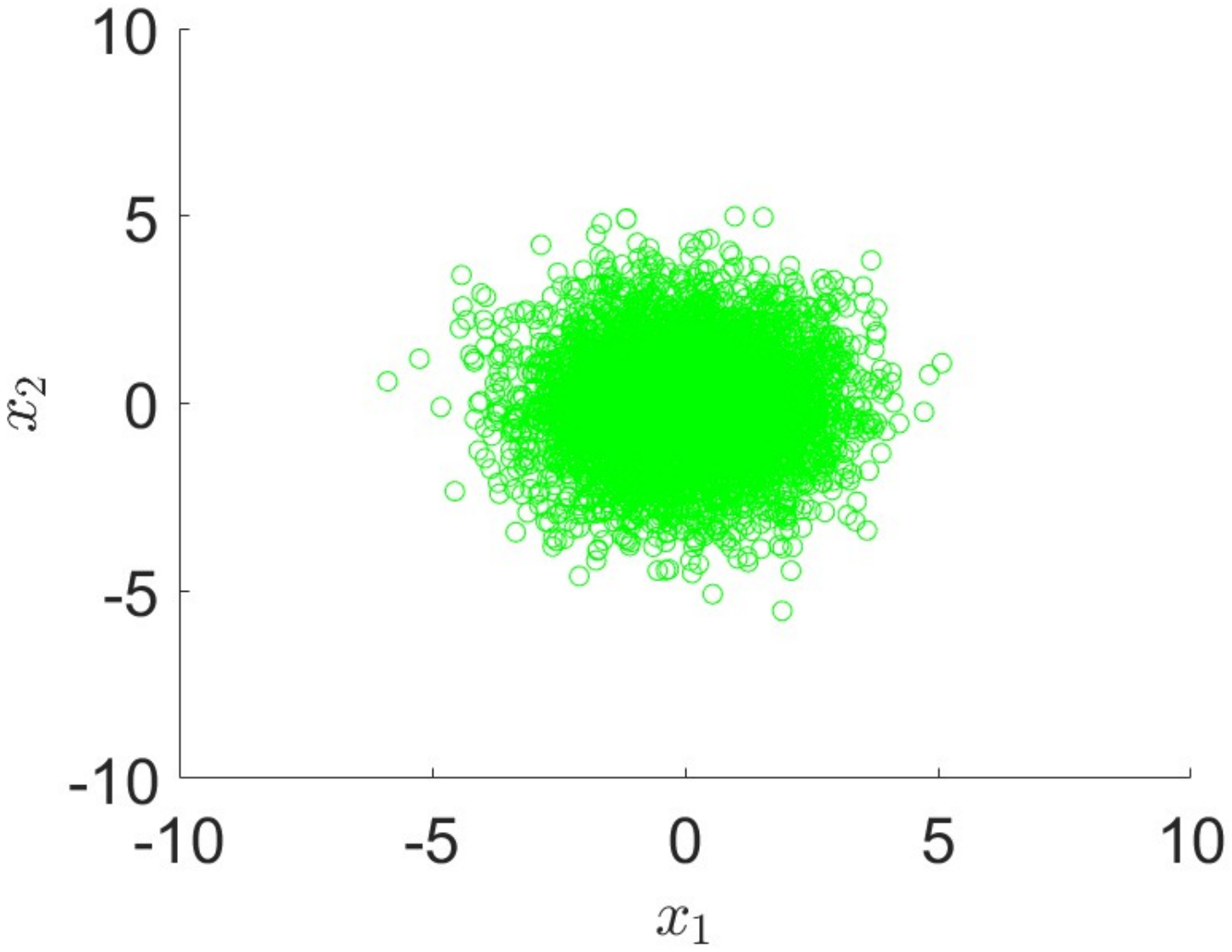}  
    \caption{}
    \label{20D_Many_Well_MALA_Reweight}
\end{subfigure}
\begin{subfigure}{.24\textwidth}
    \centering
    \includegraphics[width=.95\linewidth]{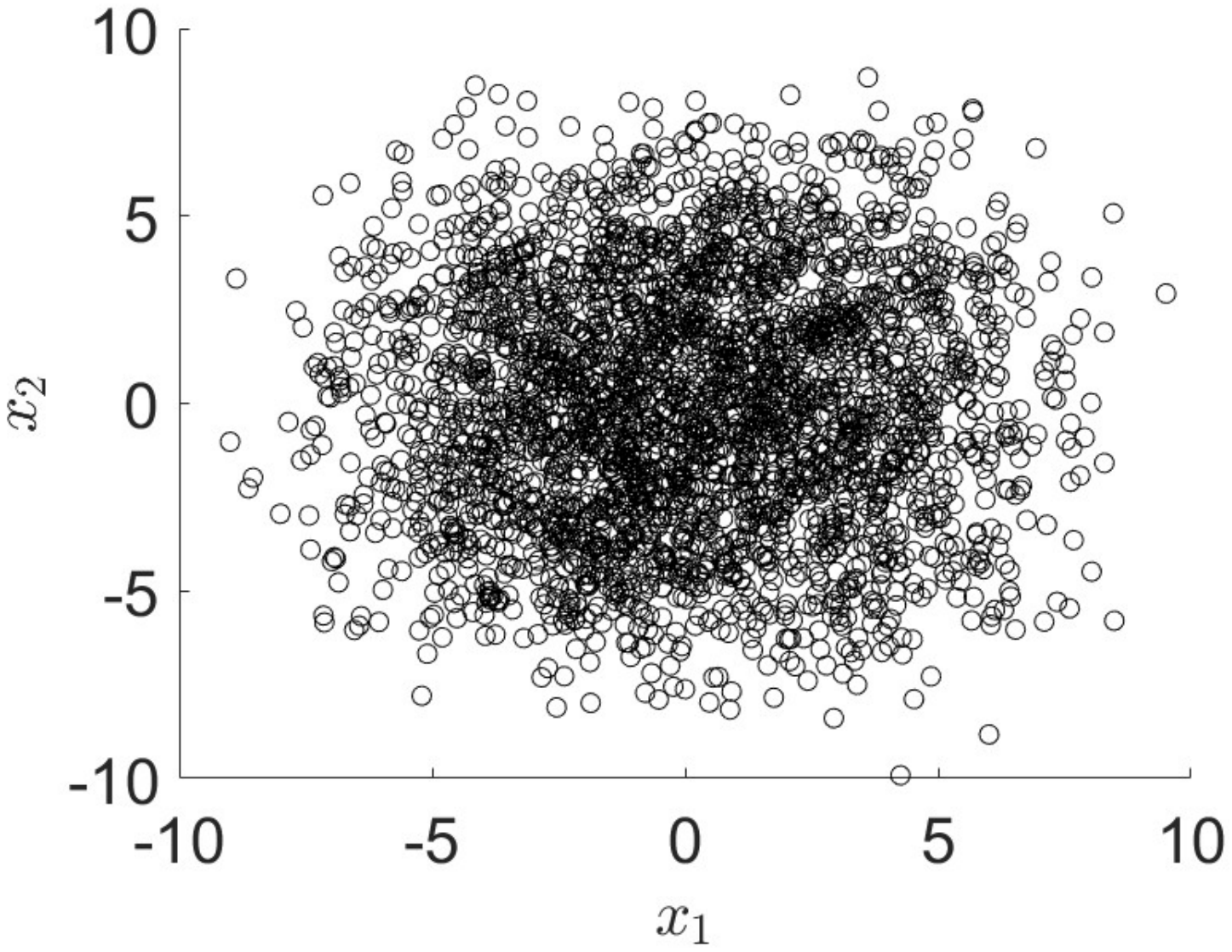}
    \caption{}
    \label{20D_Many_Well_Gaussian_MH_Reweight}
\end{subfigure}
\caption{Scatter plots of the marginal distribution at $(x_1, x_2)$ returned by different testing algorithms; (a) MALA + Snooker + BD; (b) MALA + BD; (c) MALA + Reweight; (d) Gaussian MH + Reweight.}
\label{fig: 20D Many Well scatter plot}
\end{figure}

Following the notations used in the main text, each tested algorithm is referred to as an abbreviation formed by its components. When testing the performance of the four algorithms on this example, which is of relatively higher dimension and contains more local modes, we set the inverse temperature parameter, number of samplers, and time steps to be $\beta = 0.001$, $N=3000$ and $L = 3000$, respectively. Moreover, for the standard AIS whose transition kernel is given by the MH Algorithm, we choose its Gaussian proposal density to be $\mathcal{N}(\boldsymbol{0}, \boldsymbol{I}_{20})$. The starting distribution is selected to be the normal distribution $\mathcal{N}(\boldsymbol{0},\boldsymbol{I}_{20})$ in $\mathbb{R}^{20}$. Again, we track values of the empirical KL loss function evaluated at weighted samples generated by the four testing algorithms with respect to time and plot them in Figure (\ref{fig: 20D Many Well Empirical KL loss}). For the purpose of testing the four algorithms' abilities to discover local modes, we also plot the marginal distribution of $(x_1, x_2)$ for the empirical samples returned by each algorithm, which are listed in Figure (\ref{fig: 20D Many Well scatter plot}).

From the plot of the empirical KL loss functions, we find that Interacting AIS with exploration (Algorithm~\ref{alg: AIS + snooker} in the main text) converges faster and generates samples closer to the target distribution than the other three algorithms. More importantly, from the four scattered plots in Figure (\ref{fig: 20D Many Well scatter plot}) we can see that only IAIS with exploration manages to find all the four local modes $(\pm 5\sqrt{2}, \pm 5\sqrt{2})$ within the marginal distribution of $(x_1,x_2)$, which reveals that the exploration part helps improve the AIS algorithm's ability to discover modes. 

\bibliographystyle{siamplain}
\bibliography{references}

\end{document}